\documentclass{ws-ijmpa}
\usepackage[super,compress]{cite}
\usepackage{graphicx}
\usepackage{amsmath}
\usepackage{hyperref}
\usepackage{mathtools}
\usepackage{slashed}
\usepackage{amssymb}
\usepackage{braket}
\usepackage{scalerel,stackengine}
\usepackage{dsfont}
\usepackage[caption=false]{subfig}

\newcommand*\diff{\mathop{}\!\mathrm{d}}

\newcommand{\nn}{\nonumber}

\newcommand{\be}{\begin{eqnarray}}
\newcommand{\ee}{\end{eqnarray}}
\newcommand{\ma}{\mathrm}
\newcommand{\ml}{\mathcal}
\newcommand{\bs}{\boldsymbol}
\newcommand{\Tr}{\mathrm{Tr}}

\DeclareMathOperator{\sign}{sign}
\newcommand\pig[1]{\scalerel*[5pt]{\big#1}{%
    \ensurestackMath{\addstackgap[1.5pt]{\big#1}}}}
    
\begin{document}
\markboth{Xiaojun Yao}{Open Quantum Systems for Quarkonia}

\catchline{}{}{}{}{}

\title{Open Quantum Systems for Quarkonia}

\author{Xiaojun Yao}

\address{Center for Theoretical Physics, Massachusetts Institute of Technology, 77 Massachusetts Ave, 6-304\\
Cambridge, Massachusetts 02139, United States\\
xjyao@mit.edu}

\maketitle


\begin{abstract}
I review recent applications of the open quantum system framework in the understanding of quarkonium suppression in heavy-ion collisions, which has been used as a probe of the quark-gluon plasma for decades. The derivation of the Lindblad equations for quarkonium in both the quantum Brownian motion and the quantum optical limits and their semiclassical counterparts is explained. The hierarchy of time scales assumed in the derivation is justified from the separation of energy scales in nonrelativistic effective field theories of QCD. Physical implications of the open quantum system approach are also discussed. Finally, I list some open questions for future studies.

\end{abstract}

\ccode{MIT-CTP/5273}

\tableofcontents

\section{Introduction}
Heavy quarkonia are bound states of heavy quark-antiquark pairs ($Q\bar{Q}$). Quarkonium consisting of a charm-anticharm pair ($c\bar{c}$) is named charmonium while that made up of a bottom-antibottom pair ($b\bar{b}$) is named bottomonium. Top quarks decay fast via weak interactions before the formation of strong interaction bound states. Historically, the first experimentally discovered quarkonium state is $J/\psi$~\cite{Aubert:1974js,Augustin:1974xw}, a charmonium ground state with the quantum number $J^{PC} = 1^{--}$. Since the discovery of $J/\psi$, many other quarkonium states have been discovered experimentally. The mass spectra of most ground and lower excited quarkonium states can be well described by nonrelativistic Schr\"odinger equations with simple potentials that account for the two-body interaction between the heavy quark-antiquark pair (see Refs.~\cite{Radford:2007vd,Eichten:2007qx} for recent reviews). One successful form of the potential in phenomenology is the Cornell potential that consists of a short-range Coulomb attraction and a long-range confining potential. However, the Schr\"odinger equation with a simple two-body $Q\bar{Q}$ potential fails to describe the masses of excited quarkonium states close to or above the open heavy meson threshold. These states may be described by hadronic molecules (see Ref.~\cite{Guo:2017jvc} for a recent review). In this review, we will focus on the ground and lower excited quarkonium states.

Heavy quarkonium production in high energy collisions of electrons, hadrons and/or nuclei has been used as a tool for decades to study both the perturbative and nonperturbative aspects of quantum chromodynamics (QCD), which is the theory of the strong interaction. Recent reviews can be found in Refs.~\cite{Andronic:2015wma,Chapon:2020heu}\,. In particular, quarkonium production can be used as probes of hadronic structures or nuclear media. In the latter case, quarkonium production is a particularly useful observable in relativistic heavy-ion collisions, as a probe of the hot and dense nuclear medium produced during the collision, the quark-gluon plasma (QGP). Two major scientific facilities are conducting relativistic heavy-ion collision experiments right now. One is the Relativistic Heavy Ion Collider (RHIC)~\cite{Arsene:2004fa,Adcox:2004mh,Back:2004je,Adams:2005dq} at Brookhaven National Laboratory and the other is the Large Hadron Collider (LHC)~\cite{Muller:2012zq,Aaboud:2018quy,Sirunyan:2018nsz,Acharya:2020kls} at the European Organization for Nuclear Research (CERN). The primary goal of these collider experiments is to search for the QGP and investigate its properties. The QGP created in these experiments has been shown to be an almost perfect fluid, which means it has a very small viscosity and is thus strongly-coupled. Its lifetime is about $10$ fm/c in the laboratory frame and the temperature range currently achieved is about $150-600$ MeV. Recent reviews on quarkonia in the QGP can be found in Refs.~\cite{Mocsy:2013syh,Rothkopf:2019ipj,Sharma:2021vvu}\,.

The idea of using quarkonium production as a probe of the QGP in heavy-ion collisions can be dated back to the early studies of the plasma screening effect on the $Q\bar{Q}$ bound states~\cite{Matsui:1986dk,Karsch:1987pv}. At sufficiently high temperature, the attractive potential between the heavy quark-antiquark pair is significantly suppressed: The confining part of the potential is flattened and the remaining attractive potential is too weak to support the formation of bound states. As a result, quarkonium states ``melt" inside the hot nuclear medium. Distinct quarkonium states have varying sizes and are affected by the plasma screening effect differently. Thus they have different melting temperatures, which are ordered by the binding energies or the sizes: More deeply bound states have smaller sizes and melt at higher temperatures. If the QGP is created in the collision, quarkonium states will become unbound when traversing the medium. Therefore, quarkonium suppression can be used as a signal of the QGP formation in heavy-ion collisions.

However, this simple idea of plasma screening effect can neither describe all experimental measurements, nor be self-consistent theoretically. The essential physics of the potential screening can be seen perturbatively by calculating the finite temperature correction to the quarkonium propagator, which involves the finite temperature contribution to the gluon polarization tensor. In the limit of zero energy transferred, the finite temperature contribution to the gluon polarization tensor is real and non-vanishing. Its value gives (the negative of) the gluon Debye mass which can screen the Coulomb potential and turn it into the Yukawa potential. However, in the case of finite energy transferred, the finite temperature contribution to the gluon polarization tensor can be complex, which means in addition to the screening of the real potential, an extra damping of the quarkonium state occurs~\cite{Laine:2006ns,Beraudo:2007ky,Brambilla:2008cx}. This extra damping leads to quarkonium dissociation and is originated from scattering of quarkonium states with light quarks and gluons in the QGP medium, also known as Landau damping. To distinguish the dynamical dissociation from the suppression caused by the screening of the real potential, we will name the former the dynamical screening effect while the latter the static or Debye screening effect. The static and dynamical screening effects are interrelated. They can be studied together nonperturbatively from lattice calculations of the quarkonium spectral function~\cite{Asakawa:2000tr,Asakawa:2003re,Datta:2003ww,Jakovac:2006sf,Ohno:2011zc,Ding:2012sp,Borsanyi:2014vka,Ikeda:2016czj,Aarts:2014cda,Kim:2014iga,Kim:2018yhk,Larsen:2019bwy,Larsen:2019zqv} or the real-time potential~\cite{Rothkopf:2011db,Burnier:2014ssa,Bazavov:2014kva,Burnier:2015tda,Burnier:2016kqm,Petreczky:2017aiz}.

Quarkonium dissociation can happen when the QGP temperature is below the melting temperature of a quarkonium state, when enough energy is transferred from the medium to the quarkonium state. At the same time, the inverse process of dissociation, recombination, can also happen: An unbound heavy quark-antiquark pair, may radiate out enough energy and forms a bound state. The physical importance of (re)combination was first realized in Refs.~\cite{Thews:2000rj,Andronic:2003zv,Andronic:2007bi} and has been shown later to be crucial to explain experimental measurements (see e.g., Figure 83 in Ref.~\cite{Andronic:2015wma}). Many years of experimental measurements showed that as the collision energy increases, the charmonium state $J/\psi$ becomes less suppressed. Naively, one would expect that the QGP is hotter at higher collision energies and the screening effects are stronger, which leads to more suppression in quarkonium production. The solution to this puzzle is the enhancement of (re)combination. As the collision energy increases, more charm quarks are produced in a single hard collision event. Those unbound charm and antiquark quarks that have little chance to combine into a charmonium state when produced initially in the hard collision, may come close to each other in phase space during their evolution in the QGP medium and bind together. The (re)combination contribution depends on the square of the charm quark density and thus grows fast as the collision energy increases and more charm quarks are produced.

Some phenomenological studies solved a Schr\"odinger equation with a complex potential~\cite{Strickland:2011mw,Strickland:2011aa,Krouppa:2015yoa,Krouppa:2016jcl,Krouppa:2017jlg,Bhaduri:2020lur}, which has no recombiantion contribution. Many other studies used semiclassical transport equations and modeled the recombination contribution~\cite{Grandchamp:2003uw,Grandchamp:2005yw,Yan:2006ve,Young:2008he,Young:2009tj,Liu:2009nb,Zhao:2010nk,Liu:2010ej,Zhao:2011cv,Song:2011xi,Song:2011nu,Sharma:2012dy,Zhou:2014kka,Nendzig:2014qka,Du:2015wha,Petreczky:2016etz,Zhou:2016wbo,Chen:2017duy,Zhao:2017yan,Du:2017qkv,Aronson:2017ymv,Yao:2017fuc,Yao:2018zze,Ferreiro:2018vmr,Du:2019tjf,Chen:2019qzx}. Model dependence in the implementation of recombination results in large systematic uncertainty in the calculations. To overcome this issue, a consistent theoretical framework accounting for both plasma screening effects and recombination is necessary. The open quantum system approach provides such a framework, which is the main topic of this review. We will explain in detail the application of the open quantum system framework to study quarkonium transport in a hot nuclear environment, i.e., the QGP. Lindblad equations~\cite{gorini1976completely,lindblad1976generators} in both the quantum Brownian motion and quantum optical limits will be derived and discussed. Recently there has been a review on this topic that covers many aspects~\cite{Akamatsu:2020ypb}. Here we will try to be complementary with a focus on the field theoretical aspects of the approach. We will show how the separation of energy scales and nonrelativistic effective field theories (EFT) deepen our understanding of quarkonium transport in the hot nuclear medium. One highlight will be an all-order (in the coupling constant) construction of the Lindblad equation at leading (nontrivial) power in the EFT power counting. The relation between the Lindblad equations for quarkonium and their semiclassical counterparts such as the Boltzmann equation and the Fokker-Planck equation will also be elucidated. Most importantly, new physical insights gained from studies using the open quantum systems will be discussed, as well as a new experimental observable proposed to test the insight. The impact of a theoretical framework would be limited, if no new physics could be learned from it.

This review is organized as follows: In Section~\ref{sect:open}, the basics of the open quantum system framework will be introduced. The Lindblad equations in both the quantum Brownian motion and quantum optical limits will be shown. The assumptions of hierarchical time scales in these two limits will also be illuminated. Then in Section~\ref{sect:eft}, we will explain the separation of energy scales in quarkonium and introduce the nonrelativistic effective field theories to be used in the following sections. The Lindblad equations for quarkonium will be discussed in Section~\ref{sect:lindblad}, together with their semiclassical limits and other phenomenological approaches. Furthermore, we will discuss the phenomenological implications of the open quantum system approach and the physical insights gained in Section~\ref{sect:pheno}. Finally, a brief summary is given in  Section~\ref{sect:conclusion}, with a discussion of some open questions.

\section{Open Quantum Systems}
\label{sect:open}
In this section, we will review the basics of the open quantum system framework. Detailed coverage of the topic can be found, for example in Refs.~\cite{Breuer:2002pc,Weiss,Schaller:2014}\,. We are considering a subsystem interacting with an environment. The dynamics of the whole system, consisting of the subsystem and the environment is governed by the total Hamiltonian
\be
H = H_S + H_E + H_I \,,
\ee
where $H_S$ represents the subsystem Hamiltonian, $H_E$ denotes the environment Hamiltonian, and $H_I$ contains the interaction between the subsystem and the environment. We assume the interaction Hamiltonian is of the form: 
\be
\label{eqn:HI}
H_I = \sum_{\alpha} O^{(S)}_{\alpha} \otimes O^{(E)}_{\alpha} \,,
\ee
where $\alpha$ denotes all continuous and discrete variables that the operators depend on.\footnote{For quantum field theory, the operators are fields that are functions of spatial coordinates. So the variable symbol $\alpha$ includes both discrete quantum numbers such as spin and/or color and continuous spatial coordinates. For the latter, the summation over $\alpha$ should be understood as an integration over positions. In short, for a quantum field theory in $d$ dimension, the form of the interaction Hamiltonian can be written as
\be
H_I =  \sum_{\alpha'} \int \diff^d x\,  O^{(S)}_{\alpha'}({\bs x})
O^{(E)}_{\alpha'}({\bs x})\,,\nn
\ee
where $\alpha'$ denotes only the discrete variables.} The tensor product emphasizes that the subsystem and environment operators act on states in different spaces and can be dropped when the meaning is clear.

The time evolution of the whole system is governed by the von-Neumann equation, which in the Schr\"odinger picture is written as:
\be
\frac{\diff \rho(t)}{\diff t} = -i \big[ H, \rho(t) \big] \,,
\ee
where $\rho(t)$ is the density matrix of the whole system at time $t$. A more convenient picture for time ordered perturbation theory is the interaction picture,\footnote{One way to derive the Feynman rules for quantum field theory is to use the interaction picture and Wick's theorem. We will not discuss the subtlety of Haag's theorem involved in the interaction picture here.} which is defined by
\be
\rho^{(\text{int})}(t) &=& e^{i (H_S+H_E) t} \rho(t) e^{-i (H_S+H_E) t} \\
H^{(\text{int})}_I(t) &=& e^{i (H_S+H_E) t} H_I e^{-i (H_S+H_E) t} \,.
\ee
In the interaction picture, the time evolution equation becomes
\be
\frac{\diff \rho^{(\text{int})}(t)}{\diff t} = -i \big[ H_I^{(\text{int})}, \rho^{(\text{int})}(t) \big] \,.
\ee
The formal solution is given by
\be
\label{eqn:evolution}
\rho^{(\text{int})}(t) &=& U(t) \rho^{(\text{int})}(0) U^{\dagger}(t) \\
U(t) &=& \ml{T} \exp\Big(-i \int_0^t H_I^{(\text{int})}(t') \diff t'\Big) \,,
\ee
in which $\ml{T}$ is the time-ordering operator. The time evolution of the subsystem can be obtained from Eq.~(\ref{eqn:evolution}) by tracing out the environment degrees of freedom:
\be
\label{eqn:reduced}
\rho_S^{(\text{int})}(t) = \Tr_E \pig( \rho^{(\text{int})}(t) \pig) = \Tr_E \pig( U(t) \rho^{(\text{int})}(0) U^{\dagger}(t) \pig) \,.
\ee
Before we carry out detailed calculations, we first investigate some properties of the reduced evolution equation of $\rho_S^{(\text{int})}(t)$. First, the reduced evolution equation (\ref{eqn:reduced}) preserves the trace of $\rho^{(\text{int})}_S(t)$: $\Tr\rho_S^{(\text{int})}(t)=\Tr\rho_S^{(\text{int})}(0) = \Tr\rho^{(\text{int})}(0)$, which means the total probability of all states in the subsystem is conserved. This has important physical implications for quarkonium transport in nuclear media. As discussed in the Introduction, dissociation leads to probability loss of quarkonium states. But after the dissociation of a quarkonium state, the unbound heavy quark-antiquark pair stays as an active degree of freedom and must be included in the following time evolution to preserve the total probability of the subsystem. Here the total probability is equivalent to the total number of heavy quark-antiquark pairs.\footnote{\label{ft:anni}The annihilation rate of the heavy quark-antiquark pair in quarkonium is on the order of $100$ keV and the annihilation effect can be neglected during the lifetime of the QGP, which is about $\sim10$ fm/c in the laboratory frame.} Many phenomenological studies do not keep track of the unbound $Q\bar{Q}$ pair after dissociation, which makes it difficult for these studies to take into account recombination consistently.

The second property of Eq.~(\ref{eqn:reduced}) is its time irreversibility, which is closely related to the partial trace $\Tr_E$. This connection can be explained by using the relative entropy between two quantum states. A short explanation can be found in~\ref{app:rel_entro} (see Ref.~\cite{Breuer:2002pc} for more details). 

Now we expand $U(t)$ to second order in $H^{(\text{int})}_I$ to obtain
\be
\label{eqn:finite_t}
\rho^{(\text{int})}_S(t) &=& \rho^{(\text{int})}_S(0) - i \int_0^t  \diff t' \,\Tr_E \pig( \big[ H^{(\text{int})}_I(t'), \rho^{(\text{int})}(0) \big] \pig) \nn\\
&+& \int_0^t \diff t_1 \int_0^t \diff t_2 \,\Tr_E \pig(H^{(\text{int})}_I(t_1)\rho^{(\text{int})}(0)H^{(\text{int})}_I(t_2) \nn\\
&-& \theta(t_1-t_2)H^{(\text{int})}_I(t_1)H^{(\text{int})}_I(t_2)\rho^{(\text{int})}(0) \nn\\
&-& \theta(t_2-t_1)\rho^{(\text{int})}(0)H^{(\text{int})}_I(t_1)H^{(\text{int})}_I(t_2)\  \pig) + \ml{O}\big((tH^{(\text{int})}_I)^3\big) \,.
\ee
So far we only expand the formal solution and have not made any assumption yet. The expansion does not require weak coupling (i.e., $H^{(\text{int})}_I$ is small) at this stage since the expansion parameter is in fact $tH^{(\text{int})}_I$. One can always find a sufficiently small time step $t$ such that the expansion is valid. Later we will show that the finite-difference equation (\ref{eqn:finite_t}) will turn to a well-defined differential equation in the weak-coupling limit.

The first assumption we make is the factorization of the initial density matrix:
\be
\label{eqn:factorize}
\rho^{(\text{int})}(t=0) = \rho_S^{(\text{int})}(t=0) \otimes \rho_E^{(\text{int})}(t=0) \,,
\ee
which means no correlation between the subsystem and the environment initially. The differential equation for the time evolution of $\rho_S(t)$ derived under the condition (\ref{eqn:factorize}) is rigorously only valid at $t=0$. To infer the applicability of the differential equation for later times $t>0$, one needs to assume that the factorization (\ref{eqn:factorize}) is not only valid at $t=0$, but also when $t>0$. The factorization (\ref{eqn:factorize}) when $t>0$ is valid if the subsystem-environment coupling ($H_I$) is weak and the environment is large so that the change of the environment is negligible.

Since our main interest is quarkonium transport in the QGP created in heavy-ion collisions, which is a hot nuclear medium close to thermal equilibrium for most of its lifetime, we will assume the environment density matrix is thermal and time-independent:\footnote{The QGP created in heavy-ion collisions is expanding and cooling. Thus the QGP temperature varies with spacetime. We will assume the spacetime variation of the QGP temperature is much slower than the in-medium dynamics of quarkonium. Locally, the QGP medium can be treated as a static environment.}
\be
\rho_E^{(\text{int})}(t) = \rho_E = \frac{e^{-\beta H_E}}{\Tr_E e^{-\beta H_E}} \,.
\ee
Here $\beta = 1/T$ and $T$ is the temperature of the thermal environment.
Using $2\theta(t) = 1+\sign(t)$ and
\be
H_I^{(\text{int})}(t) = e^{i(H_S+H_E)t} \Big( \sum_\alpha O^{(S)}_\alpha \otimes O_\alpha^{(E)} \Big) e^{-i(H_S+H_E)t} = \sum_\alpha O^{(S)}_\alpha(t) \otimes O^{(E)}_\alpha(t) \,,\ \ \ \ 
\ee
and defining the environment correlators (one- and two-point functions)
\be
D_\alpha(t) &=& \Tr_E \pig( \rho_E\, O^{(E)}_\alpha(t) \pig) \\
\label{eqn:2point}
D_{\alpha\beta}(t_1,t_2) &=& \Tr_E \pig( \rho_E\, O^{(E)}_\alpha(t_1) O^{(E)}_\beta(t_2) \pig)\,,
\ee
we find Eq.~(\ref{eqn:finite_t}) can be written as
\be
\label{eqn:pre-lindblad}
\rho_S^{(\text{int})}(t) &=& \rho_S^{(\text{int})}(0) - i\int_0^t \diff t' \sum_\alpha D_\alpha(t')
\pig[ O^{(S)}_\alpha(t'), \rho^{(\text{int})}_S(0) \pig]  \nn\\
&-& \int_0^t\diff t_1 \int_0^t\diff t_2 \frac{\sign(t_1-t_2)}{2} \sum_{\alpha,\beta} D_{\alpha\beta}(t_1,t_2) \pig[O^{(S)}_\alpha(t_1)O^{(S)}_\beta(t_2), \rho_S^{(\text{int})}(0) \pig]  \nn\\
&+& \int_0^t\diff t_1 \int_0^t\diff t_2 \sum_{\alpha,\beta} D_{\alpha\beta}(t_1,t_2) \Big( O^{(S)}_\beta(t_2) \rho_S^{(\text{int})}(0) O^{(S)}_\alpha(t_1) \nn\\
&-& \frac{1}{2} \pig\{O^{(S)}_\alpha(t_1)O^{(S)}_\beta(t_2), \rho_S^{(\text{int})}(0) \pig\} \Big) + \ml{O}\big((tH^{(\text{int})}_I)^3\big)  \,.
\ee
The one-point function $D_\alpha(t)$ is vanishing in many cases of interest. For example, if the environment operator is the gauge field in QCD, $O^{(E)}_\alpha(t) = A_\mu^a(t,{\bs x})$ (here in this example, the $\alpha$ symbol contains the adjoint color index $a$, the Lorentz index $\mu$ and the spatial coordinate ${\bs x}$) and the thermal QGP is overall color neutral, the one-point function $\Tr_E(\rho_E A_\mu^a(t,{\bs x}) )$ vanishes. For cases with nonvanishing one-point functions $D_\alpha(t)\neq0$, we can remove the term containing the one-point function by redefining 
\be
O^{(E)}_\alpha &\to& O^{(E)}_\alpha - \Tr_E(\rho_E O^{(E)}_\alpha) \\
\label{eqn:redefine}
H_S &\to& H_S + \sum_\alpha \Tr_E(\rho_E O^{(E)}_\alpha) O^{(S)}_\alpha \,.
\ee
Then the definition of the interaction picture is different due to the extra term in Eq.~(\ref{eqn:redefine}). From now on, we will drop the term containing the one-point function in Eq.~(\ref{eqn:pre-lindblad}).

The expression (\ref{eqn:pre-lindblad}) is a finite-difference equation rather than a differential equation. To convert it into a differential equation, we need to divide Eq.~(\ref{eqn:pre-lindblad}) by $t$ and investigate whether the $t\to0$ limit can be well-defined. Since we have dropped the term containing the one-point function, the nontrivial parts of the right hand side (the last three lines) of Eq.~(\ref{eqn:pre-lindblad}) seem to scale as $t^2$. After the division by $t$, it seems that the last three lines vanish linearly as $t\to0$. In the following we will discuss two examples where the existence of the $t\to0$ limit can be shown under certain approximations. These approximations are valid under hierarchies of time scales. We will now explain the relevant time scales that show up in Eq.~(\ref{eqn:pre-lindblad}).

\subsection{Relevant Time Scales}
\label{sect:time_scale}
The expansion parameter in Eq.~(\ref{eqn:pre-lindblad}) is $t H_I^{(\text{int})} $. For the validity of the expansion, we require the time step $t$ to be much smaller than the inverse of the interaction rate between the subsystem and the environment or equivalently, the mean free time.
Since the environment is thermal, the interaction gradually drives the subsystem to equilibrium. So the {\it subsystem} has a typical {\it relaxation time} $\tau_R$.\footnote{We follow the standard in the literature to use the term ``relaxation time". But what we really mean here is the inverse of the interaction rate between the subsystem and the environment.} At second order in perturbation theory, we expect
\be
\tau_R \sim \frac{T}{\big(H_I^{(\text{int})} \big)^2}\,,
\ee
where the environment temperature $T$ is inserted for the correct dimension. The factor $T$ can be generated from the phase space integration of a thermal distribution that comes from $D_{\alpha\beta}(t_1,t_2)$.

The second relevant time scale, the {\it subsystem intrinsic time} $\tau_S$, is determined by the typical energy gaps between states in the subsystem:
\be
\tau_S \sim\frac{1}{H_S}\,,
\ee
where $H_S$ should be interpreted as the typical eigenenergy (gap) of the subsystem.

Finally, we study the time scale involved in the environment correlator $D_{\alpha\beta}(t_1,t_2)$, defined in Eq.~(\ref{eqn:2point}). We assume the environment is invariant under time translation and define the Fourier transform\footnote{The environment correlator $D(\omega)$ is Hermitian in the frequency space $D^*_{\alpha\beta}(\omega) = D_{\beta\alpha}(\omega)$. The factor ``$i$" in the definition of $\Sigma_{\alpha\beta}(\omega)$ is introduced such that $\Sigma^*_{\alpha\beta}(\omega) = \Sigma_{\beta\alpha}(\omega)$.}
\be
\label{eqn:fourier}
D_{\alpha\beta}(t_1,t_2) = D_{\alpha\beta}(t_1-t_2) &=& \int \frac{\diff\omega}{2\pi} \, e^{-i\omega(t_1-t_2)} D_{\alpha\beta}(\omega) \\
\label{eqn:fourier2}
\sign(t_1-t_2) D_{\alpha\beta}(t_1-t_2) &=& i\int \frac{\diff\omega}{2\pi} \, e^{-i\omega(t_1-t_2)} \Sigma_{\alpha\beta}(\omega) \,.
\ee
The definition of $D_{\alpha\beta}(t_1,t_2)$ includes a thermal environment density matrix. The typical value of $\omega$ is naturally $T$, the environment temperature.\footnote{Other thermal scales such as the Debye mass can be generated and alter the scaling of the environment correlation time. We will discuss this situation in Section.~\ref{sect:lindblad}.} This gives another time scale, the {\it environment correlation time} $\tau_E$, which can be estimated by
\be
\tau_E \sim \frac{1}{T} \,.
\ee
Its physical meaning can be seen by considering the situation with $t_1-t_2\gg \tau_E \sim\frac{1}{\omega}$. The phase in Eq.~(\ref{eqn:fourier}) oscillates rapidly and thus the environment correlator vanishes by the Riemann–Lebesgue lemma.

We have introduced three time scales: the subsystem relaxation time, the subsystem intrinsic time and the environment correlation time. We will estimate these time scales for quarkonium in the QGP later in Section~\ref{sect:lindblad}. The two approximations that we are going to consider below are specified by the separation of these three time scales. The limit of {\it quantum Brownian motion} is specified by
\be
\tau_R &\gg& \tau_E \\
\tau_S &\gg& \tau_E \,,
\ee
while the {\it quantum optical} limit is described by
\be
\tau_R &\gg& \tau_E \\
\tau_R &\gg& \tau_S \,.
\ee
One hierarchy is common in these two limits: $\tau_R \gg \tau_E$, which is valid if $T\gg H_I^{(\text{int})}$. This hierarchy corresponds to the Markovian approximation: During the typical time period of the subsystem relaxation, the environment correlation has been lost. As we will see later, the quantum evolution equations derived in these two limits only depend on the current state of the subsystem. The Markovian condition is generally true when the subsystem and the environment are weakly coupled.
These two limits are not mutually exclusive. In the case with $\tau_R \gg \tau_S \gg \tau_E$, both limits are valid. However, they provide different ways of approximating Eq.~(\ref{eqn:pre-lindblad}). We will now explain each limit in detail.

\subsection{Quantum Brownian Motion}
\label{sect:brown}
With the Fourier transforms (\ref{eqn:fourier}, \ref{eqn:fourier2}), Eq.~(\ref{eqn:pre-lindblad}) can be written as
\be
\label{eqn:pre-brown}
\rho_S^{(\text{int})}(t) &=& \rho_S^{(\text{int})}(0) \nn\\
&-& \int_0^t\diff t_1 \int_0^t\diff t_2 \int\frac{\diff\omega}{2\pi}\,e^{-i\omega(t_1-t_2)} \frac{i}{2} \sum_{\alpha,\beta} \Sigma_{\alpha\beta}(\omega) \pig[O^{(S)}_\alpha(t_1)O^{(S)}_\beta(t_2), \rho_S^{(\text{int})}(0) \pig]  \nn\\
&+& \int_0^t\diff t_1 \int_0^t\diff t_2 \int\frac{\diff\omega}{2\pi}\,e^{-i\omega(t_1-t_2)} \sum_{\alpha,\beta} D_{\alpha\beta}(\omega) \Big( O^{(S)}_\beta(t_2) \rho_S^{(\text{int})}(0) O^{(S)}_\alpha(t_1) \nn\\
&-& \frac{1}{2} \pig\{O^{(S)}_\alpha(t_1)O^{(S)}_\beta(t_2), \rho_S^{(\text{int})}(0) \pig\} \Big) \,.
\ee
We will investigate the time integrals. One explicit example is:
\be
\label{eqn:time_integral}
\int_0^t \diff t_1 \,e^{-i\omega t_1} O_\alpha^{(S)}(t_1) = \int_0^t \diff t_1 \,e^{-i\omega t_1} e^{iH_St_1} O_\alpha^{(S)}(0)\, e^{-iH_St_1} \,.
\ee
Now we use one of the two hierarchies specifying the quantum Brownian motion limit: $\tau_S \gg \tau_E$. Since $T$ is the typical value of $\omega$ and 
\be
\tau_E \sim \frac{1}{T} \sim \frac{1}{\omega} \ \ \ \ \ \ \ \ \ \ 
\tau_S \sim \frac{1}{H_S}\,,
\ee
we have $H_S \ll \omega $.
Therefore we can expand $H_S$ when compared with $\omega$ in Eq.~(\ref{eqn:time_integral}). At leading order (zeroth order in $H_S$), the time integral (\ref{eqn:time_integral}) becomes
\be
\label{eqn:time_integral1}
O_\alpha^{(S)}(0) \int_0^t \diff t_1 \,e^{-i\omega t_1}  = O_\alpha^{(S)}(0) \,2 e^{-i \omega t /2 } \frac{\sin( \frac{\omega t}{2})}{\omega}\,.
\ee
The next-leading order (linear order in $H_S$) term in the time integral (\ref{eqn:time_integral}) is
\be
\label{eqn:time_integral2}
i\big[H_S, O_\alpha^{(S)}(0) \big] \int_0^t t_1\diff t_1 \,e^{-i\omega t_1}
= - \big[H_S, O_\alpha^{(S)}(0) \big] \frac{\partial}{\partial\omega}\Big( 2 e^{-i \omega t /2 } \frac{\sin( \frac{\omega t}{2})}{\omega} \Big)
\,.
\ee
The next-leading order term represents the effect of quantum dissipation~\cite{Akamatsu:2020ypb}. For heavy quarks in a weakly-coupled QGP, the dissipation is originated from their recoil when they scatter with light quarks and gluons in the QGP.

Next we will use the second hierarchy $\tau_R \gg \tau_E$, i.e., the Markovian approximation. Because of $\tau_R \gg \tau_E$, we can always choose a time step $t$ in Eq.~(\ref{eqn:pre-brown}) such that $\tau_R \gg t \gg \tau_E$. Since $\tau_E \sim \frac{1}{T} \sim \frac{1}{\omega}$, we have $t \gg \frac{1}{\omega}$ and can evaluate Eqs.~(\ref{eqn:time_integral1}, \ref{eqn:time_integral2}) in the limit $t\to+\infty$. Using
\be
\label{eqn:sinc}
\lim_{a\to+\infty}\frac{\sin(ax)}{x} = \pi\delta(x)\,,
\ee
we find Eq.~(\ref{eqn:time_integral1}) turns to
\be
\label{eqn:delta0}
\lim_{t\to+\infty}\int_0^t \diff t_1 \,e^{-i\omega t_1} = 2\pi \delta(\omega) \,,
\ee
which means only the zero frequency limit of $\Sigma_{\alpha\beta}(\omega)$ and $D_{\alpha\beta}(\omega)$ contribute in the quantum Brownian motion limit. Now we are ready to write out the Lindblad equation in the quantum Brownian motion limit.

\subsubsection{Leading Order in $H_S$}
Taking the leading order term in the expansion of $H_S$ and using Eq.~(\ref{eqn:delta0}), we find Eq.~(\ref{eqn:pre-brown}) can be written as
\be
\label{eqn:pre-brown_lo}
\rho_S^{(\text{int})}(t) &=& \rho_S^{(\text{int})}(0) - t \frac{i}{2} \sum_{\alpha,\beta} \Sigma_{\alpha\beta}(\omega=0) \pig[O^{(S)}_\alpha(0)O^{(S)}_\beta(0), \rho_S^{(\text{int})}(0) \pig]  \nn\\
&+& t \sum_{\alpha,\beta} D_{\alpha\beta}(\omega=0) \Big( O^{(S)}_\beta(0) \rho_S^{(\text{int})}(0) O^{(S)}_\alpha(0) \nn\\
&-& \frac{1}{2} \pig\{O^{(S)}_\alpha(0)O^{(S)}_\beta(0), \rho_S^{(\text{int})}(0) \pig\} \Big) \,,
\ee
where the factor of $t$ is generated trivially from the integral over $t_2$ after we plug Eq.~(\ref{eqn:delta0}) into Eq.~(\ref{eqn:pre-brown}) (one can also see this by applying Eq.~(\ref{eqn:delta0}) to the integral over $t_2$ and use $2\pi\delta(0) = t$). Dividing Eq.~(\ref{eqn:pre-brown_lo}) by $t$ and taking the limit $t\to0$, we find
\be
\label{eqn:brown}
\frac{\diff \rho_S^{(\text{int})}(t)}{\diff t}\bigg|_{t=0} &=& -\frac{i}{2} \pig[ \sum_{\alpha,\beta} \Sigma_{\alpha\beta}(\omega=0) O^{(S)}_\alpha(0) O^{(S)}_\beta(0) , \rho_S^{(\text{int})}(0)\pig] \nn\\
&+& \sum_{\alpha,\beta} D_{\alpha\beta}(\omega=0) \Big( O^{(S)}_\beta(0) \rho_S^{(\text{int})}(0) O^{(S)}_\alpha(0) \nn\\
&-& \frac{1}{2} \pig\{O^{(S)}_\alpha(0)O^{(S)}_\beta(0), \rho_S^{(\text{int})}(0) \pig\} \Big) \,.
\ee
It seems we are making contradictory approximations: In Eq.~(\ref{eqn:delta0}) we take the limit $t\to+\infty$ while in Eq.~(\ref{eqn:brown}) we take the limit $t\to0$. In fact, these two limits are not contradictory. As discussed earlier, under the Markovian approximation, (i.e., the hierarchy $\tau_R \gg \tau_E$), we can always choose a time step $t$ such that $\tau_R \gg t \gg \tau_E$. In Eq.~(\ref{eqn:delta0}), we are using the part $t \gg \tau_E$ to approximate the time integral, while in Eq.~(\ref{eqn:brown}), we are using the part $\tau_R \gg t$. The derived Lindblad equation is a coarse-grained evolution in time. 

The starting time $t=0$ is just a choice. We can choose an arbitrary starting time $t$. If we assume the total density matrix is factorized at the starting time $t$ (see Eq.~(\ref{eqn:factorize}) and the discussions below it), we can derive the leading order Lindblad equation for the quantum Brownian motion at an arbitrary time $t$.

Going back to the Schr\"odinger picture is easy since all the operators on the right hand side of Eq.~(\ref{eqn:brown}) are at the same time. For completeness, we write out the evolution equation in the Schr\"odinger picture explicitly:
\be
\frac{\diff \rho_S(t)}{\diff t} &=& -i \pig[ H_S + \frac{1}{2}\sum_{\alpha,\beta} \Sigma_{\alpha\beta}(\omega=0) O^{(S)}_\alpha O^{(S)}_\beta , \rho_S(t)\pig] \nn\\
&+& \sum_{\alpha,\beta} D_{\alpha\beta}(\omega=0) \Big( O^{(S)}_\beta \rho_S(t) O^{(S)}_\alpha - \frac{1}{2} \pig\{O^{(S)}_\alpha O^{(S)}_\beta, \rho_S(t) \pig\} \Big) \,.
\ee

\subsubsection{Next-Leading Order in $H_S$}
At next leading order in $H_S$, we need to deal with the following integral generated from Eq.~(\ref{eqn:time_integral2}) in the Markovian limit ($t\to+\infty$)
\be
\int \diff\omega \frac{\partial \delta(\omega)}{\partial \omega} \delta(\omega) D(\omega) \,.
\ee
Integration by parts leads to
\be
\int \diff\omega \frac{\partial \delta(\omega)}{\partial \omega} \delta(\omega) D(\omega) 
= - \int \diff\omega \delta(\omega) \frac{\partial \big( \delta(\omega) D(\omega) \big) }{\partial \omega} = -\frac{1}{2} \delta(0) \frac{\partial D(\omega)}{\partial \omega} \bigg|_{\omega=0} \,.
\ee
Repeating the procedures in the leading-order calculation, we find the next-leading order contributions (the leading order term is omitted here)
\be
\frac{\diff \rho_S^{(\text{int})}(t)}{\diff t} &=&
-\frac{i}{4} \sum_{\alpha,\beta} \frac{\partial \Sigma_{\alpha\beta}(\omega=0)}{\partial\omega}  \Big[ \pig[ H_S, O^{(S)}_\alpha(t) \pig] O^{(S)}_\beta(t) , \rho_S^{(\text{int})}(t)\Big] \nn\\
&+& \frac{i}{4} \sum_{\alpha,\beta} \frac{\partial \Sigma_{\alpha\beta}(\omega=0)}{\partial\omega}  \Big[  O^{(S)}_\alpha(t) \pig[ H_S, O^{(S)}_\beta(t)\pig] , \rho_S^{(\text{int})}(t)\Big] \nn\\
&+& \frac{1}{2} \sum_{\alpha,\beta} \frac{\partial D_{\alpha\beta}(\omega=0)}{\partial\omega} \bigg( - \pig[ H_S, O^{(S)}_\beta(t)\pig] \rho_S^{(\text{int})}(t) O^{(S)}_\alpha(t)
\nn\\
&+& O^{(S)}_\beta(t) \rho_S^{(\text{int})}(t) \pig[ H_S, O^{(S)}_\alpha(t) \pig]
- \frac{1}{2} \Big\{\pig[ H_S, O^{(S)}_\alpha(t) \pig]O^{(S)}_\beta(t), \rho_S^{(\text{int})}(t) \Big\}
\nn\\
&+& \frac{1}{2} \Big\{O^{(S)}_\alpha(t) \pig[ H_S, O^{(S)}_\beta(t)\pig], \rho_S^{(\text{int})}(t) \Big\} \bigg) \,.
\ee
The combination of the leading and next-leading order results, as the Caldeira-Leggett equation~\cite{CALDEIRA1983587}, cannot be written as a Lindblad equation. But we can include some of the next-next-leading order (in $H_S$) terms to make the evolution equation Lindbladian:
\be
\label{eqn:brown_final}
\frac{\diff \rho_S(t)}{\diff t} &=& -i \pig[H_S+\Delta H_S,  \rho_S(t) \pig] + \sum_{\alpha,\beta} D_{\alpha\beta}(\omega=0) \Big( \widetilde{O}^{(S)}_\beta \rho_S(t) \widetilde{O}^{(S)\dagger}_\alpha \nn\\
&-& \frac{1}{2} \pig\{\widetilde{O}^{(S)\dagger}_\alpha \widetilde{O}^{(S)}_\beta, \rho_S(t) \pig\} \Big) \\
\Delta H_S &\equiv& \frac{1}{2}\sum_{\alpha,\beta} \Sigma_{\alpha\beta}(\omega=0) O^{(S)}_\alpha O^{(S)}_\beta \\
&+& \frac{1}{4}\sum_{\alpha,\beta} \frac{\partial \Sigma_{\alpha\beta}(\omega=0)}{\partial\omega} \Big( \pig[ H_S, O^{(S)}_\alpha \pig] O^{(S)}_\beta -  O^{(S)}_\alpha \pig[ H_S, O^{(S)}_\beta \pig] \Big) \nn\\
\label{eqn:brown_final_O}
\widetilde{O}^{(S)}_\alpha &\equiv& O^{(S)}_\alpha - \frac{1}{2} \sum_{\beta,\gamma} D_{\alpha\beta}^{-1}(\omega=0) \frac{\partial D_{\beta\gamma}(\omega=0)}{\partial \omega} \pig[H_S, O^{(S)}_\gamma \pig] \\
\widetilde{O}^{(S)\dagger}_\alpha &\equiv& O^{(S)}_\alpha + \frac{1}{2} \sum_{\beta,\gamma} \pig[H_S, O^{(S)}_\gamma \pig]  \frac{\partial D_{\gamma\beta}(\omega=0)}{\partial \omega} D_{\beta\alpha}^{-1}(\omega=0) \,,
\ee
where we have transformed back to the Schr\"odinger picture.

\subsection{Quantum Optical Limit}
\label{sect:optical}

The quantum optical limit can be formulated conveniently in the basis of the subsystem eigenstates $\{ |n\rangle \}$:
\be
H_S |n\rangle = E_n |n\rangle \,.
\ee
All the eigenstates form a complete set in the subsystem space: $\sum_n |n\rangle \langle n| = \mathds{1}_S$.\footnote{The labeling $n$ can be a continuous variable. Then the summation over $n$ will be replaced by an integral.} The reason why the basis of eigenstates is convenient is one of the two hierarchies specifying the quantum optical limit: $\tau_R \gg \tau_S$. For the dynamics of quarkonium, the subsystem intrinsic time scale $\tau_S$ can be interpreted as the period of the $Q\bar{Q}$ pair revolving around each other. The hierarchy indicates that during the time when quarkonium dissociation and/or recombination is happening, the $Q\bar{Q}$ pair related to the process has revolved each other for many periods. The $Q\bar{Q}$ pair has been influenced by the potential generated via each other for a long time and thus the eigenstates solved from the Schr\"odinger equation with the potential can serve as a good basis for the calculation.

Now we take Eq.~(\ref{eqn:pre-lindblad}) and insert complete sets of eigenstates to obtain
\be
\label{eqn:pre-optical}
\rho_S^{(\text{int})}(t) &=& \rho_S^{(\text{int})}(0) - \int_0^t\diff t_1 \int_0^t\diff t_2 \frac{\sign(t_1-t_2)}{2} \sum_{\alpha,\beta} D_{\alpha\beta}(t_1,t_2) \nn\\
&\times& \sum_{n,m,k} \pig[|n \rangle \langle n| O^{(S)}_\alpha(t_1) |m \rangle \langle m| O^{(S)}_\beta(t_2) |k \rangle \langle k|, \rho_S^{(\text{int})}(0) \pig]  \nn\\
&+& \int_0^t\diff t_1 \int_0^t\diff t_2 \sum_{\alpha,\beta} D_{\alpha\beta}(t_1,t_2) \nn\\
&\times& \sum_{n,m,k,l}\Big( |n \rangle \langle n| O^{(S)}_\beta(t_2) |m \rangle \langle m| \rho_S^{(\text{int})}(0) |k \rangle \langle k| O^{(S)}_\alpha(t_1) |l \rangle \langle l| \nn\\
&-& \frac{1}{2} \pig\{ |k \rangle \langle k| O^{(S)}_\alpha(t_1) |l \rangle \langle l
|n \rangle \langle n| O^{(S)}_\beta(t_2) |m \rangle \langle m|, \rho_S^{(\text{int})}(0) \pig\} \Big) \,.
\ee
Using Eqs.~(\ref{eqn:fourier}, \ref{eqn:fourier2}) and
\be
\langle n| O_\alpha^{(S)}(t) |m\rangle = \langle n| e^{iH_S t} O_\alpha^{(S)}  e^{-iH_S t} |m\rangle = e^{i(E_n-E_m)t} \langle n| O_\alpha^{(S)} |m\rangle\,,
\ee
we can evaluate the time integrals to obtain
\be
\rho_S^{(\text{int})}(t) &=& \rho_S^{(\text{int})}(0) - \frac{i}{2} \int\frac{\diff\omega}{2\pi} \sum_{\alpha,\beta} \Sigma_{\alpha\beta}(\omega) \sum_{n,m,k} 
\langle n| O^{(S)}_\alpha |m \rangle \langle m| O^{(S)}_\beta |k \rangle \nn\\
&\times&  2e^{-i(\omega -E_n +E_m) t/2} \frac{\sin(\frac{(\omega -E_n +E_m)t}{2})}{\omega -E_n +E_m} \,
2e^{i(\omega +E_m - E_k) t/2} \frac{\sin(\frac{(\omega +E_m - E_k)t}{2})}{\omega +E_m - E_k} \nn\\
&\times& \Big[ |n\rangle \langle k|, \rho_S^{(\text{int})}(0) \Big] \nn\\
&+& \int\frac{\diff\omega}{2\pi} \sum_{\alpha,\beta} D_{\alpha\beta}(\omega) \sum_{n,m,k,l}  \langle k| O^{(S)}_\alpha |l \rangle \langle n| O^{(S)}_\beta |m \rangle \nn\\
&\times& 2e^{-i(\omega -E_k +E_l) t/2} \frac{\sin(\frac{(\omega -E_k +E_l)t}{2})}{\omega -E_k +E_l} \,
2e^{i(\omega +E_n - E_m) t/2} \frac{\sin(\frac{(\omega +E_n - E_m)t}{2})}{\omega +E_n - E_m} \nn\\
\label{eqn:pre-optical2}
&\times& \bigg( |n\rangle \langle m| \rho_S^{(\text{int})}(0) |k\rangle \langle l| - \frac{1}{2}\Big\{ |k\rangle \langle l|n\rangle \langle m| , \rho_S^{(\text{int})}(0) \Big\} \bigg) \,.
\ee
Now we apply the second hierarchy of time scales $\tau_R \gg \tau_E$, the Markovian approximation, which was also used in the case of quantum Brownian motion. To use Eq.~(\ref{eqn:sinc}), we also need the other hierarchy $\tau_R\gg\tau_S$, since the energy gap of the subsystem $E_n-E_m$ also shows up in the phase, together with $\omega$. Choosing the time step $t$ such that $\tau_R \gg t \gg \tau_E,\tau_S$, we can set $t\to+\infty$ in one of the two sine functions, since $\omega \sim \frac{1}{\tau_E}$ and $E_n-E_m\sim \frac{1}{\tau_S}$. Then we find
\be
&&\bigg(\lim_{t\to+\infty}
2e^{-i(\omega -E_k +E_l) t/2} \frac{\sin(\frac{(\omega -E_k +E_l)t}{2})}{\omega -E_k +E_l} \bigg)\,
2e^{i(\omega +E_n - E_m) t/2} \frac{\sin(\frac{(\omega +E_n - E_m)t}{2})}{\omega +E_n - E_m} \nn\\
\label{eqn:delta}
&=& 2\pi \delta(\omega -E_k +E_l)\, 
e^{i(E_k-E_l+E_n-E_m)t/2}\, \frac{2\sin(\frac{(E_k-E_l+E_n-E_m)t}{2})}{E_k-E_l+E_n-E_m} \,.
\ee

\subsubsection{Discrete Eigenenergies}
If the subsystem eigenenergies are discrete, we can further simplify Eq.~(\ref{eqn:delta}). If $E_k -E_l +E_n -E_m=0$, Eq.~(\ref{eqn:delta}) becomes $2\pi \delta(\omega -E_k +E_l)\,t$. If $E_k -E_l +E_n -E_m\neq0$, it can be estimated as $E_k -E_l +E_n -E_m \sim \frac{1}{\tau_S}$. Since we have chosen the time step $t$ such as $t\gg \tau_S$, we can set $t\to+\infty$ in the second line of Eq.~(\ref{eqn:delta}) and find it vanishes. In a nutshell, Eq.~(\ref{eqn:delta}) can be written as
\be
\label{eqn:kronecker}
2\pi \delta(\omega -E_k +E_l)\, t\, \delta_{E_k -E_l , E_m -E_n} \,,
\ee
where the delta function with arguments in the parentheses is a Dirac delta function while the delta function with arguments in the subscript is a Kronecker delta function.
Plugging Eq.~(\ref{eqn:kronecker}) into Eq.~(\ref{eqn:pre-optical2}), dividing the whole equation by $t$ and taking the $t\to0$ limit, (which is allowed since $\tau_R\gg t \gg \tau_E,\tau_S$, as explained in the case of quantum Brownian motion,) we obtain the quantum master equation in the quantum optical limit at $t=0$
\be
\frac{\diff \rho_S^{(\text{int})}(t)}{\diff t}\bigg|_{t=0} &=& -i \sum_{n,k} \sigma_{nk} \Big[ |n\rangle \langle k|, \rho_S^{(\text{int})}(0) \Big] \\
&+& \sum_{n,m,k,l} \gamma_{nm,kl} \Big( |n\rangle \langle m| \rho_S^{(\text{int})}(0) |k\rangle \langle l| - \frac{1}{2}\pig\{ |k\rangle \langle l|n\rangle \langle m| , \rho_S^{(\text{int})}(0) \pig\} \Big) \nn\\
\sigma_{nk} &=& \frac{1}{2} \sum_{\alpha,\beta} \sum_m  \Sigma_{\alpha\beta}(E_n-E_m) \delta_{E_n,E_k}  \langle n| O^{(S)}_\alpha |m \rangle \langle m| O^{(S)}_\beta |k \rangle \\
\gamma_{nm,kl} &=& \sum_{\alpha,\beta} D_{\alpha\beta}(E_m-E_n) \delta_{E_k-E_l,E_m-E_n}
\langle k| O^{(S)}_\alpha |l \rangle \langle n| O^{(S)}_\beta |m \rangle
\,,
\ee
where in the second-to-last and last lines $(E_n-E_m)$ and $(E_m-E_n)$ are the arguments of $\Sigma_{\alpha\beta}(\omega)$ and $D_{\alpha\beta}(\omega)$ respectively. Transforming back to the Schr\"odinger picture and assuming the factorization of the total density matrix at an arbitrary time $t$ (as also done in the case of quantum Brownian motion), we obtain the Schr\"odinger-picture Lindblad equation in the quantum optical limit at an arbitrary time $t$
\be
\frac{\diff \rho_S(t)}{\diff t} &=& -i \Big[H_S + \sum_{n,k} \sigma_{nk}  |n\rangle \langle k|, \rho_S(t) \Big] \nn\\
\label{eqn:L_optical}
&+& \sum_{n,m,k,l} \gamma_{nm,kl} \Big( |n\rangle \langle m| \rho_S(t) |k\rangle \langle l| - \frac{1}{2}\pig\{ |k\rangle \langle l|n\rangle \langle m| , \rho_S(t) \pig\} \Big) \,.
\ee

\subsubsection{Continuous Eigenenergies}
\label{sect:continuous_eigen}
If the eigenenergies of the subsystem are continuous, we cannot simply write Eq.~(\ref{eqn:delta}) as Eq.~(\ref{eqn:kronecker}). The subtlety is in the subsystem intrinsic time scale $\tau_S$, which is estimated from the typical energy gap in the subsystem. If the subsystem eigenenergy is a continuum, the energy gap can be tiny and $\tau_S$ can be large so that the hierarchy $\tau_R\gg\tau_S$ breaks down. On the other hand, in a thermal environment, the typical energy transferred is on the order of $T$. So most transitions in the subsystem are between states with a typical energy gap on the order of $T$. For subsystems that are sufficiently weakly-coupled with the environment, the hierarchy $\tau_R\gg\tau_S$ may still be valid. Under the assumption of $\tau_R\gg\tau_S,\tau_E$, terms with $|E_k -E_l +E_n -E_m|\gtrsim\frac{1}{\tau_S}$ or $|E_k -E_l +E_n -E_m|\gtrsim\frac{1}{\tau_E}$ are still vanishing in Eq.~(\ref{eqn:delta}), as argued previously. But terms with $ |E_k -E_l +E_n -E_m| \lesssim \frac{1}{\tau_R}$ are nonvanishing, since the time step $t$ satisfies $\tau_R\gg t\gg \tau_E,\tau_S$. (In the case of discrete eigenenergies, $|E_k -E_l +E_n -E_m|$ is either zero or gapped by $\frac{1}{\tau_S}$.) So a more careful treatment is necessary. 
We note that for $ |E_k -E_l +E_n -E_m| \lesssim \frac{1}{\tau_R}$, Eq.~(\ref{eqn:delta}) can be written as
$2\pi \delta(\omega -E_k +E_l)\,t$
when we take the limit $t\to0$ to obtain a differential equation from the finite-difference equation, which is justified by $\tau_R \gg t$. (Similar discussions can also be found in the case of quantum Brownian motion.) One can still write a Lindblad-like equation for subsystems with continuous eigenenergies in the quantum optical limit, which has the same form as Eq.~(\ref{eqn:L_optical}) but $\sigma_{nk}$ and $\gamma_{nm,kl}$ are given by
\be
\sigma_{nk} &=& \begin{cases}
\frac{1}{2} \sum_{\alpha,\beta} \sum_m  \Sigma_{\alpha\beta}  \langle n| O^{(S)}_\alpha |m \rangle \langle m| O^{(S)}_\beta |k \rangle & \text{if } |E_n - E_k|<\Delta E \\
0 & \text{otherwise}
\end{cases}\\
\label{eqn:gamma_continuous}
\gamma_{nm,kl} &=& \begin{cases} \sum_{\alpha,\beta} D_{\alpha\beta}
\langle k| O^{(S)}_\alpha |l \rangle \langle n| O^{(S)}_\beta |m \rangle & \text{if } |E_k - E_l +E_n-E_m| < \Delta E \\
0 & \text{otherwise}\,,
\end{cases} \ \ \ \ \ 
\ee
where the arguments of $\Sigma_{\alpha\beta}$ and $D_{\alpha\beta}$ are $(E_n-E_m)$ and $(E_m-E_n)$ respectively. Here $\Delta E$ ($\ll\frac{1}{\tau_S},\frac{1}{\tau_E}$) can be thought of as a parameter that defines an approximation of the exact equation. Its value can be chosen by comparing with the exact solution of the subsystem evolution equation.

If we assume the subsystem density matrix is diagonal: $\langle n| \rho_S(t)|m\rangle \propto \langle n|m\rangle$, a rate equation can be well-defined for both subsystems with discrete and continuous eigenenergies in the quantum optical limit:\footnote{If the subsystem has degenerate eigenenergies, we will assume each eigenstate subspace corresponding to the same eigenenergy has been diagonalized.}
\be
\label{eqn:rate_optical_diag}
\frac{\diff \langle n|\rho_S(t)|n\rangle }{\diff t}
= \sum_{m} \gamma_{nm,mn}  \langle m| \rho_S(t) |m\rangle  - \sum_{m} \gamma_{mn,nm}  \langle n| \rho_S(t) |n\rangle \,.
\ee

Later in Section~\ref{sect:boltzmann}, we will use the quantum optical limit and derive the semiclassical Boltzmann equation to describe quarkonium transport in the QGP. We will deal with subsystems with continuous eigenenergies (the relative kinetic energy of an unbound $Q\bar{Q}$ pair is continuous). When we write down the evolution equation of the density matrix elements that describe the quarkonium dynamics, we will find two of the eigenenergies $E_k,E_l,E_m,E_n$ are discrete (for bound states) and the other two are continuous (for unbound states). We will show only the case with $E_{k} - E_l + E_n - E_{m}=0$ in Eq.~(\ref{eqn:gamma_continuous}) contributes in the semiclassical limit.

We have introduced the general framework of open quantum systems and derived the Lindblad equations in both the quantum Brownian motion and the quantum optical limits. To apply the general construction to study quarkonium evolution in the QGP, we need an explicit theory to describe the interaction between quarkonium and the QGP. Scrutinizing the hierarchies of time scales that define the two limits is also important. It turns out that the hierarchy of time scales is closely related with the separation of energy scales in nonrelativistic heavy $Q\bar{Q}$ pairs. The separation of energy scales allows the construction of effective field theories, which can significantly simplify the calculations. We will briefly review nonrelativistic effective field theories of QCD and the separation of energy scales in the next section.



\section{Effective Field Theories for Quarkonium}
\label{sect:eft}
\subsection{Separation of Energy Scales}
\label{sect:separation}
In vacuum, the standard hierarchy of scales that is relevant for quarkonium is $M \gg Mv \gg Mv^2$, where $M$ is the heavy quark mass and $v$ denotes the typical relative velocity between the $Q\bar{Q}$ pair~\cite{Bodwin:1994jh}. The physical meanings of $Mv$ and $Mv^2$ are the typical relative momentum between the $Q\bar{Q}$ pair and the binding energy respectively. We will label the modes with the energy scale $M$, $Mv$, $Mv^2$ as the hard, soft and ultrasoft modes. The four momenta of these modes scale as
\be
p_h^\mu &\sim&    M(1,1,1,1)  \\
p_s^\mu &\sim&    M(v,v,v,v) \\
p_{us}^\mu &\sim& M(v^2,v^2,v^2,v^2) \,.
\ee
For deeply bound quarkonium states, the relative velocity can be estimated by an attractive Coulomb potential
\be
M v^2 \sim \frac{C_F\alpha_s(Mv)}{r}\,,
\ee
where $C_F = \frac{N_c^2-1}{2N_c}$ and the strong coupling constant $\alpha_s$ is estimated at the scale $\frac{1}{r}\sim Mv$. Replacing $\frac{1}{r}$ by $Mv$, we find
\be
v \sim C_F \alpha_s(Mv) \,.
\ee
Estimates showed $v^2\sim 0.3$ for charmonium and $v^2\sim 0.1$ for bottomonium~\cite{Bodwin:1994jh}. The hard, soft and ultrasoft scales are listed in Table~\ref{tab:scale} for both charmonium and bottomonium. In addition to the hard, soft and ultrasoft modes, there is also a Coulomb mode which scales as
\be
p^\mu_c\sim M(v^2, v, v, v)\,.
\ee
We will discuss the importance of the Coulomb mode in Section~\ref{sect:structure_f}.

\begin{table}
\centering
 \begin{tabular}{| c | c | c |} 
 \hline
 \rule{0pt}{2.5ex}
 Scale & Charmonium($c\bar{c}$) & Bottomonium ($b\bar{b}$) \\ 
 \hline
 $M$ & $1.5$ GeV & $4.5$ GeV\\
 \hline
 $Mv$ & $0.9$ GeV & $1.5$ GeV\\
 \hline
 $Mv^2$ & $0.5$ GeV & $0.5$ GeV\\
 \hline
 \end{tabular}
 \caption{Separation of scales for charmonium and bottomonium in vacuum. The numbers are taken from Ref.~\cite{Braaten:1996ix}\,.}
 \label{tab:scale}
\end{table}

When it comes to quarkonium states evolving inside the QGP, thermal scales such as the plasma temperature $T$ show up.\footnote{At high temperature, other thermal scales such as the Debye mass $m_D \sim g(T)T \ll T$ can show up. At low temperature, it is expected $T\sim m_D$ and thus $T$ is the only thermal scale.} In current heavy-ion collision experiments, the temperature range achieved is about $150-600$ MeV. The assumed hierarchy $M \gg T$ is well justified for heavy quarks, especially for the bottom quark. Under this hierarchy, thermal production of heavy $Q\bar{Q}$ pairs is exponentially suppressed. As mentioned earlier in Footnote~\ref{ft:anni}, the annihilation of $Q\bar{Q}$ pairs is negligible during the lifetime of the QGP. So during the evolution inside the QGP, the heavy quark number is approximately conserved. After its formation, the QGP expands fast and cools down. Furthermore, the binding energy of quarkonium in the QGP can be largely modified. So both $T>Mv^2$ and $Mv^2>T$ are possible. The scale $Mv^2$ provides an estimate of the subsystem intrinsic time scale
\be
\label{eqn:tau_s}
\tau_S \sim \frac{1}{Mv^2} \,.
\ee
Rigorously speaking, the above estimate is only valid for transitions between different $Q\bar{Q}$ bound states or between bound and unbound states. The energy spectrum of bound states is discrete and below the threshold while that of unbound states is a continuum above the threshold. The energy gap in an unbound-unbound transition is a continuum and ranges from zero to infinity. If the energy gap is large, the hierarchy $\tau_S \gg \tau_E$ in the quantum Brownian motion will be violated. If it is small, the hierarchy $\tau_R \gg \tau_S$ in the quantum optical limit will break down. When a large/tiny amount of energy is transferred in an unbound-unbound transition, the interaction can take a very short/long time and breaks one of the hierarchies of time scales. However, since our main interest is quarkonium in the end of the evolution, we can still use $Mv^2$ to estimate the typical energy gap in the $Q\bar{Q}$ relative motion. Later in Section~\ref{sect:boltzmann}, when we derive the semiclassical Boltzmann equation for quarkonium by using the quantum optical limit, we will deal with the situation with continuous energy gaps more carefully.

\subsection{NRQCD versus pNRQCD}
With $M\gg T$, one can construct nonrelativistic QCD (NRQCD)~\cite{Bodwin:1994jh} as in vacuum by integrating out modes with energy scales larger than $M$ from QCD. This step is not affected by the thermal effects. Since $M\gg \Lambda_{\text{QCD}}$, the integrating-out procedure can be done perturbatively. The NRQCD Lagrangian for heavy quarks is then obtained by a nonrelativistic expansion. The power counting parameter is $v$. At leading power in $v$,\footnote{To distinguish the expansion order in the power counting parameter of the EFT from the expansion order in the strong coupling constant, we will use terms such as ``leading power" and ``next-leading power" for the EFT power counting.} the Lagrangian can be written as
\be
\ml{L}_{\rm{NRQCD}}=\psi^\dagger \Big( iD_0 + \frac{\nabla^2}{2M} \Big) \psi + \chi^\dagger \Big( iD_0 - \frac{\nabla^2}{2M} \Big) \chi\,,
\ee
where $\psi^\dagger$($\chi$) and $\psi$($\chi^\dagger$) are the creation and annihilation operators for a heavy quark (antiquark). The covariant derivative is defined by $D_\mu = \partial_\mu - igA_\mu$. At leading power in $v$, the spatial derivative is ordinary~\cite{Braaten:1996ix}. The four-fermion operators are omitted here since these operators mainly account for the generation or the annihilation of $Q\bar{Q}$ pairs. The number of heavy quarks is almost conserved during the in-medium evolution, which is the main topic of the discussions here. The gauge field and light quark parts of the NRQCD Lagrangian are just QCD with momenta $\lesssim M$.

We can construct the potential NRQCD (pNRQCD) by further integrating out the modes between the hard and the soft scales from NRQCD~\cite{Brambilla:1999xf,Brambilla:2004jw}. If $T \gtrsim Mv$, thermal effects must be accounted for in the construction~\cite{Brambilla:2008cx,Brambilla:2010vq,Brambilla:2011sg,Brambilla:2013dpa}. If $Mv\gg T$, the soft modes are not affected by thermal effects and the construction is similar to that in vacuum. Since we will use pNRQCD in Section~\ref{sect:lindblad} for quarkonium transport in a low temperature QGP, we will focus on the case $Mv\gg T$ here. For $Mv\gg \Lambda_{\text{QCD}}$, the construction can be done perturbatively and the pNRQCD Lagrangian can be obtained from the NRQCD Lagrangian by systematic nonrelativistic and multipole expansions. The power counting parameters are $v$ and $r\sim\frac{1}{Mv}$, the typical size of quarkonium states. The pNRQCD Lagrangian (for $Mv\gg \Lambda_{\text{QCD}}$) can be written as
\be\nn
\ml{L}_\ma{pNRQCD} &=& \int \diff^3r\, \Tr\Big(  \ma{S}^{\dagger}(i\partial_0-H_s)\ma{S} +\ma{O}^{\dagger}( iD_0-H_o )\ma{O} + V_A( \ma{O}^{\dagger}\bs r \cdot g{\bs E} \ma{S} + \ma{h.c.})  \\
\label{eq:lagr}
&&+ \frac{V_B}{2}\ma{O}^{\dagger}\{ \bs r\cdot g\bs E, \ma{O}  \} +\cdots \Big) \, ,
\ee
where h.c. is the abbreviation for Hermitian conjugate and higher order terms in the power counting are omitted. In the Lagrangian, ${\bs E}$ represents the chromoelectric field and $D_0\ma{O} = \partial_0\ma{O} -ig [A_0, \ma{O}]$. The gauge field and light quark parts of the pNRQCD Lagrangian are just QCD with momenta $\lesssim Mv$. The degrees of freedom in the heavy $Q\bar{Q}$ sector are the color singlet $\ma{S}(\bs R, \bs r, t)$ and octet $\ma{O}(\bs R, \bs r, t)$ with the center-of-mass (c.m.) and relative positions ${\bs R}$ and ${\bs r}$. The trace acts in the color space. The matrix elements of both the color singlet and octet fields are 
\be
\ma{S}_{ij}(\bs R, \bs r, t) &=& \frac{\delta_{ij}}{\sqrt{N_c}}S(\bs R, \bs r, t) \\
\ma{O}_{ij}(\bs R, \bs r, t) &=& \frac{1}{\sqrt{T_F}}O^a(\bs R, \bs r, t)(T_F^a)_{ij} \,,
\ee
where $T_F^a$ is the generator of the fundamental representation of SU(3) and is normalized by $\Tr(T_F^a T_F^b) = T_F\delta^{ab}$ with $T_F=\frac{1}{2}$.

The color singlet and octet Hamiltonians are organized by powers of $\frac{1}{M}$ or equivalently, $v$:
\be
H_{s} &=& \frac{(i\bs \nabla_\ma{cm})^2}{4M} + \frac{(i\bs \nabla_\ma{rel})^2}{M} + V_{s}^{(0)} + \frac{V_{s}^{(1)}}{M} + \frac{V_{s}^{(2)}}{M^2} + \cdots\\
\label{eqn:Ho}
H_{o} &=& \frac{(i\bs D_\ma{cm})^2}{4M} + \frac{(i\bs \nabla_\ma{rel})^2}{M} + V_{o}^{(0)} + \frac{V_{o}^{(1)}}{M} + \frac{V_{o}^{(2)}}{M^2} + \cdots\,.
\ee
By the virial theorem, $\bs p_\ma{rel}^2/M \sim V_{s,o}^{(0)}\sim Mv^2$. Higher-order terms of the potentials including the relativistic corrections, spin-orbital and spin-spin interactions are suppressed by extra powers of $v$. 
The pNRQCD is a theory for the modes below the soft scale, so the c.m. kinetic terms are subleading in powers of $v$. Therefore, at leading power in $v$:
\be
\label{eqn:hamiltonian}
H_{s,o}  =  \frac{(i\bs \nabla_\ma{rel})^2}{M} + V_{s,o}^{(0)}\,.
\ee
The relative motions of the color singlet and octet are coupled via the chromoelectric dipole vertex. The chromomagnetic vertices are suppressed by powers of $v$. The potentials and the Wilson coefficients $V_{A,B}$ in the chromoelectric dipole vertices can be obtained by matching pNRQCD with NRQCD at the soft scale $Mv$. Perturbatively, at leading order in $\alpha_s(Mv)$ we have~\cite{Brambilla:2004jw}
\be
\label{eqn:match}
V_{s}^{(0)} = -C_F\frac{\alpha_s}{r}\,,\ \ \ \ \ \ \ V_{o}^{(0)} = \frac{1}{2N_c}\frac{\alpha_s}{r}\,,\ \ \ \ \ \ \ V_A=V_B=1\,.
\ee
The potential is Coulomb, since we assume $Mv\gg T$ and $Mv\gg \Lambda_{\text{QCD}}$. One can improve the potentials by computing high order corrections or doing a nonperturbative matching calculation.

In the next section, we will show quarkonium transport equations derived at leading power in $v$ and linear power in $r$. Now we will discuss the major simplifications gained at these powers.

\subsubsection{Leading Power in $v$}
At leading power in $v$, the Hamiltonians of the singlet and octet fields (\ref{eqn:hamiltonian}) are simple. Furthermore, the quarkonium wavefunction is simple. In NRQCD, the quarkonium wavefunction $|H\rangle$ can be expanded in the Fock space as
\be
|H\rangle = |Q\bar{Q}\rangle + |Q\bar{Q}g\rangle +  |Q\bar{Q}q\bar{q}\rangle + \cdots \,,
\ee
where $q, \bar{q}$ and $g$ denote a light quark, a light antiquark and a gluon respectively. The Fock states with dynamical gluons and light quark-antiquark pairs are suppressed by powers of $v$ with respect to the state $|Q\bar{Q}\rangle$~\cite{Bodwin:1994jh}. Therefore, at leading power in $v$, the quarkonium wavefunction is just a $Q\bar{Q}$ pair in the color singlet
\be
|H\rangle = |Q\bar{Q}\rangle \,.
\ee
Since the potential is attractive for a color singlet $Q\bar{Q}$ while repulsive for a color octet, a color singlet can be either bound or unbound while a color octet is always unbound. The eigenenergies of bound and unbound states are negative and positive respectively. Quarkonium dissociation and regeneration occur as a singlet-octet transition (bound singlet only) via the dipole vertex $ \ma{O}^{\dagger}\bs r \cdot g{\bs E} \ma{S} + \ma{h.c.}$. Some amount of energy has to be transferred from the environment to the $Q\bar{Q}$ pair in dissociation and vice versa in recombination. These aspects are crucial when we derive the transport equation for quarkonium in the quantum optical limit.

\subsubsection{Leading Nontrivial Power in $r$}
\label{sect:leading_r}
The linear power in $r$ is the leading nontrivial power since below the linear power, there is no interaction between the relative motion of the $Q\bar{Q}$ pair and the gauge field in the medium and thus no modification on the quarkonium wavefunction. At linear power in $r$, the interaction between the relative motion of a $Q\bar{Q}$ pair in the color singlet and the gauge field is a dipole interaction and is weakly-coupled since $rT\ll1$. As we have discussed in Section~\ref{sect:open}, the Markovian approximation used in both the quantum Brownian motion and the quantum optical limits is valid if the subsystem and the environment are weakly-coupled. With $rT\ll 1$, the weak-coupling is justified and the weak-coupling expansion works better with decreasing temperatures. So the pNRQCD provides a good tool to study quarkonium transport in a strongly-coupled QGP. However, we need to resum all other interaction vertices that are not suppressed by $rT$, such as the interaction in $\Tr(\ma{O}^\dagger iD_0 \ma{O})$, to all orders in $g(T)$. The calculations can be made simple by field redefinitions
\be
\label{eqn:O_redef}
\ma{O}(\bs R, \bs r, t) &\to& W_{[(\bs R,t),(\bs R, t_0)]} \widetilde{\ma{O}}(\bs R, \bs r, t) (W_{[(\bs R,t),(\bs R, t_0)]})^\dagger \\
\label{eqn:E_redef}
E_i(\bs R, t) &\to& W_{[(\bs R,t),(\bs R, t_0)]} \widetilde{E}_i(\bs R, \bs r, t) (W_{[(\bs R,t),(\bs R, t_0)]})^\dagger 
\ee
where ${W}_{[(\bs R,t),(\bs R, t_0)]}$ is a Wilson line in the fundamental representation
\be
\label{eqn:wilson_line}
 {W}_{[(\bs R, t_f),(\bs R, t_i)]} = \ml{P}\exp\Big( ig\int^{t_f}_{t_i} \diff s \, A^a_0 ({\bs R}, s) T_F^a\Big) \,.
\ee
Here $\ml{P}$ denotes path ordering and $t_0$ is an arbitrary constant, which will be canceled when we compute the matrix element of the transition~\cite{Yao:2020eqy}. With the field redefinition, the original covariant derivative $D_0$ of the octet field becomes an ordinary derivative $\partial_0$ and all the octet and chromoelectric fields in the pNRQCD Lagrangian are replaced by the fields with the tilde.

\section{Quantum and Semiclassical Transport Equations for Quarkonium}
\label{sect:lindblad}
\begin{figure}
\centering
\includegraphics[height=3.0in]{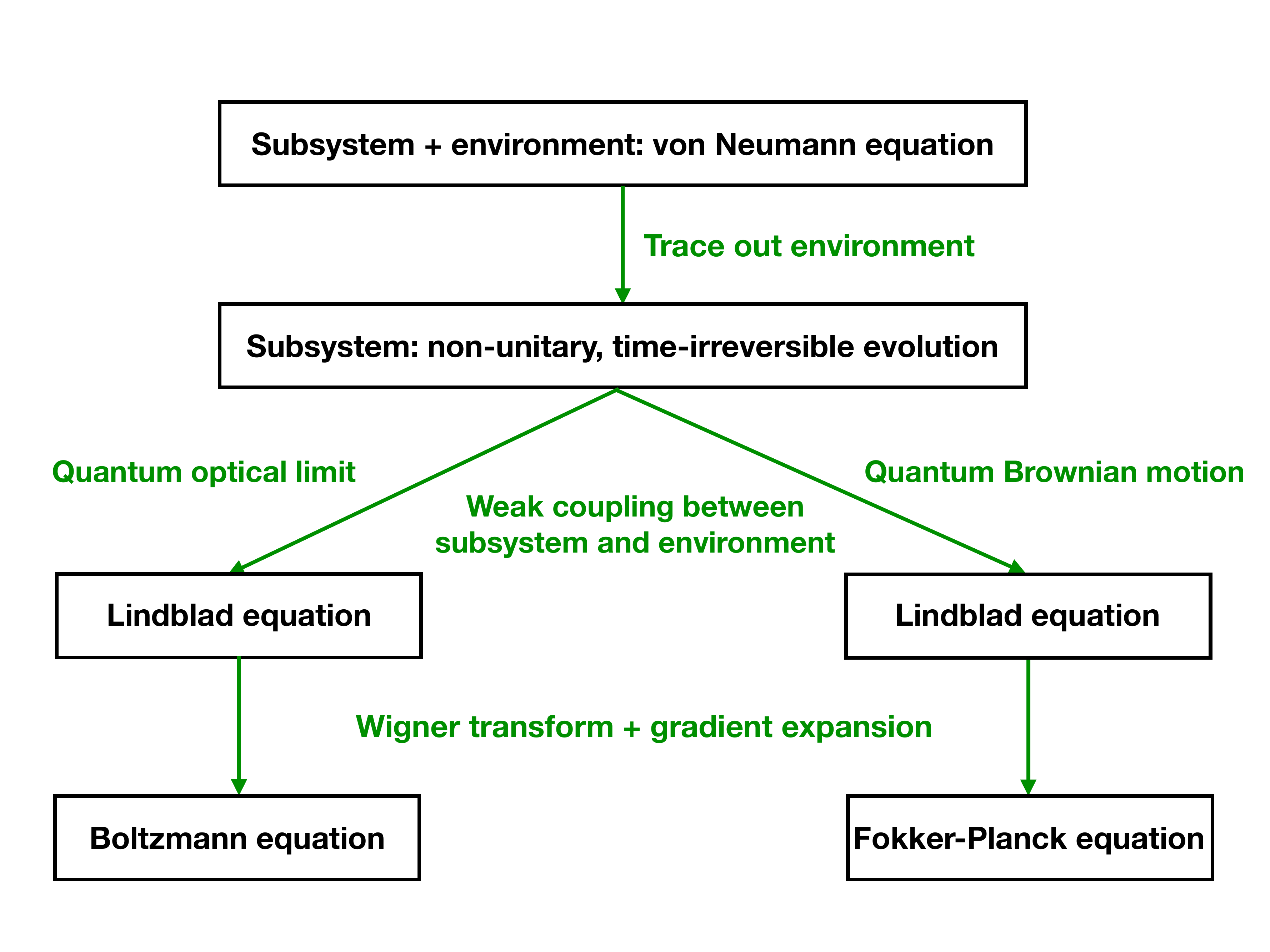}
\caption{Schematic diagram for the derivation of quantum master equations and semiclassical transport equations.}
\label{fig:scheme}
\end{figure}

In this section, we will explain the derivation of the quantum transport equations and their semiclassical correspondents for quarkonium in both the high temperature and low temperature limits. The two limits are determined by whether the multipole expansion is valid, i.e. whether the quarkonium size is small $rT \ll 1$. Before we move to detailed discussions of each limit, we outline the general procedure in the derivation, which is shown in Fig.~\ref{fig:scheme}. The first three steps have been discussed in Section~\ref{sect:open}. In the last step, we apply a Wigner transform and a gradient expansion to convert the Lindblad equations into the semiclassical transport equations. The Wigner transform is defined by
\be
f({\bs x},{\bs k},t) = \int \diff^3 x' e^{-i{\bs k}\cdot{\bs x}'} \pig\langle {\bs x}+\frac{{\bs x}'}{2} \pig| \rho(t) \pig| {\bs x}-\frac{{\bs x}'}{2} \pig\rangle \,,
\ee
which connects the quantum density matrix on the right hand side with the phase space distribution on the left hand side. It can also be defined by projecting onto the momentum states
\be
f({\bs x},{\bs k},t) =\int \diff^3 x' e^{i{\bs k}'\cdot{\bs x}} \pig\langle {\bs k}+\frac{{\bs k}'}{2} \pig| \rho(t) \pig| {\bs k}-\frac{{\bs k}'}{2} \pig\rangle\,.
\ee
The distribution defined from the Wigner transform is not positive definite and some smearing is required for the positive definiteness. A Gaussian smearing is discussed in~\ref{app:smear}. The Wigner transform is only defined for continuous variables such as the position and the momentum. For discrete variables such as the quantum numbers specifying the color and spin, we will take the diagonal elements of the density matrix to obtain the semiclassical transport equations.

In general, applying only the Wigner transform is not enough to convert the Lindblad equation into a semiclassical evolution in phase space. Operators of the forms $O\rho$ and $O_1\rho\, O_2$ appear in the Lindblad equation, which makes the Wigner transform more complicated. To see the complication, we consider the Wigner transform of two operators $A$ and $B$
\be
&& \int \diff^3 x' e^{-i{\bs k}\cdot{\bs x}'} \pig\langle {\bs x}+\frac{{\bs x}'}{2} \pig| A B \pig| {\bs x}-\frac{{\bs x}'}{2} \pig\rangle \\
&=& \int \diff^3 x' \int \diff^3 y \,e^{-i{\bs k}\cdot{\bs x}'} \pig\langle {\bs x}+\frac{{\bs x}'}{2} \pig| A \pig| {\bs y} \pig\rangle \pig\langle {\bs y}\pig| B \pig| {\bs x}-\frac{{\bs x}'}{2} \pig\rangle \,.
\ee
We see that the integral over ${\bs x}'$ cannot be simply done to obtain the phase space representation. To simplify the integral, we define ${\bs x}_1 = {\bs x} + \frac{{\bs x}'}{2}$, ${\bs x}_2 = {\bs x} - \frac{{\bs x}'}{2}$ and $f({\bs x},{\bs x}') = \langle {\bs x}_1 | f | {\bs x}_2 \rangle$. Then we can show
\be
\label{eqn:gradient_AB}
&& \int \diff^3 x' \int \diff^3 y \,e^{-i{\bs k}\cdot{\bs x}'} \pig\langle {\bs x}+\frac{{\bs x}'}{2} \pig| A \pig| {\bs y} \pig\rangle \pig\langle {\bs y}\pig| B \pig| {\bs x}-\frac{{\bs x}'}{2} \pig\rangle \nn\\
&=& \int \diff^3 (x_1-y) \,e^{-i{\bs k}\cdot({\bs x}_1-{\bs y})} \int \diff^3 (y-x_2) \,e^{-i{\bs k}\cdot({\bs y}-{\bs x}_2)} \nn\\
&\times&
A\pig({\bs x}+\frac{{\bs y}-{\bs x}_2}{2} , {\bs x}_1-{\bs y} \pig) B\pig( {\bs x}+\frac{{\bs y}-{\bs x}_1}{2} , {\bs y}-{\bs x}_2 \pig) \,.
\ee
Expanding $A$ in powers of $\frac{{\bs y}-{\bs x}_2}{2}$
\be
A\pig({\bs x}+\frac{{\bs y}-{\bs x}_2}{2} , {\bs x}_1-{\bs y} \pig) = \sum_{n=0}^\infty \frac{1}{n!}\Big( \frac{{\bs y}-{\bs x}_2}{2} \Big)^n \frac{\partial^n}{\partial {\bs x}^n}
A\big({\bs x}, {\bs x}_1-{\bs y} \big)
\ee
and similarly for $B$ in powers of $\frac{{\bs y}-{\bs x}_1}{2}$, we find Eq.~(\ref{eqn:gradient_AB}) turns to
\be
A({\bs x},{\bs k}) B({\bs x},{\bs k}) 
+ \frac{i}{2}\Big( \frac{\partial A({\bs x},{\bs k})}{\partial{\bs x}}
\frac{\partial B({\bs x},{\bs k})}{\partial{\bs k}}
- \frac{\partial A({\bs x},{\bs k})}{\partial{\bs k}}
\frac{\partial B({\bs x},{\bs k})}{\partial{\bs x}}
\Big) +\cdots\,.
\ee
This is the gradient expansion and corresponds to the semiclassical expansion (for more details, see e.g. Ref.~\cite{haug2008quantum}).

\subsection{High Temperature Limit}
\label{sect:highT}
First we consider the high temperature limit specified by $M \gg T \gg m_D \gg \Lambda_{\text{QCD}}$, where the Debye mass scales as $m_D\sim g(T)T$. In this limit, the QGP is a weakly-coupled plasma and the strong coupling constants at the scales $T$ and $m_D$ satisfy $g(T) \ll 1$, $g(m_D) \ll 1$ respectively. Now we can estimate the relevant time scales introduced in Section~\ref{sect:time_scale}. The range of the environment correlation time can be estimated by
\be
\frac{1}{T} \lesssim \tau_E \lesssim \frac{1}{m_D}\,,
\ee
for soft and hard interactions in which the typical energy-momentum transferred is $m_D$ and $T$ respectively. As explained in Section~\ref{sect:separation}, the subsystem intrinsic time scale is $\tau_S\sim (Mv^2)^{-1}$. The subsystem relaxation time can be approximated by the inverse of the interaction rate between a heavy quark and the QGP.\footnote{Some studies (see Ref.~\cite{Akamatsu:2020ypb} and references therein) used the momentum drag coefficient as an estimate of the relaxation time,
$
\tau_R \sim \frac{M}{g^4 T^2}
$.
Considering that the expansion in Eq.~(\ref{eqn:finite_t}) is essentially expanding in $tH_I^{(\text{int})}$, we use the inverse of the interaction rate (or the mean free time) to estimate the subsystem relaxation time, which is also explained in Section~\ref{sect:time_scale}.} The interaction rate can be estimated as~\cite{Moore:2004tg}
\be
\tau_R \sim \frac{m_D^2}{g^4T^3} \sim \frac{1}{g^2 T}
\ee
The Markovian condition $\tau_R \gg \tau_E$ is satisfied in a weakly-coupled QGP. For the quantum optical limit, we also require $\tau_R\gg \tau_S$ which leads to $Mv^2 \gg g^2T$.
For the quantum Brownian motion limit, the condition $\tau_S \gg \tau_E$ is equivalent to
$g T \gg Mv^2$.
If we take the estimate in Table~\ref{tab:scale}, $\tau_S \gg \tau_E$ is valid when the temperature is on the order of GeV. If we take into account the decrease of the quarkonium binding energy in the QGP due to the static plasma screening effect, the condition is probably valid at lower temperature. However, with the decrease in the binding energy, the heavy quark pair in the loosely-bound quarkonium state behaves more like an unbound pair in the QGP. Then the estimate of the intrinsic time scale of the subsystem $\tau_S\sim({Mv^2})^{-1}$ may break down, since it is obtained by considering transitions involving a well-defined bound state. As discussed below Eq.~(\ref{eqn:tau_s}), only bound $Q\bar{Q}$ pairs are of our interest in the end and we will assume this estimate still holds.

\subsubsection{Lindblad Equation}
Many studies worked in this high temperature limit and derived the Lindblad equation for quarkonium in the quantum Brownian motion limit~\cite{Akamatsu:2012vt,Akamatsu:2014qsa,Blaizot:2015hya,DeBoni:2017ocl,Blaizot:2017ypk}. Some studies used the path integral formalism~\cite{Young:2010jq,Akamatsu:2012vt,Akamatsu:2014qsa,Blaizot:2015hya,DeBoni:2017ocl}, which involves the calculation of the influence functional~\cite{feynman2000theory} and is different from the approach introduced in Section~\ref{sect:open}. The work in Ref.~\cite{Blaizot:2018oev} studied the Lindblad equation in the quantum optical limit and the entropy production. Most of these studies used a nonrelativistic quantum mechanical treatment of the $Q\bar{Q}$ pair, motivated by the NRQCD Lagrangian at leading power in $v$. The total Hamiltonian is written as
\be
\frac{\hat{\bs p}_Q^2}{2M} + 
\frac{\hat{\bs p}_{\bar{Q}}^2}{2M} + H_{q+A} + 
\int\diff^3x \pig( \delta^3({\bs x} - \hat{\bs x}_Q) T_F^a -
\delta^3({\bs x} - \hat{\bs x}_{\bar{Q}}) T_F^{*a} 
\pig) g A_0^a({\bs x})\,,
\ee
where $H_{q+A}$ denotes the Hamiltonian of the environment that describes the light quark and gauge fields. Here $T_F^a$ with $a=1,2,\cdots,8$ is the generator of the fundamental (triplet) representation of SU(3) and $-T_F^{*a}$ is the generator of the anti-triplet representation. The mapping between this explicit form of the Hamiltonian and the general expression (\ref{eqn:HI}) is given by
\be
O_\alpha^{(S)} &\to& \delta^3({\bs x} - \hat{\bs x}_Q) T_F^a -
\delta^3({\bs x} - \hat{\bs x}_{\bar{Q}}) T_F^{*a} = O^a({\bs x}) \\
O_\alpha^{(E)} &\to& g A_0^a({\bs x}) \\
\sum_\alpha &\to& \sum_a \int\diff^3x \,.
\ee
From now on, we will use the convention that repeated indexes are summed over. In the interaction picture, the gauge field becomes time-dependent. One relevant environment correlator is given by
\be
D_{\alpha\beta}(t_1,t_2) \to g^2\Tr_E\big( \rho_E A_0^a(t_1,{\bs x}_1) 
A_0^b(t_2,{\bs x}_2)
\big) \,.
\ee
With a thermal environment density matrix, this is just the Wightman functions ($x=(t,{\bs x})$):
\be
g^2 \Tr_E\big( \rho_E A_0^a(t_1,{\bs x}_1) A_0^b(t_2,{\bs x}_2)
\big) &=& g^2 \langle A_0^a(t_1,{\bs x}_1) A_0^b(t_2,{\bs x}_2) \rangle_T \nn\\
&=& D^{>ab}(x_1,x_2)
 = D^{<ba}(x_2,x_1) \,.
\ee
The other relevant environment correlator is
\be
\Sigma_{\alpha\beta}(t_1,t_2) &\to& -i g^2 \sign(t_1-t_2) \Tr_E\big( \rho_E A_0^a(t_1,{\bs x}_1) 
A_0^b(t_2,{\bs x}_2) 
\big) \nn\\
&=& -i g^2 \sign(t_1-t_2) \langle A_0^a(t_1,{\bs x}_1) A_0^b(t_2,{\bs x}_2) \rangle_T =  \Sigma^{ab}(x_1,x_2) \,.
\ee
For a translationally invariant environment, we can Fourier transform both correlators into energy-momentum space. We can also use a mixed space representation, which gives  $D^{>ab}(q_0,{\bs x}_1-{\bs x}_2)$ and $\Sigma^{ab}(q_0,{\bs x}_1-{\bs x}_2)$ when only the time variable is Fourier transformed.

Using the results shown in~\ref{app:correlator}, we obtain the Lindblad equation for a $Q\bar{Q}$ pair in the limit of quantum Brownian motion\footnote{The self-energy term $\frac{1}{2(N_c^2-1)}\Sigma(q_0=0,{\bs 0})  \big(T_F^a T_F^a + T_F^{*a} T_F^{*a} \big)  $ of $\Delta H_S$ is dropped since it commutes with $\rho_S(t)$.}
\be
\label{eqn:lindblad_aka1}
\frac{\diff\rho_S(t)}{\diff t} &=&  -i \big[ H_S+\Delta H_S, \rho_S(t)\big] + \frac{1}{N_c^2-1}\int\diff^3x \int \diff^3y \, D^{>}(q_0=0,{\bs x}-{\bs y}) \nn\\
&\times& \Big( \widetilde{O}^a({\bs y}) \rho_S(t) \widetilde{O}^{a\dagger}({\bs x})  - \frac{1}{2} \big\{  \widetilde{O}^{a\dagger}({\bs x}) \widetilde{O}^a({\bs y}) , \rho_S(t) \big\}
\Big) \\
H_S +\Delta H_S &=& \frac{\hat{\bs p}_Q^2}{2M} + \frac{\hat{\bs p}_{\bar{Q}}^2}{2M} 
+ \frac{1}{N_c^2-1} \bigg( \frac{-1}{2}\Sigma(q_0=0,\hat{\bs x}_Q - \hat{\bs x}_{\bar{Q}}) \big( T_F^a T_F^{*a} + T_F^{*a} T_F^a\big) \nn\\
&+& T \int\diff^3x\diff^3y\frac{\partial\Sigma(q_0=0,{\bs x}-{\bs y})}{\partial q_0}\pig( O^a({\bs x}) \widetilde{O}^a({\bs y}) -\widetilde{O}^a({\bs x}) O^a({\bs y})  \pig)  \bigg) \\
\widetilde{O}^a(\bs x) &=& \delta^3({\bs x} - \hat{\bs x}_Q) T_F^a -
\delta^3({\bs x} - \hat{\bs x}_{\bar{Q}}) T_F^{*a} +\frac{1}{8MT}\nn\\
&\times& \Big( \big( \nabla_{{\bs x}_Q}^2 \delta^3({\bs x}-\hat{\bs x}_Q) \big) T_F^a + 2 \big(\nabla_{{\bs x}_Q} \delta^3({\bs x}-\hat{\bs x}_Q) \big) \cdot \nabla_{{\bs x}_Q} T_F^a \nn\\
&-&  \big( \nabla_{{\bs x}_{\bar{Q}}}^2 \delta^3({\bs x}-\hat{\bs x}_{\bar{Q}}) \big) T_F^{*a} - 2 \big( \nabla_{{\bs x}_{\bar{Q}}} \delta^3({\bs x}-\hat{\bs x}_{\bar{Q}}) \big) \cdot \nabla_{{\bs x}_{\bar{Q}}} T_F^{*a} \Big)\,,\ \ \ \ \
\ee
where we have used $(N_c^2-1)D^{>ab} = \delta^{ab}D^>$, $(N_c^2-1)\Sigma^{ab} =  \delta^{ab}\Sigma$, $D^>(q_0,{\bs x})=D^>(q_0,-{\bs x})$ and $\Sigma(q_0,{\bs x})=\Sigma(q_0,-{\bs x})$. Here by the notation $(\nabla_{{\bs x}_Q} \cdots)$, we mean the operator $\nabla_{{\bs x}_Q} $ only acts inside the parentheses. If there are no parentheses, $\nabla_{{\bs x}_Q}$ acts on everything on its right. The dot product $\cdot$ is originated from the contraction in $(\nabla_i \cdots)\nabla_i$. In momentum space, the expression is simpler
\be
\frac{\diff\rho_S(t)}{\diff t} &=&  -i \big[ H_S+\Delta H_S, \rho_S(t)\big] + \frac{1}{N_c^2-1}\int\frac{\diff^3q}{(2\pi)^3} \, D^{>}(q_0=0,{\bs q}) \nn\\
&\times& \Big( \widetilde{O}^a({\bs q}) \rho_S(t) \widetilde{O}^{a\dagger}({\bs q})  - \frac{1}{2} \big\{  \widetilde{O}^{a\dagger}({\bs q}) \widetilde{O}^a({\bs q}) , \rho_S(t) \big\}
\Big) \\
\widetilde{O}^a({\bs q}) &=& e^{\frac{i}{2}{\bs q}\cdot \hat{\bs x}_Q} \Big( 1-\frac{{\bs q}\cdot \hat{\bs p}_Q }{4MT}  \Big) e^{\frac{i}{2}{\bs q}\cdot \hat{\bs x}_Q} T_F^a - e^{\frac{i}{2}{\bs q}\cdot \hat{\bs x}_{\bar{Q}}} \Big( 1-\frac{{\bs q}\cdot \hat{\bs p}_{\bar{Q}} }{4MT}  \Big) e^{\frac{i}{2}{\bs q}\cdot \hat{\bs x}_{\bar{Q}}} T_F^{*a} \,,
\ee
where we have used $D^{>}(q_0=0,{\bs q}) = D^{>}(q_0=0,-{\bs q})$. We can project the Lindblad equation (\ref{eqn:lindblad_aka1}) onto the position space and color space. The basis of the position space can be written as $| {\bs r}_1, {\bs r}_2 \rangle$,
where ${\bs r}_1$(${\bs r}_2$) denotes the position of the $Q$($\bar{Q}$). The basis of the color space can be chosen as the color singlet ($|s\rangle$) and the octet ($|a\rangle$, $a=1,2,\cdots,N_c^2-1$)
\be
|s\rangle &=& \frac{\delta_{ij}}{\sqrt{N_c}} |i\rangle_Q |j\rangle_{\bar{Q}}\\
|a\rangle &=& \frac{(T_F^a)_{ij}}{T_F} |i\rangle_Q |j\rangle_{\bar{Q}} \,,
\ee
where $T_F=\frac{1}{2}$ and $i$($j$) is the index of the (anti-)triplet representation. Details of the projection onto the color space can be found in Refs.~\cite{Blaizot:2017ypk,Akamatsu:2020ypb} while details for the position space can be found in Refs.~\cite{Blaizot:2017ypk}\,.
We provide some useful formulas for the projection onto the position space in~\ref{app:projection}. Transitions among color singlets and octets in the static limit ($M\to+\infty$) have also been discussed in Refs.~\cite{Akamatsu:2020ypb,Escobedo:2020tuc}\,.

\subsubsection{Stochastic Schr\"odinger and Schr\"odinger-Langevin Equations}
Solving the Lindblad equation can be computationally expensive, since it is an evolution equation of the density matrix. Thus it is worth investigating if the Lindblad equation is equivalent to some evolution equation of the wavefunction, which may be computationally cheap to solve.
The recoil-less limit of the Lindblad equation (\ref{eqn:lindblad_aka1}) is equivalent to a stochastic Schr\"odinger equation with the Hamiltonian
\be
\label{eqn:stochastic}
H_{\text{sto}}(t) &=& H_S + \Delta H_S + \int\diff^3x \,\Theta^a(t,{\bs x}) O^a({\bs x}) \\
\langle \Theta^a(t,{\bs x})  \Theta^b(s,{\bs y})\rangle &=& \delta^{ab} \delta(t-s) D^>(q_0=0,{\bs x}-{\bs y}) \,,
\ee
where the $\Theta^a(t,{\bs x})$ term is stochastic.
The stochastic Schr\"odinger equation has been studied in Refs.~\cite{Akamatsu:2011se,Rothkopf:2013kya,Kajimoto:2017rel,Islam:2020bnp}\,. The main discovery of these studies is that the stochastic term leads to decoherence of the quarkonium wavefunction. This provides one microscopic interpretation of quarkonium dissociation.

In the recoil-less limit, the dissipation effect is neglected. To account for the dissipation effect, a damping term can be added to the stochastic Schr\"odinger equation. Ref.~\cite{Katz:2015qja} used this approach and studied a Schr\"odinger-Langevin equation that has both stochastic and damping terms.

A more systematic way of including the recoil effect is to apply the quantum state diffusion method to rewrite the Lindblad equation as a nonlinear stochastic Schr\"odinger equation. This has been studied for the case of one heavy quark~\cite{Akamatsu:2018xim} and a $Q\bar{Q}$ pair~\cite{Miura:2019ssi} in the QGP. It has been shown that the dissipation effect is crucial for the thermalization of the relative motion of the $Q\bar{Q}$~\cite{Miura:2019ssi}. The dissipation effect also slows down the suppression of the ground state~\cite{Miura:2019ssi}. A generalization to incorporate noise that is correlated at a finite time length was studied in Ref.~\cite{Sharma:2019xum}\,.

\subsubsection{Fokker-Planck and Langevin Equations}
The semiclassical limit of the Lindblad equation (\ref{eqn:lindblad_aka1}) can be obtained by first applying a Wigner transform to the equation that is projected onto the position space, followed by the gradient expansion. For a heavy $Q\bar{Q}$ pair in Quantum Electrodynamics (QED), where the heavy quark pair has opposite electric charge, the semiclassical transport equation can be written as~\cite{Blaizot:2017ypk,Akamatsu:2020ypb}
\be
0&=&\bigg( \frac{\partial}{\partial t} + \frac{{\bs p}_1\cdot\nabla_{{\bs x}_1} + {\bs p}_2\cdot\nabla_{{\bs x}_2} }{M}
+\nabla_{{\bs r}} \Sigma(q_0=0,{\bs r} ) \cdot \big(\nabla_{{\bs p}_1} - \nabla_{{\bs p}_2} \big) \\
&+& \frac{1}{2} \frac{\partial^2 D^>(q_0=0,{\bs r}={\bs 0})}{\partial r_{i} \partial r_{j}} \Big( \frac{\partial^2}{\partial p_{1i}\partial p_{1j} } + \frac{\partial}{\partial p_{1i}} \frac{p_{1j}}{MT} + \frac{\partial^2}{\partial p_{2i}\partial p_{2j} } + \frac{\partial}{\partial p_{2i}} \frac{p_{2j}}{MT}  \Big) \nn\\
&-& \frac{1}{2} \frac{\partial^2 D^>(q_0=0,{\bs r})}{\partial r_{i} \partial r_{j}} \Big( 2\frac{\partial^2}{\partial p_{1i}\partial p_{2j} } + \frac{p_{2j}}{MT} \frac{\partial}{\partial p_{1i}}  +  \frac{p_{1j}}{MT}\frac{\partial}{\partial p_{2i}} \Big) \bigg) 
f_{Q\bar{Q}}({\bs x}_1, {\bs x}_2, {\bs p}_1, {\bs p}_2,t) \,,\nn
\ee
where ${\bs r}={\bs x}_1-{\bs x}_2$ and $f_{Q\bar{Q}}({\bs x}_1, {\bs x}_2, {\bs p}_1, {\bs p}_2,t)$ is the phase space distribution of a $Q\bar{Q}$ pair. The $-\Sigma(q_0=0,{\bs r})$ term can be interpreted as the potential between the $Q\bar{Q}$ pair. This equation is a Fokker-Planck equation (or a Boltzmann equation with collision terms of the Fokker-Planck type) and thus is equivalent to a Langevin equation:~\cite{Blaizot:2017ypk,Akamatsu:2020ypb}
\be
\frac{\diff }{\diff t}\begin{pmatrix}
{\bs x}_1\\{\bs x}_2
\end{pmatrix} &=& \frac{1}{M} \begin{pmatrix}
{\bs p}_1\\{\bs p}_2
\end{pmatrix}\\
\frac{\diff }{\diff t}\begin{pmatrix}
p_{1i}\\p_{2i}
\end{pmatrix}
&=& \begin{pmatrix}
\partial_{x_{1i}}\\\partial_{x_{2i}}
\end{pmatrix} \Sigma(q_0=0,{\bs r}) +\frac{1}{2MT} \Gamma_{ij} \begin{pmatrix}
p_{1j}\\p_{2j}
\end{pmatrix} +  \begin{pmatrix}
\Theta_{1i}\\\Theta_{2i}
\end{pmatrix} \\
\Gamma_{ij} &=& \begin{pmatrix}
\partial_{r_{i}} \partial_{r_{j}} D^>(q_0=0,{\bs r}=0) & -\partial_{r_{i}} \partial_{r_{j}} D^>(q_0=0,{\bs r}) \\
-\partial_{r_{i}} \partial_{r_{j}} D^>(q_0=0,{\bs r}) & \partial_{r_{i}} \partial_{r_{j}} D^>(q_0=0,{\bs r}=0) 
\end{pmatrix} \\
\langle \Theta_{1i}(t_1) \Theta_{1j}(t_2) \rangle &=&  \langle \Theta_{2i}(t_1) \Theta_{2j}(t_2) \rangle = -\partial_{r_{i}} \partial_{r_{j}} D^>(q_0=0,{\bs r}=0) \delta(t_1-t_2)\ \ \ \  \\
\langle \Theta_{1i}(t_1) \Theta_{2j}(t_2) \rangle &=& \partial_{r_{i}} \partial_{r_{j}} D^>(q_0=0,{\bs r}) \delta(t_1-t_2) \,.
\ee

For a $Q\bar{Q}$ pair in QCD, the Lindblad equation in the basis of the color singlet and octet is more complicated. The off-diagonal elements in the color space are decoupled from the diagonal elements. One usually assumes the density matrix is diagonal in the color space. Then the Lindblad equation becomes coupled evolution equations for $\langle s| \rho_S |s\rangle $ and $\langle a| \rho_S |a\rangle $. It is not easy to convert the coupled equations into Langevin equations in the semiclassical limit, since there is no classical analog of color. However, one can approximately rewrite the semiclassical limit of the coupled equations as Langevin equations for both the color singlet and octet $Q\bar{Q}$ pairs plus rate equations governing the transition between the singlet and octet pair~\cite{Blaizot:2017ypk,Akamatsu:2020ypb}. Another strategy is to treat the singlet-octet transition as perturbation to the state that is in color equilibrium~\cite{Blaizot:2017ypk}.

\subsection{Low Temperature Limit}
\label{sect:lowT}
As explained in the beginning of Section~\ref{sect:lindblad}, low temperature here means the temperature fits into the hierarchy $M\gg Mv \gg T$. Under this hierarchy, the size of quarkonium is small enough that a multipole expansion can be carried out in NRQCD. So pNRQCD is a valid description of the dynamics of quarkonium. We will further consider two different hierarchies of the remaining scales. In both cases, we have $Mv \gg T$ and the leading interaction between quarkonium and the QGP is a dipole interaction that scales as $rT\sim\frac{T}{Mv}$. In this way, the weak coupling between the subsystem and the environment is justified. Thus, the Markovian condition $\tau_R \gg \tau_E$ is valid.

\subsubsection{Hierarchy 1: $M\gg Mv \gg T \gg Mv^2$}
\label{sect:brown_pnrqcd}
We first focus on the case where the temperature is in the nonperturbative regime, which means $g(T)\sim1$ and $T\sim m_D \gg Mv^2$. This hierarchy has been studied in Refs.~\cite{Brambilla:2016wgg,Brambilla:2017zei,Brambilla:2020qwo}\,. Estimating the time scales we find $\tau_S \gg \tau_E$. So the Lindblad equation in the limit of the quantum Brownian motion is a valid description. Neglecting the c.m. motion of the $Q\bar{Q}$ pair in pNRQCD, the Hamiltonian can be written as
\be
H_S &=& \frac{{\bs p}_{\text{rel}}^2}{M} - \frac{C_F\alpha_s}{r} |s\rangle \langle s| + \frac{\alpha_s}{2N_cr} |a\rangle \langle a| \\
H_I &=& r_i\bigg( \sqrt{\frac{T_F}{N_c}}\pig(|s\rangle\langle a| + |a\rangle \langle s| \pig) + \frac{1}{2} d^{abc}|b\rangle \langle c| \bigg) g\widetilde{E}_i^a({\bs R}=0) \,,
\ee
where $d_{abc} = 2\Tr(T_F^a\{ T^b_F, T_F^c\})$.
The mapping to the general theory discussed in Section~\ref{sect:open} is given by
\be
O^{(S)}_\alpha &\to& r_i\bigg( \sqrt{\frac{T_F}{N_c}}\pig(|s\rangle\langle a| + |a\rangle \langle s| \pig) + \frac{1}{2} d^{abc}|b\rangle \langle c| \bigg) = O_i^a \\
O^{(E)}_\alpha &\to& g \widetilde{E}_i^a({\bs R}=0)\\
\sum_{\alpha} &\to& \sum_a \sum_i \,.
\ee
The correction to the subsystem Hamiltonian is
\be
\label{eqn:dH_pnrqcd}
\Delta H_S &=& \frac{1}{2}\Sigma_{ij}^{ab}(\omega=0,{\bs R}=0) \, O_i^a O_j^b \nn\\
&=& \frac{1}{2(N_c^2-1)} \Sigma(\omega=0,{\bs R}=0)
\, r^2 \Big( C_F |s\rangle \langle s| + \frac{N^2_c-2}{4N_c} |a\rangle \langle a|
\Big)\,,
\ee
where we have used $(N_c^2-1)\Sigma_{ij}^{ab}(q) = \Sigma(q)\, \delta_{ij} \delta^{ab}$ and dropped the term involving $\frac{\partial}{\partial q_0}\Sigma(q_0=0)$. The definition of the environment correlator $\Sigma$ will be given in Section~\ref{sect:structure_f}. With Eq.~(\ref{eqn:2T}), the subsystem operator $\widetilde{O}^{(S)}_\alpha$ defined in Eq.~(\ref{eqn:brown_final}) can be evaluated as
\be
\widetilde{O}_i^a &=& O_i^a - \frac{1}{4T} \big[ H_S,\ O_i^a \big] = \sqrt{\frac{T_F}{N_c}} \Big(r_i + \frac{1}{2MT}\nabla_i + \frac{N_c}{8T}\frac{\alpha_s r_i}{r} \Big) |s\rangle\langle a| \nn\\
&+& \sqrt{\frac{T_F}{N_c}} \Big(r_i + \frac{1}{2MT}\nabla_i - \frac{N_c}{8T}\frac{\alpha_s r_i}{r} \Big) |a\rangle \langle s|  + \frac{1}{2}
\Big(r_i + \frac{1}{2MT}\nabla_i \Big) d^{abc}|b\rangle \langle c| \,.\ \ \ 
\ee
With these terms given, the Lindblad equation can be written as
\be
\frac{\diff\rho_S(t)}{\diff t} &=& -i\big[ H_S + \Delta H_S,\, \rho_S(t) \big] \nn\\
&+& \frac{D(\omega=0,{\bs R}=0)}{N_c^2-1} \Big( \widetilde{O}_i^a \rho_S(t) \widetilde{O}_i^{a\dagger} -\frac{1}{2}\pig\{ \widetilde{O}_i^{a\dagger} 
\widetilde{O}_i^a ,\, \rho_S(t)\pig\} \Big)
\,,
\ee
where we have used $(N_c^2-1)D_{ij}^{ab}(q) = D(q)\delta_{ij}\delta^{ab}$.  Again, the definition of the environment correlator $D$ will be given in Section~\ref{sect:structure_f}. Assuming the density matrix is diagonal in the color space,
\be
\label{eqn:rho_color_diag}
\rho_S(t) = \begin{pmatrix}
\rho_S^{(s)}(t) & 0 \\
0 & \rho_S^{(o)}(t)
\end{pmatrix} = \begin{pmatrix}
\langle s|\rho_S(t)|s\rangle & 0 \\
0 & \langle a |\rho_S(t) |a\rangle 
\end{pmatrix}\,,
\ee
the Lindblad equation can be rewritten as
\be
\label{eqn:pheno_lindblad}
\frac{\diff\rho_S(t)}{\diff t} &=& -i\big[ H_S + \Delta H_S,\, \rho_S(t) \big] \nn\\
&+& \frac{D(\omega=0,{\bs R}=0)}{N_c^2-1} \Big( L_{\alpha i} \rho_S(t) L^\dagger_{\alpha i} - \frac{1}{2}\pig\{  L^\dagger_{\alpha i}L_{\alpha i},\, \rho_S(t)\pig\} \Big) \\
\Delta H_S &=& \frac{\Sigma(\omega=0,{\bs R}=0)}{2(N_c^2-1)}r^2 \begin{pmatrix}
C_F & 0\\
0 & \frac{N_c^2-2}{4N_c}
\end{pmatrix} \\
L_{1i} &=& \sqrt{C_F}\Big(r_i + \frac{1}{2MT}\nabla_i - \frac{N_c}{8T}\frac{\alpha_s r_i}{r} \Big)\begin{pmatrix}
0 & 0\\
1 & 0
\end{pmatrix} \\
L_{2i} &=& \sqrt{\frac{T_F}{N_c}}\Big(r_i + \frac{1}{2MT}\nabla_i + \frac{N_c}{8T}\frac{\alpha_s r_i}{r} \Big)\begin{pmatrix}
0 & 1\\
0 & 0
\end{pmatrix} \\
L_{3i} &=& \sqrt{\frac{N_c^2-4}{4N_c}}
\Big(r_i + \frac{1}{2MT}\nabla_i \Big)\begin{pmatrix}
0 & 0\\
0 & 1
\end{pmatrix}
\ee
The recoil-less limit of this Lindblad equation has been derived in Refs.~\cite{Brambilla:2016wgg,Brambilla:2017zei} and numerically studied in Refs.~\cite{Brambilla:2017zei,Brambilla:2020qwo} for heavy-ion phenomenology, by coupling the Lindblad equations with the bulk dynamics of the QGP described by 1+1D and 3+1D hydrodynamics. The Lindblad equation solves the coupled dynamics of the color singlet and octet density matrices and conserves the total number of $Q\bar{Q}$ pairs. The equation is characterized by two nonperturbative transport coefficients: $\Sigma(q_0=0,{\bs R}=0)$ and $D(q_0=0,{\bs R}=0)$, which are gauge invariant and can be calculated nonperturbatively. (The $\Sigma(q_0=0,{\bs R}=0)$ and $D(q_0=0,{\bs R}=0)$ defined here correspond to the $\gamma$ and $\kappa$ used in Refs.~\cite{Brambilla:2019tpt,Brambilla:2020qwo} respectively. The normalization differs by a factor of $\frac{T_F}{N_c}$.) The ranges of these two parameters are estimated from existing $2+1$ flavor lattice QCD calculation results~\cite{Brambilla:2019tpt} (the value of $\gamma$ is estimated from the lattice results of the quarkonium mass shift at finite temperature)
\be
\label{eqn:gamma}
-3.8 \lesssim \frac{\gamma}{T^3} \lesssim -0.7 \\
\label{eqn:kappa}
0.24 \lesssim \frac{\kappa}{T^3} \lesssim 4.2 \,.
\ee
These two parameters characterize the quarkonium in-medium dynamics in the recoil-less limit of the quantum Brownian motion. 
We will write down their explicit definitions in Section~\ref{sect:structure_f} and discuss them in a more general setup.

We have been discussing the case with the temperature as a nonperturbative scale. If the temperature is a perturbative scale, we can calculate the relevant environment correlators perturbatively. The case with $T\gg Mv^2 \gg m_D,\Lambda_{\text{QCD}}$ has been studied in Ref.~\cite{Brambilla:2017zei}\,. 
The hierarchy $Mv^2 \gg m_D$ may break the assumption of $\tau_S \gg \tau_E$ which is required for the validity of the quantum Brownian motion. However, it has been shown that under the hierarchy $T \gg Mv^2 \gg m_D$, the dominant contribution to the quarkonium dynamical evolution is gluon absorption and emission, and the Landau damping contribution is small~\cite{Brambilla:2010vq,Brambilla:2011sg,Brambilla:2013dpa}. In the former case, i.e., gluon absorption and emission, the environment correlation time can be estimated as $\tau_E\sim \frac{1}{T}$ while in the latter case, i.e., Landau damping, we have $\tau_E\sim\frac{1}{m_D}$. So for the gluon absorption and emission, the hierarchy $\tau_S \gg \tau_E$ is still valid. We can still describe the dynamics of a quarkonium state as a quantum Brownian motion under this hierarchy. Since only the zero energy limit of the environment correlator contributes, one can use the Hard Thermal Loop (HTL) effective theory~\cite{Braaten:1989mz,Braaten:1989kk,Braaten:1991gm} to calculate these correlators. In this case, the second order (in $H_S$) terms in $e^{iH_St} O_\alpha^{(S)} e^{-iH_St}$ are the leading contributions. Details about this perturbative construction can be found in the recent review~\cite{Akamatsu:2020ypb}.

\subsubsection{Hierarchy 2: $M\gg Mv \gg Mv^2, T $}
\label{sect:boltzmann}
Here the temperature can be in the perturbative or nonperturbative regime and the following discussion works for both regimes. In the nonperturbative regime, we expect $g(T)\sim1$ so $T$ represents all relevant thermal scales. This hierarchy is different from the previous case in the relative size between $Mv^2$ and $T$. Here $Mv^2$ and $T$ are not widely separated, and the hierarchy of time scales $\tau_S \sim \frac{1}{Mv^2}\gg \tau_E\sim\frac{1}{T}$ may no longer be valid. So we will consider the quantum optical limit rather than the quantum Brownian motion limit. The interaction bewteen quarkonium and the QGP is weak since it scales as $rT\sim\frac{T}{Mv}$. In the weak coupling limit (only the weak coupling between quarkonium and the QGP is required, the QGP can be strongly-coupled), the Markovian condition $\tau_R\gg\tau_E$ is valid. We now investigate the second hierarchical condition for the quantum optical limit. The subsystem relaxation time can be estimated as
\be
\tau_R \sim \frac{1}{(rT)^2 T} \sim \frac{M^2v^2}{T^3} \,,
\ee
and the subsystem intrinsic time scale is given by Eq.~(\ref{eqn:tau_s}), 
$\tau_S\sim\frac{1}{Mv^2}$.
We can immediately see that $\tau_R \gg \tau_S$ is well justified if $T\lesssim Mv^2$ and marginally valid if $T\sim Mv^{3/2}$.

Restoring the c.m. motion of the $Q\bar{Q}$ pair in the pNRQCD Lagrangian, we find the mapping between the operators in pNRQCD and those in the general theory introduced in Section~\ref{sect:open} is 
\be
O^{(S)}_{\alpha} &\to& \sqrt{\frac{T_F}{N_c}}\int\diff^3r\, S^\dagger(\bs R, \bs r) r_i \widetilde{O}^a(\bs R, \bs r)\,,\ \ \sqrt{\frac{T_F}{N_c}}\int\diff^3r \, \widetilde{O}^{a\dagger}(\bs R, \bs r) r_i S(\bs R, \bs r)\ \ \ \ \nn\\
&& \frac{d_{abc}}{2} \int\diff^3r\, \widetilde{O}^{b\dagger}(\bs R, \bs r) r_i\widetilde{O}^{c}(\bs R, \bs r) \\
O^{(E)}_{\alpha} &\to& g \widetilde{E}_i^{a\dagger}(\bs R)\,, \ \ g \widetilde{E}_i^{a}(\bs R) \\
\sum_{\alpha} &\to& \sum_i \sum_a \int \diff^3 R \,.
\ee

It is not illuminating to write out the complete Lindblad equation (\ref{eqn:L_optical}) here, which includes both singlet-octet and octet-octet transitions. We will focus on deriving the semiclassical Boltzmann equation for quarkonium from the Lindblad equation. The derivation was first worked out in Ref.~\cite{Yao:2018nmy} for a weakly-coupled QGP and in Ref.~\cite{Yao:2020eqy} for a strongly-coupled QGP. We will write down the evolution equation of the density matrix elements that describe the bound states, which are of the form $\langle {\bs k}_1,nl, s | \rho_S(t) | {\bs k}_2, nl, s \rangle$. Here ${\bs k}$ is the c.m. momentum of the bound $Q\bar{Q}$ pair, $nl$ denotes the radial and orbital angular momentum quantum numbers of the quarkonium state, and $s$ indicates the state is a color singlet. The derivation uses the fact that at leading power in $v$, the eigenenergies of the $Q\bar{Q}$ pair are independent of their c.m. momentum. Also at leading nontrivial power in $r$, the quarkonium state is coupled with the QGP only via the singlet-octet dipole vertex. The term representing dissociation in the Lindblad equation~(\ref{eqn:L_optical}) is $-\frac{1}{2}\gamma_{nm,kl} \{ |k\rangle \langle l|n\rangle \langle m| , \rho_S \}$. As explained in Section~\ref{sect:continuous_eigen}, for subsystems with continuous eigenenergies in the quantum optical limit, transitions with nonzero but small values of $E_k-E_l+E_n-E_m$ will contribute to the term $\gamma_{nm,kl}$. Here we show this does not happen for the evolution of the density matrix elements describing quarkonium. The dissociation term $-\frac{1}{2}\gamma_{nm,kl} \{ |k\rangle \langle l|n\rangle \langle m| , \rho_S \}$ when sandwiched between $\langle {\bs k}_1,nl,s|$ and $| {\bs k}_2,nl,s \rangle$ can be calculated by setting $|k\rangle = |{\bs k}_1,nl, s\rangle$, $|l\rangle = |n \rangle = | {\bs p}_{\ma{cm}}, {\bs p}_{\ma{rel}}, a\rangle$ (an octet pair with color $a$, c.m. and relative momenta ${\bs p}_{\ma{cm}}$, ${\bs p}_{\ma{rel}}$) and $|m\rangle = |{\bs k}_3,n'l', s\rangle$. The eigenenergies are given by $E_k = E_{nl}$, $E_m =  E_{n'l'}$ and $E_l = E_n = p_{\text{rel}}^2/M$ at leading power in $v$ (see Section~\ref{sect:eft}), where $E_{nl}<0$ denotes the binding energy of a quarkonium state with the quantum number $nl$ and $p_{\text{rel}}^2/M$ is the eigenenergy of an unbound $Q\bar{Q}$ pair. Here the states $|l\rangle$ and $|n\rangle$ have the same relative momentum and thus the same energy due to the term $\langle l | n \rangle$ and the fact that the wavefunction of the relative motion satisfies $\langle {\bs p}_{\text{rel}}| {\bs p}_{\text{rel}}'\rangle \propto \delta^3({\bs p}_{\text{rel}} - {\bs p}_{\text{rel}}')$. The fact that the states $|l\rangle$ and $|n\rangle$ have the same relative momentum is true to all orders in the coupling constant at leading (nontrivial) power in $r$. The dipole interaction between two color octets can alter the wavefunction of the relative motion, but this happens beyond the leading (nontrivial) power in $r$ for the evolution of $\langle {\bs k}_1,nl, s | \rho_S(t) | {\bs k}_2, nl, s \rangle$. Then we find $E_k-E_l+E_n-E_m = E_{nl} - E_{n'l'}$, which is discrete. Thus, we can use Eq.~(\ref{eqn:kronecker}) to approximate Eq.~(\ref{eqn:delta}) in the quantum optical limit. Because of the Kronecker delta function for $E_{nl}-E_{n'l'}$ and the Dirac delta function for momentum conservation, only terms with ${\bs k}_3 = {\bs k}_1$ and $n'l'=nl$ contribute if we assume the eigenenergies are nondegenerate. Then one can simplify the dissociation term and obtain~\cite{Yao:2020eqy}
\be
\label{eqn:step_d2}
&-& t \frac{T_F}{2N_c} \int\frac{\diff^3p_{\ma{cm}}}{(2\pi)^3} \frac{\diff^3p_{\ma{rel}}}{(2\pi)^3} \frac{\diff^4q}{(2\pi)^4} (2\pi)^4 \Big(\delta^3({\bs k_1} - {\bs p}_\ma{cm} +{\bs q}) + 
\delta^3({\bs k_2} - {\bs p}_\ma{cm} +{\bs q}) 
\Big) \nn\\
&\times&  
\delta(E_{nl}-E_p+q^0)
\langle \psi_{nl} | r_{i_1} | \Psi_{{\bs p}_\ma{rel}} \rangle 
\langle \Psi_{{\bs p}_\ma{rel}} | r_{i_2} | \psi_{nl} \rangle 
D_{i_1i_2}(q^0,{\bs q})
  \langle {\bs k}_1, nl, s| \rho_S | {\bs k}_2, nl, s\rangle \,,\nn\\
\ee
where $| \psi_{nl} \rangle $ and $| \Psi_{{\bs p}_\ma{rel}} \rangle $ are the wavefunctions of the bound and unbound pair respectively and $D_{i_1i_2}(q^0,{\bs q})$ denotes the environment correlator (\ref{eqn:2point}) and its explicit definition will be given in Section~\ref{sect:structure_f}. After the Wigner transform and proper shifting of ${\bs p}_{\text{cm}}$, the dissociation term becomes
\be
&-& t \frac{T_F}{N_c} \sum_{i_1,i_2} \int\frac{\diff^3p_{\ma{cm}}}{(2\pi)^3} \frac{\diff^3p_{\ma{rel}}}{(2\pi)^3} \frac{\diff^4q}{(2\pi)^4} (2\pi)^4\delta^3({\bs k} - {\bs p}_\ma{cm} + {\bs q}) \delta(E_{nl}-E_p + q^0) \nn\\
\label{eqn:step_d3}
&\times& \langle \psi_{nl} | r_{i_1} | \Psi_{{\bs p}_\ma{rel}} \rangle 
\langle \Psi_{{\bs p}_\ma{rel}} | r_{i_2} | \psi_{nl} \rangle
D_{i_1i_2}(q^0,{\bs q})
f_{nl}({\bs x}, {\bs k}) \equiv -t\, \ml{C}_{nl}^-({\bs x}, {\bs k}) \,.
\ee

The recombination term $\gamma_{nm,kl} |n\rangle \langle m| \rho_S |k\rangle \langle l| $ in Eq.~(\ref{eqn:L_optical}) when sandwiched between $\langle {\bs k}_1,nl,s|$ and $| {\bs k}_2,nl,s \rangle$ can be calculated by setting $|n\rangle= |{\bs k}_1,nl,s\rangle$, $|l\rangle=| {\bs k}_2,nl,s \rangle$, $|m\rangle = |{\bs p}_{1\ma{cm}}, {\bs p}_{1\ma{rel}}, a_1\rangle $ and $|k\rangle = |{\bs p}_{2\ma{cm}}, {\bs p}_{2\ma{rel}}, a_2\rangle $. Here $E_k-E_l+E_n-E_m = p_{2\text{rel}}^2/M - p_{1\text{rel}}^2/M$ is continuous and not gapped from zero, which corresponds to the case discussed in Section~\ref{sect:continuous_eigen}. Quantum transitions with ${\bs p}_{1\text{rel}} \neq {\bs p}_{2\text{rel}}$ can occur in the quantum optical limit. However, as we will show in the following, in the semiclassical limit, only the transition with ${\bs p}_{1\text{rel}} = {\bs p}_{2\text{rel}}$ will contribute. The case with ${\bs p}_{1\text{rel}} \neq {\bs p}_{2\text{rel}}$ can be included as corrections to the semiclassical limit, which are originated from the higher order terms in the gradient expansion. One can show the recombination term $\gamma_{nm,kl} |n\rangle \langle m| \rho_S |k\rangle \langle l| $ leads to~\cite{Yao:2020eqy}
\be
&& \frac{T_F}{N_c} 
\int\frac{\diff^3p_{1\ma{cm}}}{(2\pi)^3} \frac{\diff^3p_{1\ma{rel}}}{(2\pi)^3} 
\frac{\diff^3p_{2\ma{cm}}}{(2\pi)^3} \frac{\diff^3p_{2\ma{rel}}}{(2\pi)^3}  
\int \diff^3 R_1 \int \diff^3R_2 \int_{0}^{t} \diff t_1 \int_{0}^{t} \diff t_2 \nn \\ 
&\times&   e^{i(E_{nl}t_1 - {\bs k}_1\cdot {\bs R}_1 )  -i(E_{p_1}t_1 - {\bs p}_{1\ma{cm}} \cdot {\bs R}_1) } 
       e^{-i(E_{nl}t_2 - {\bs k}_2\cdot {\bs R}_2 )  + i(E_{p_2}t_2 - {\bs p}_{2\ma{cm}} \cdot {\bs R}_2) }  
\langle \psi_{nl} | r_{i_1} | \Psi_{{\bs p}_{1\ma{rel}}} \rangle 
       \nn\\
&\times&
\langle \Psi_{{\bs p}_{2\ma{rel}}} | r_{i_2} | \psi_{nl} \rangle 
D^{a_2a_1}_{i_2i_1}(t_2, t_1, {\bs R}_2, {\bs R}_1)  
\big\langle {\bs p}_{1\ma{cm}}, {\bs p}_{1\ma{rel}}, a_1 \big| \rho_S \big| {\bs p}_{2\ma{cm}}, {\bs p}_{2\ma{rel}}, a_2 \big\rangle \,,
\ee
where $E_p=p^2/M$.
Since the color is intrinsically quantum, we will assume the octet density matrix is diagonal in the color space to connect the quantum transport equation with the semiclassical one
\be
\label{eqn:color_semi}
 \big\langle {\bs p}_{1\ma{cm}}, {\bs p}_{1\ma{rel}}, a_1 \big| \rho_S \big| {\bs p}_{2\ma{cm}}, {\bs p}_{2\ma{rel}}, a_2 \big\rangle
\approx \frac{\delta^{a_1a_2}}{N_c^2-1}
\big\langle {\bs p}_{1\ma{cm}}, {\bs p}_{1\ma{rel}} \big| \rho^{(o)}_S \big| {\bs p}_{2\ma{cm}}, {\bs p}_{2\ma{rel}} \big\rangle \,.
\ee
As done before, we can set $t\to+\infty$ in the Markovian condition. Then the two time integrals can be evaluated as in Eq.~(\ref{eqn:delta}). But we cannot write the result as one delta function multiplied by the time length $t$, since in general terms with ${\bs p}_{1\text{rel}} \neq {\bs p}_{2\text{rel}}$ and thus $E_{p1} \neq E_{p2}$ are nonvanishing. But in the semiclassical limit, we can still obtain one delta function multiplied by the time length $t$. The key is the gradient expansion of the Wigner transform, which corresponds to the semiclassical expansion.
To apply the Wigner transform, we set ${\bs k}_1 = {\bs k} + {\bs k}'/2$ and ${\bs k}_2 = {\bs k} - {\bs k}'/2$. Changing variables ${\bs p}_{1\ma{cm}} \to {\bs p}_{\ma{cm}} + {\bs p}'_{\ma{cm}}/2$ and ${\bs p}_{2\ma{cm}} \to {\bs p}_{\ma{cm}} - {\bs p}'_{\ma{cm}}/2$, we obtain
\be
\label{eqn:step_r3}
&& \frac{T_F}{N_c}
\int\frac{\diff^3p_{\ma{cm}}}{(2\pi)^3} \frac{\diff^3p_{1\ma{rel}}}{(2\pi)^3} 
 \frac{\diff^3p_{2\ma{rel}}}{(2\pi)^3}
\frac{\diff^4q}{(2\pi)^4}
\int_{0}^{t} \diff t_1 \int_{0}^{t} \diff t_2 \, e^{i(E_{nl}-E_{p_1}-q^0)t_1}  \nn \\
&\times&  
e^{-i(E_{nl}-E_{p_2}-q^0)t_2}
(2\pi)^3\delta^3({\bs k} - {\bs p}_{\ma{cm}} - {\bs q}) 
 \langle \psi_{nl} | r_{i_1} | \Psi_{{\bs p}_{1\ma{rel}}} \rangle
\langle \Psi_{{\bs p}_{2\ma{rel}}} | r_{i_2} | \psi_{nl} \rangle
 D_{i_2i_1}(q^0,{\bs q}) \nn\\
 &\times& \int \diff^3 x_\ma{rel}\, e^{-i({\bs p}_{1\ma{rel}} - {\bs p}_{2\ma{rel}}) \cdot {\bs x}_\ma{rel} } \frac{1}{N_c^2-1} f_{Q\bar{Q}}^{(o)} \Big({\bs x}_\ma{cm}, {\bs p}_{\ma{cm}}, {\bs x}_\ma{rel}, \frac{ {\bs p}_{1\ma{rel}}+{\bs p}_{2\ma{rel}} }{2} \Big) \,,
\ee
where $f_{Q\bar{Q}}^{(o)} ({\bs x}_\ma{cm}, {\bs p}_{\ma{cm}}, {\bs x}_\ma{rel}, \frac{ {\bs p}_{1\ma{rel}}+{\bs p}_{2\ma{rel}} }{2} )$ is the phase space distribution of a color octet $Q\bar{Q}$ pair with c.m. and relative positions and momenta ${\bs x}_\ma{cm}, {\bs p}_{\ma{cm}}, {\bs x}_\ma{rel}, \frac{ {\bs p}_{1\ma{rel}}+{\bs p}_{2\ma{rel}} }{2}$. 
Expanding $f_{Q\bar{Q}}^{(o)}$ around ${\bs x}_{\ma{rel}}=0$ leads to
\be
\label{eqn:gradient}
f_{Q\bar{Q}}^{(o)}({\bs x}_\ma{cm}, {\bs p}_{\ma{cm}}, {\bs x}_\ma{rel}, \frac{ {\bs p}_{1\ma{rel}}+{\bs p}_{2\ma{rel}} }{2}, t ) 
= f_{Q\bar{Q}}^{(o)}({\bs x}_\ma{cm}, {\bs p}_{\ma{cm}}, 0, \frac{ {\bs p}_{1\ma{rel}}+{\bs p}_{2\ma{rel}} }{2} ,t ) \nn\\
+  {\bs x}_\ma{rel}  \cdot \nabla_{{\bs x}_\ma{rel}} f_{Q\bar{Q}}^{(o)}({\bs x}_\ma{cm}, {\bs p}_{\ma{cm}}, {\bs x}_\ma{rel}, \frac{ {\bs p}_{1\ma{rel}}+{\bs p}_{2\ma{rel}} }{2},t )\Big|_{{\bs x}_\ma{rel}=0} + \cdots \,, 
\ee
which corresponds to the gradient expansion. The leading order term gives the collision term in the semiclassical Boltzmann equation while the next-leading order term gives the leading quantum correction~\cite{Yao:2020eqy}. Keeping only the leading term in the gradient expansion, the integral over ${\bs x}_{\text{rel}}$ gives $\delta^3({\bs p}_{1\text{rel}} - {\bs p}_{2\text{rel}})$, which leads to $E_{p1}=E_{p2}$. Then we can evaluate the time integrals as in the calculation of dissociation. Finally we obtain the Wigner transform of the recombination term
\be
&& t \frac{T_F}{N_c} \sum_{i_1,i_2}
\int\frac{\diff^3p_{\ma{cm}}}{(2\pi)^3} \frac{\diff^3p_{\ma{rel}}}{(2\pi)^3}
\frac{\diff^4q}{(2\pi)^4} 
(2\pi)^4\delta^3({\bs k} - {\bs p}_{\ma{cm}} - {\bs q}) 
\delta(E_{nl} - E_p - q^0) D_{i_2i_1}(q^0,{\bs q}) \nn\\
&\times& \langle \psi_{nl} | r_{i_1} | \Psi_{{\bs p}_{1\ma{rel}}} \rangle
\langle \Psi_{{\bs p}_{2\ma{rel}}} | r_{i_2} | \psi_{nl} \rangle
\frac{f_{Q\bar{Q}}^{(o)}({\bs x}_\ma{cm}, {\bs p}_{\ma{cm}}, {\bs 0}, {\bs p}_{\ma{rel}} ) }{N_c^2-1} \equiv t\, \ml{C}_{nl}^+({\bs x}_\ma{cm}, {\bs k})  \,.
\ee
Putting everything together leads to
\be
\label{eqn:Boltzmann}
\frac{\partial}{\partial t} f_{nl}({\bs x}, {\bs k}, t) + \frac{{\bs k}}{2M} \cdot \nabla_{\bs x} f_{nl}({\bs x}, {\bs k}, t)
= \ml{C}_{nl}^+({\bs x}, {\bs k}, t) - \ml{C}_{nl}^-({\bs x}, {\bs k}, t) \,,
\ee
where the free streaming term on the left hand side is originated from the Wigner transform of the commutator $[H_S+\Delta H_S,\rho_S]$ in the Lindblad equation. Details of the derivation can be found in Refs.~\cite{Yao:2018nmy,Yao:2020eqy}\,. Similar collision terms can also be found in Ref.~\cite{Yao:2018sgn}\,. The leading quantum correction to the recombination term $\ml{C}^+_{nl}$ can be found in Ref.~\cite{Yao:2020eqy}\,. If the position ${\bs x}$ and momentum ${\bs k}$ are integrated over the whole phase space, the Boltzmann equation turns to a rate equation. The early work~\cite{Borghini:2011ms} treated quarkonium as a multi-level system and studied the diagonal elements of the Lindblad equation (\ref{eqn:rate_optical_diag}), which is similar to a rate equation.

\section{Physical Implications}
\label{sect:pheno}
\subsection{Structure Functions of Quark-Gluon Plasma}
\label{sect:structure_f}
The environment correlators that appear in the Lindblad equation for the quantum Brownian motion~(\ref{eqn:lindblad_aka1})  are given by
\be
D^>(x_1-x_2) &=& g^2\langle A_0^a(x_1) A_0^a(x_2) \rangle_T \\
\Sigma(x_1-x_2) &=& -i g^2 \sign(t_1-t_2) \langle  A_0^a(x_1) A_0^a(x_2) \rangle_T \,.
\ee
They are not gauge invariant and thus cannot be interpreted as structure functions reflecting the properties of the QGP. To make them gauge invariant, resummations of higher order interactions are necessary. One possible way of resummation can be shown in the small quarkonium size limit, by using the pNRQCD Lagrangian.

\begin{figure}[t]
\centering
\subfloat[Wilson lines for dissociation.]{\includegraphics[height=1.95in]{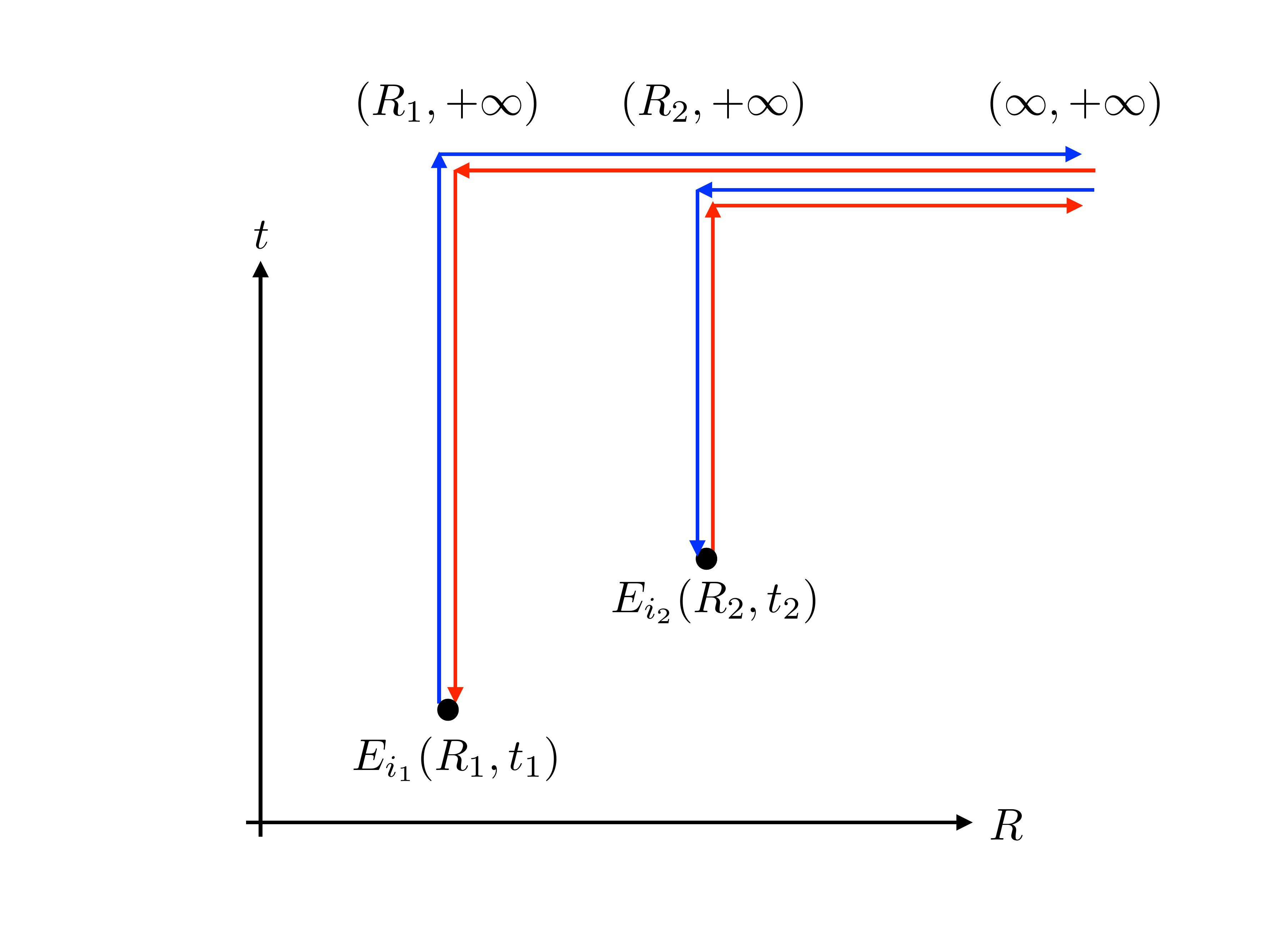}}
\subfloat[Wilson lines for recombination.]{\includegraphics[height=1.95in]{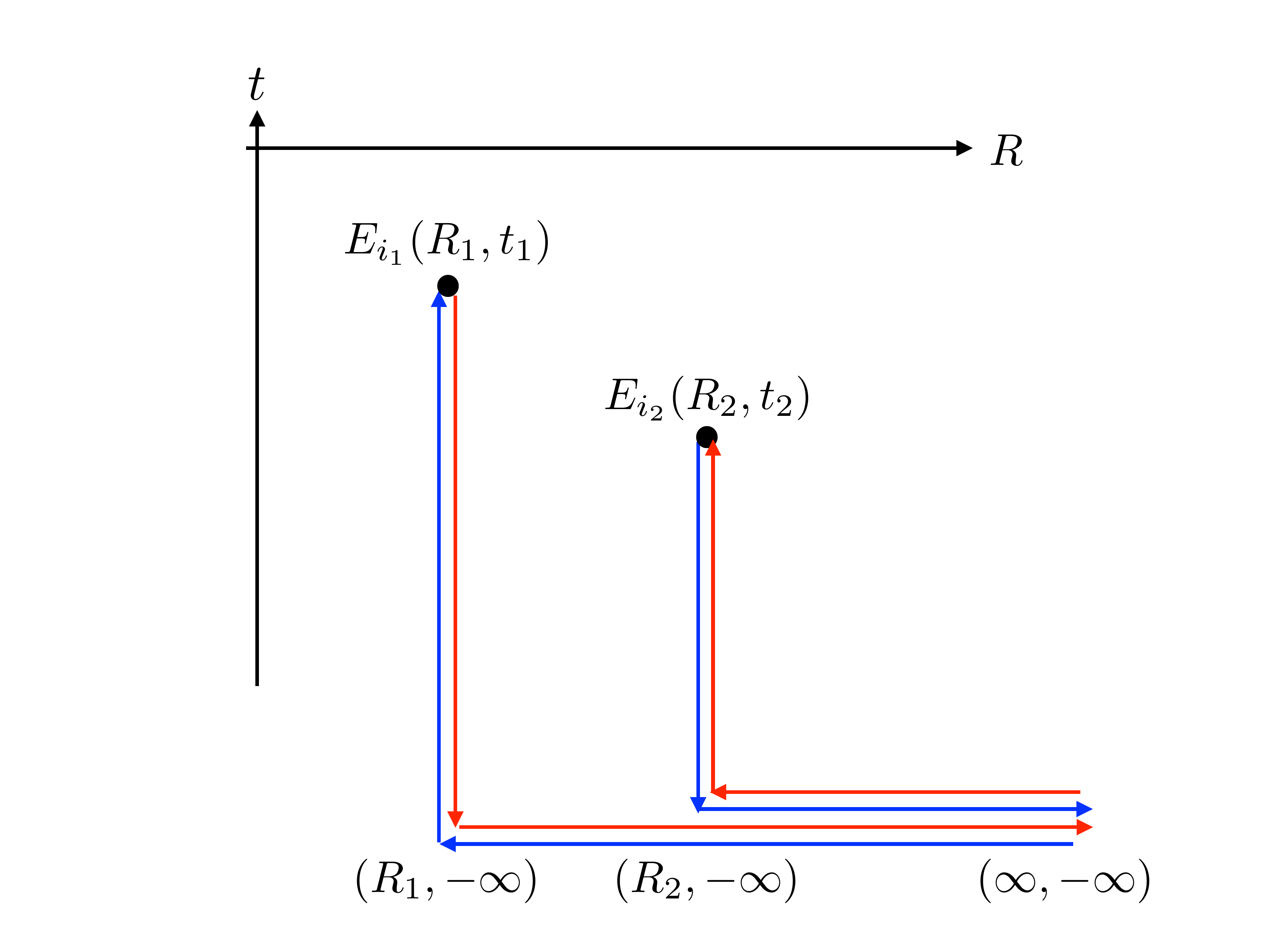}}
\caption{Staple-shape Wilson lines in the definition of the environment correlators~(\ref{eqn:g++}). The double arrow indicates the adjoint representation. The spatial Wilson lines at infinite time are generated by the Coulomb modes which mediate the interaction between the c.m. motion of an octet $Q\bar{Q}$ pair and the QGP. A series of Feynman diagrams is resummed to obtain the spatial Wilson lines~\cite{Yao:2020eqy}. The plots are taken from Ref.~\cite{Yao:2020eqy}\,.}
\label{fig:wilson}
\end{figure}

The environment correlators that show up in the Boltzmann equation derived in the quantum optical limit (see Section~\ref{sect:boltzmann}) are defined by
\be
\label{eqn:g++}
D_{i_1i_2}(t_1,{\bs R}_1, t_2,{\bs R}_2) = g^2\big\langle \big( E_{i_1}({\bs R}_1,t_1) \ml{W}_1 \ml{W}'_1 \big)^a \big( \ml{W}'_2 \ml{W}_2 E_{i_2}({\bs R}_2,t_2) \big)^a
\big\rangle_T\,,
\ee
where $\ml{W}_1$, $\ml{W}'_1$, $\ml{W}_2'$ and $\ml{W}_2$ are four Wilson lines in the adjoint representation, shown in Fig.~\ref{fig:wilson}. The Wilson lines along the time axis resum the octet-$A_0$ interactions and are obtained from the field redefinition explained in Section~\ref{sect:leading_r}. For dissociation, the Wilson lines resum final-state interactions since the octet is a final state while for recombination, the octet is an initial state and the Wilson lines resum initial-state interactions. The difference is reflected in whether the Wilson lines connect with positive or negative infinite time. The Wilson lines along the spatial direction at infinite time resum Coulomb interactions between the c.m. motion of the octet and the gauge field. The Coulomb interaction between the heavy quark and antiquark is already accounted for in the singlet and octet Hamiltonians $H_{s,o}$. The momentum scaling of the Coulomb mode is
$p_c^\mu \sim M(v^2,v,v,v) $.
For Coulomb modes, the c.m. kinetic energy term $-\frac{D_{\bs R}^2}{4M}$ in Eq.~(\ref{eqn:Ho}) is at leading power in $v$ and thus the interactions therein must be calculated to all orders for the leading power construction. This resummation was first done in Ref.~\cite{Yao:2020eqy} by a diagram-by-diagram calculation. This chromoelectric correlator is gauge invariant and thus is a structure function that encodes properties of the QGP. In energy-momentum space, this structure function is momentum dependent.

For the Lindblad equation discussed in Section~\ref{sect:brown_pnrqcd}, the c.m. motion of the $Q\bar{Q}$ pair is dropped. So both ${\bs R}_1$ and ${\bs R}_2$ in Eq.~(\ref{eqn:g++}) can be set to ${\bs 0}$. In this case, the two chromoelectric fields are located at the same position and only the Wilson lines along the time axis shown in Fig.~\ref{fig:wilson} matter\footnote{The Wilson lines along the spatial direction at infinite time can be dropped. The remaining correlator~(\ref{eqn:corre_t}) is already gauge invariant, since the two Wilson lines along the time axis end at the same spacetime point and we have the freedom to choose the global gauge at that point.}
\be
\label{eqn:corre_t}
D_{i_1i_2}(t_1, t_2) = g^2 \big\langle  \big( E_{i_1}(t_1) \ml{W}_{[t_1,\pm\infty]} \big)^a \big( \ml{W}_{[\pm\infty,t_2]} E_{i_2}(t_2) \big)^a
\big\rangle_T\,.
\ee
In energy-momentum space, this structure function is momentum independent. The other environment correlator that appears in the Lindblad equation for the quantum Brownian motion discussed in Section~\ref{sect:brown_pnrqcd} is
\be
\Sigma_{i_1i_2}(q_0=0) = g^2 \int_{-\infty}^{+\infty}\diff t\, \text{Im} \big\langle \ml{T} \big( E_{i_1}(t) \ml{W}_{[t,\pm\infty]} \big)^a \big( \ml{W}_{[\pm\infty,0]} E_{i_2}(0) \big)^a
\big\rangle_T\,,
\ee
which is also gauge invariant and encodes properties of the QGP.

In the limit of quantum Brownian motion, only the zero energy limits of the momentum independent chromoelectric structure functions $D$ and $\Sigma$ contribute. The zero energy limit of the momentum independent structure function $D(q_0=0)$ is just the heavy quark diffusion coefficient $\kappa$ (up to a normalization constant) that has been calculated perturbatively by using the HTL effective theory~\cite{CaronHuot:2007gq,CaronHuot:2008uh} and nonperturbatively on lattice~\cite{Banerjee:2011ra,Francis:2015daa,Brambilla:2020siz} and via AdS/CFT~\cite{CasalderreySolana:2006rq,Gubser:2006qh,CaronHuot:2008uh,casalderrey2014gauge}. The quenched lattice QCD calculation in Ref.~\cite{Brambilla:2020siz} calculated $\kappa$ in a wide range of temperatures (up to $T=10^4T_c$). Recent extractions of the heavy quark diffusion coefficient from experimental data can be found in Refs.~\cite{Xu:2017obm,Cao:2018ews,Xu:2018gux}\,. The other structure function $\Sigma$ in the zero energy limit has not been calculated nonperturbatively, neither has the more general momentum dependent chromoelectric structure function $D(q_0,{\bs q})$. Ref.~\cite{Brambilla:2019tpt} estimated $\Sigma(q_0=0)$ using its relation with the quarkonium mass shift in the medium. 

\begin{figure}[t]
\centering
\includegraphics[height=0.9in]{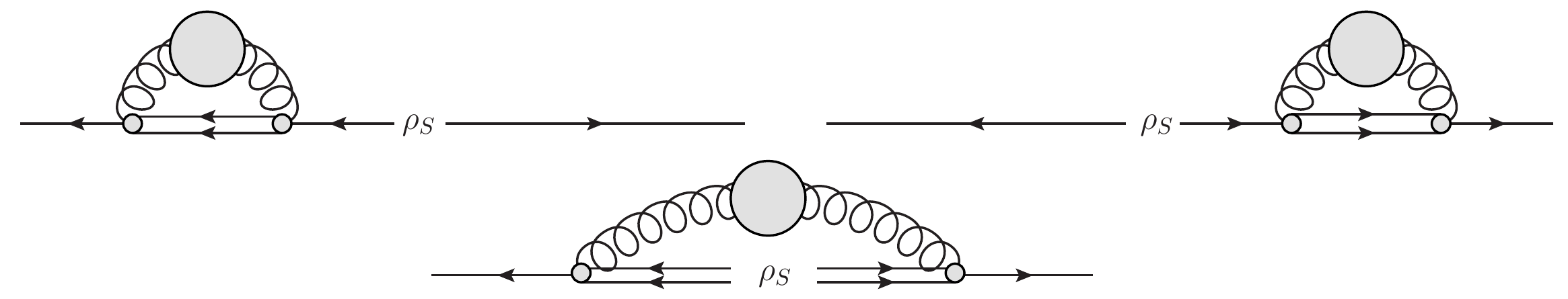}
\caption{Diagrammatic representation of the time evolution of a quarkonium state in the open quantum system framework. The single solid line represents a singlet $Q\bar{Q}$ pair while the double solid line denotes an octet pair. The grey blob denotes the environment correlators. The two diagrams in the first row describe Debye screening and dissociation. The diagram on the second row represents regeneration. This figure is similar to Figure 2 of Ref.~\cite{Yao:2020kqy}\,.}
\label{fig:unify}
\end{figure}

\subsection{Unifying Debye Screening, Dissociation and Recombination}
As discussed in the Introduction, the physical understanding of Debye screening and dissociation can be unified by studying the thermal loop correction to the quarkonium propagator. The open quantum system framework provides a way to unite Debye screening, dissociation and recombination. The time evolution of the density matrix element that involves a quarkonium state $\langle {\bs k}_1,nl, s | \rho_S(t) | {\bs k}_2, nl, s \rangle$, can be represented diagrammatically in Fig.~\ref{fig:unify}. The single solid line represents a color singlet $Q\bar{Q}$ pair while the double solid line denotes an octet pair. The grey blob denotes the environment correlators. The two diagrams in the first row describe the thermal loop correction to the color singlet propagator and thus account for Debye screening and dissociation. The diagram on the second row, which is similar to an interference between the amplitude and its complex conjugate, accounts for the transition from the octet to the singlet and thus can describe quarkonium regeneration. The trace preserving property of the Lindblad equation guarantees that the total probability of the subsystem is conserved, i.e., the total number of $Q\bar{Q}$ pairs is conserved. At equilibrium, dissociation and recombination reach detailed balance~\cite{Yao:2017fuc}. The consistency between dissociation and recombination is also reflected in the  Kubo–Martin–Schwinger relation satisfied by the environment correlator $D(x_1,x_2)$ that appears in both the dissociation and recombiantion terms in the Lindblad equation (see e.g. Section~\ref{sect:boltzmann}).

\subsection{Coupled Evolution of Heavy Quarks and Quarkonia}
In the quantum Brownian motion limit, the dynamical evolution of a quarkonium state is described as an evolution of the $Q\bar{Q}$ pair that is interacting with the QGP and with each other. It can be thought of as an evolution of open heavy quarks, plus interactions among heavy quarks. In the quantum optical limit, the dynamical evolution of a quarkonium state is described as transitions between different eigenstates that include bound singlet, unbound singlet and unbound octet states. In either limit, the dynamics of quarkonium is deeply connected with the dynamics of open heavy quarks. In the quantum Brownian motion limit, this is also reflected in that the environment correlator $D(q_0=0)$, i.e., the heavy quark diffusion coefficient, affects the dynamics of both open heavy quarks and quarkonia. In the quantum optical limit, the recombination term in the Boltzmann equation depends on the phase space distribution of unbound $Q\bar{Q}$ pairs, which needs to be separately solved from real-time dynamics. Semiclassically, the coupled dynamics of open and hidden heavy flavors can be realized by using coupled Langevin equations of color singlet and octet states~\cite{Blaizot:2017ypk,Akamatsu:2020ypb} or coupled Boltzmann equations~\cite{Yao:2018zrg,Yao:2020xzw}. In the former case, the singlet-octet transition is governed by a rate equation. In the latter case, the dynamics of the open heavy quarks is described by linearized Boltzmann equations~\cite{Svetitsky:1987gq,Gossiaux:2008jv,Gossiaux:2009mk,Uphoff:2014hza,Ke:2018tsh,Ke:2018jem} and the dynamics of quarkonia is governed by Eq.~(\ref{eqn:Boltzmann}).
\begin{figure}[t]
    \centering
    \includegraphics[height=2.0in]{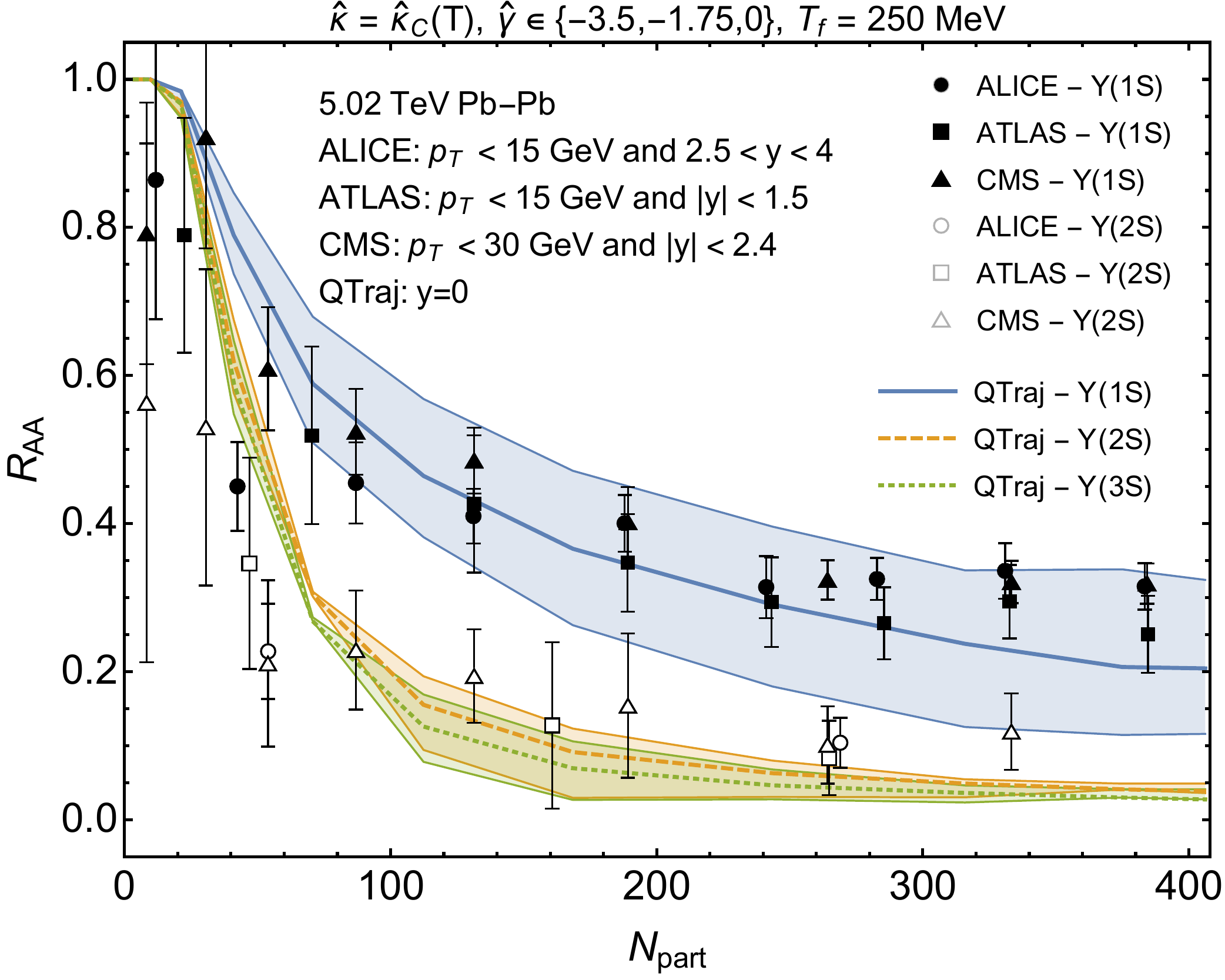}
\caption{Calculation results of the recoil-less Lindblad equation~(\ref{eqn:pheno_lindblad}) in the limit of quantum Brownian motion~\cite{Brambilla:2020qwo} on bottomonia $R_\ma{AA}$ as a function of centrality in $5.02$ TeV Pb-Pb collisions. The central value of a lattice QCD calculation result of the parameter $\kappa$ is used. The uncertainty band is generated by three different choices of the parameter $\gamma$. The freezeout temperature is $T_f=250$ MeV. Experimental data are taken from the results of the ALICE~\cite{Acharya:2020kls}, ATLAS~\cite{Lee:2021vlb} and CMS~\cite{Sirunyan:2018nsz} collaborations. The plot is provided courtesy of authors of Ref.~\cite{Brambilla:2020qwo} (version 1).}
\label{fig:raa_recoilless}
\end{figure}

Phenomenological results of the recoil-less Lindblad equation~(\ref{eqn:pheno_lindblad}) in the limit of quantum Brownian motion~\cite{Brambilla:2020qwo} on the bottomonium nuclear modification factors $R_{\text{AA}}$ (that measures how much quarkonium production is suppressed in heavy-ion collisions) are shown in Fig.~\ref{fig:raa_recoilless} as a function of centrality (or equivalently, the number of participant nucleons in one heavy-ion collision $N_{\rm{part}}$). In the plot, the central value of a lattice QCD calculation result of the parameter $\kappa$ is used. The uncertainty band is generated by three different choices of the parameter $\gamma$, listed in the plot (see also Eq.~(\ref{eqn:gamma})). The freezeout temperature is fixed at $T_f=250$ MeV. More precise nonperturbative calculations of the parameters will help to reduce the uncertainty in the calculation results. 

\begin{figure}[t]
\centering
\subfloat[$R_\ma{AA}(N_\ma{part})$ in $5.02$ TeV Pb-Pb collisions.]{\includegraphics[height=1.8in]{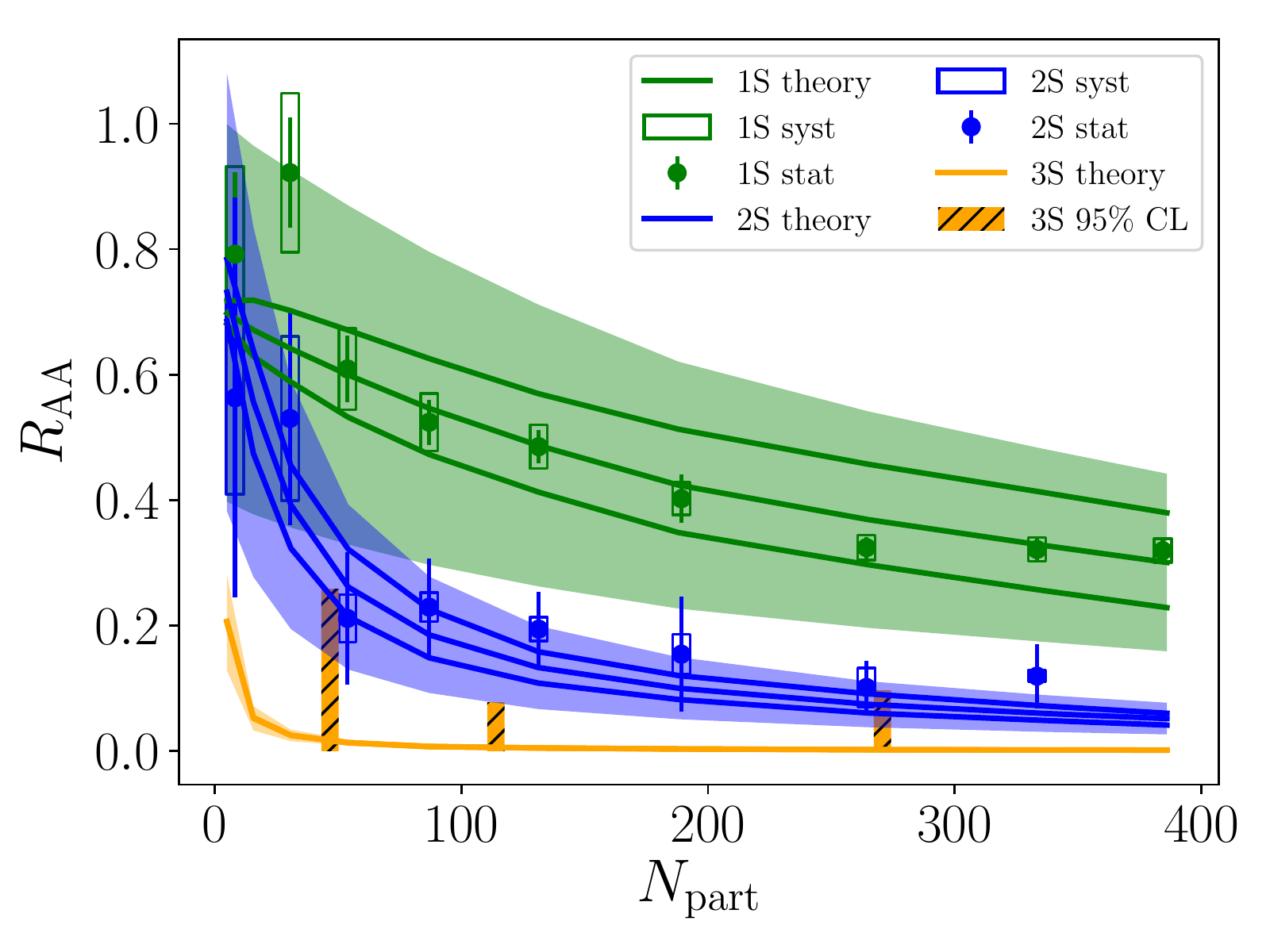}}
\subfloat[$R_\ma{AA}(p_T)$ in $5.02$ TeV Pb-Pb collisions.]{\includegraphics[height=1.8in]{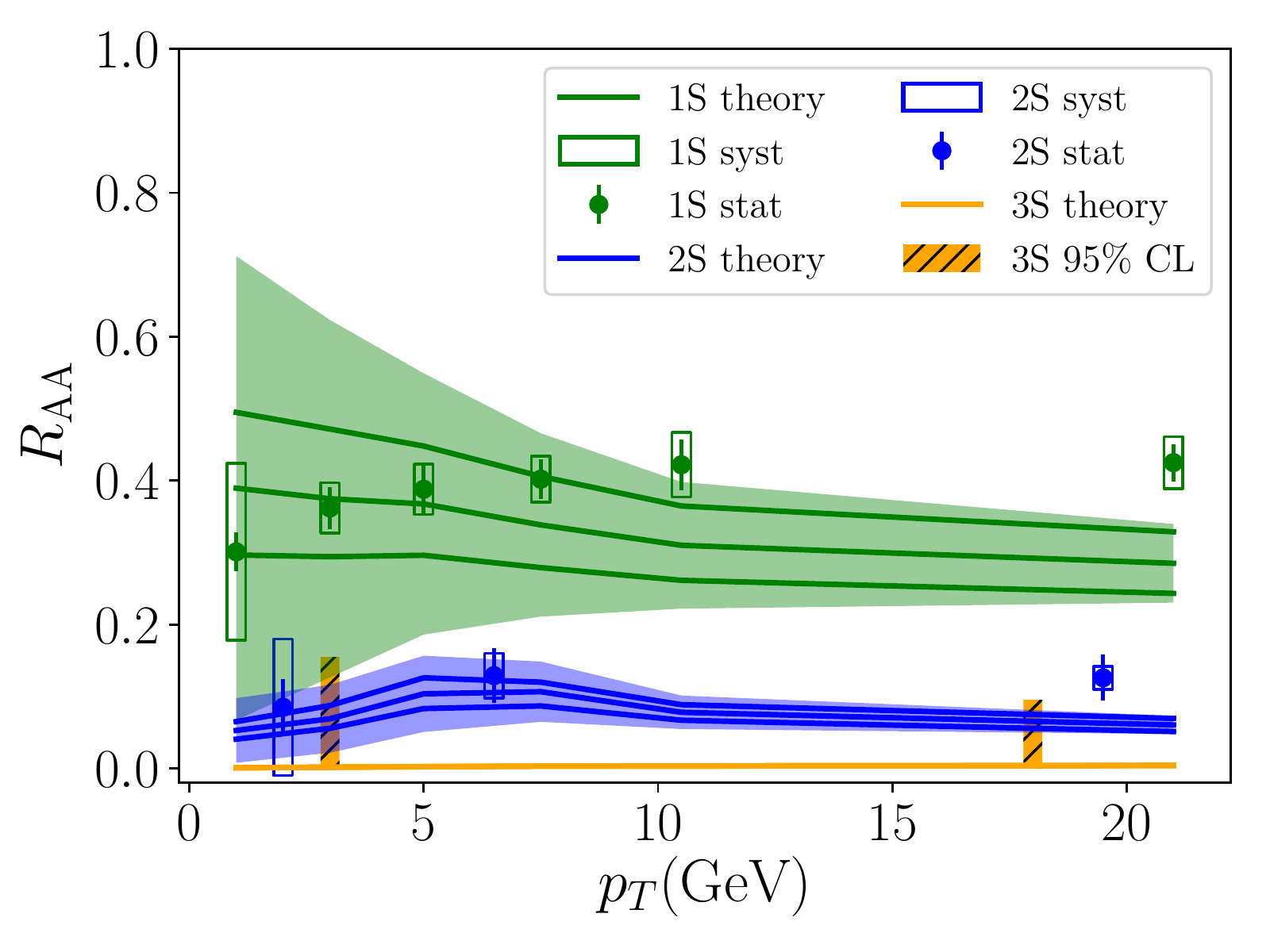}}

\subfloat[$R_\ma{AA}(y)$ in $5.02$ TeV Pb-Pb collisions.]{\includegraphics[height=1.8in]{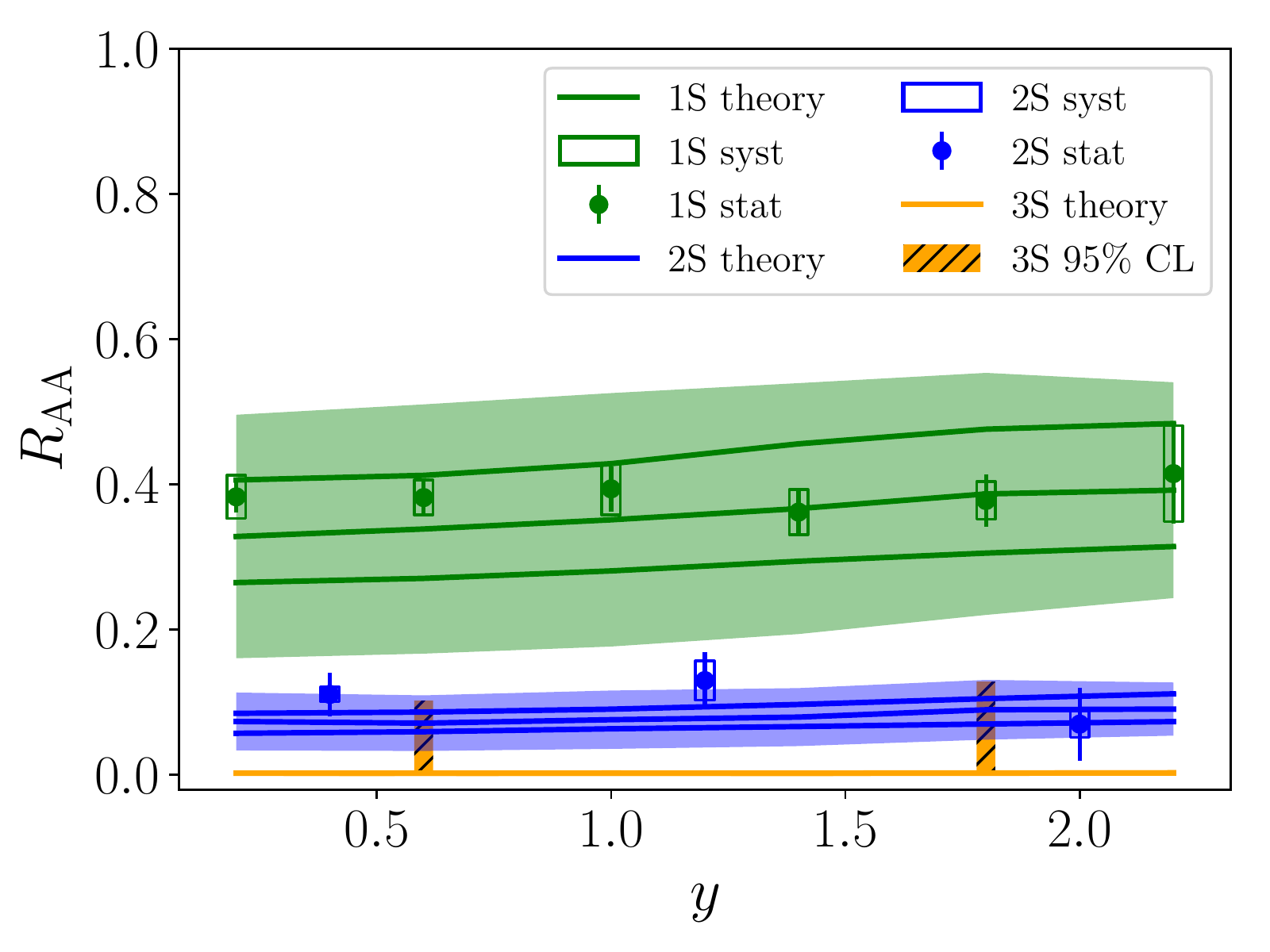}}
\subfloat[$R_\ma{AA}(N_\ma{part})$ in $2.76$ TeV Pb-Pb collisions.]{\includegraphics[height=1.8in]{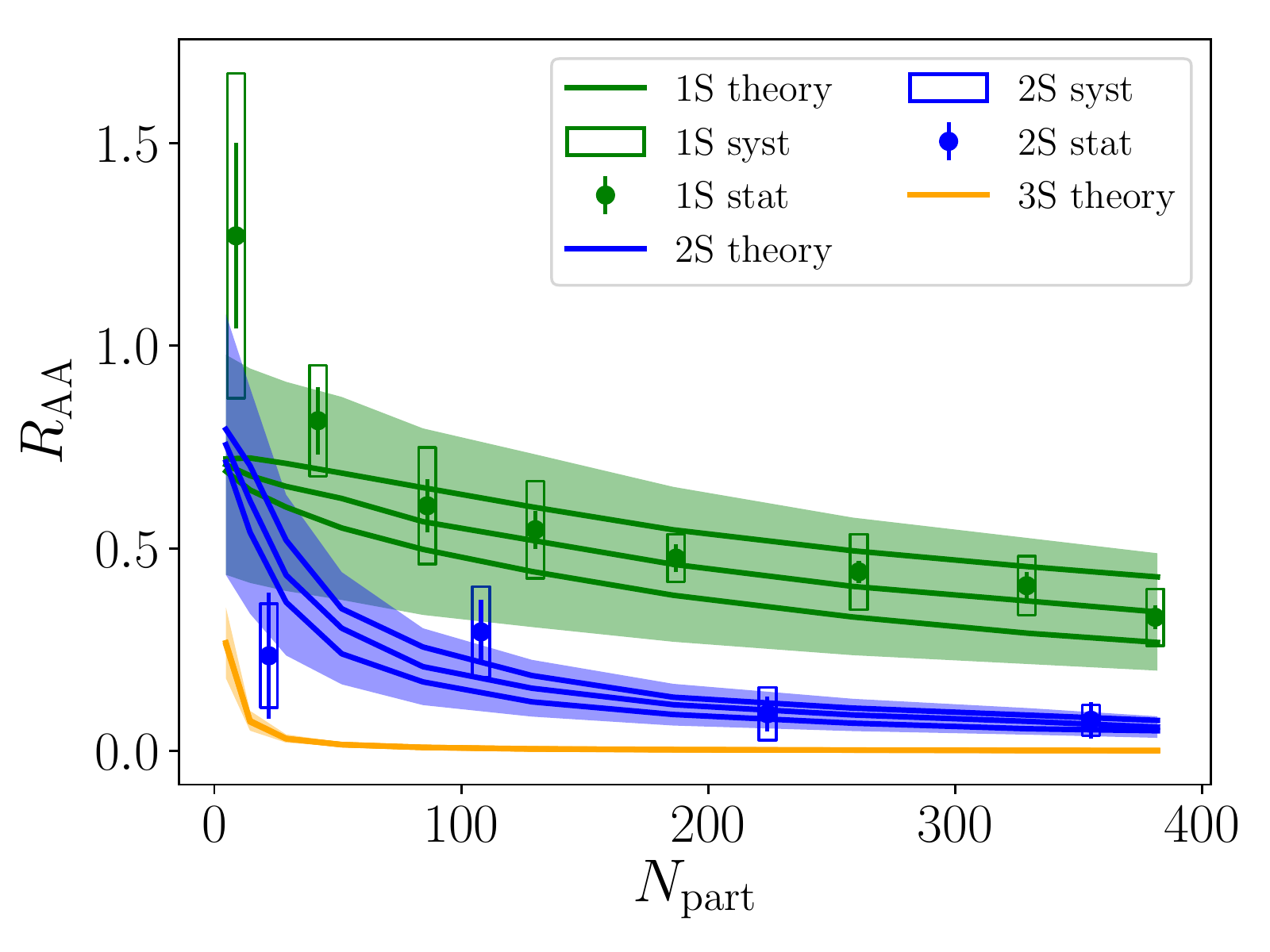}}

\subfloat[$R_\ma{AA}(p_T)$ in $2.76$ TeV Pb-Pb collisions.]{\includegraphics[height=1.8in]{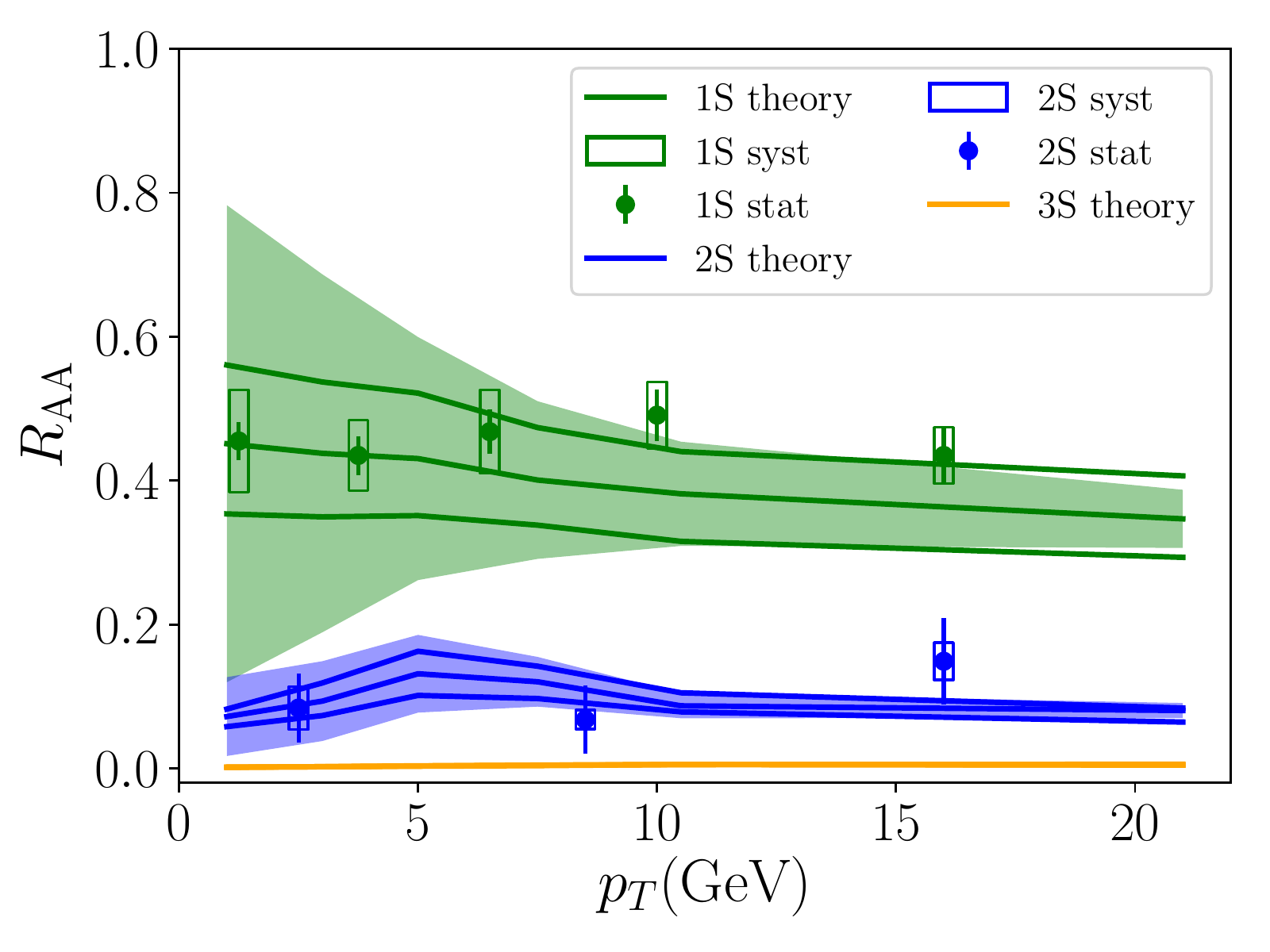}}
\subfloat[$R_\ma{AA}(N_\ma{part})$ in $200$ GeV Au-Au collisions.]{\includegraphics[height=1.8in]{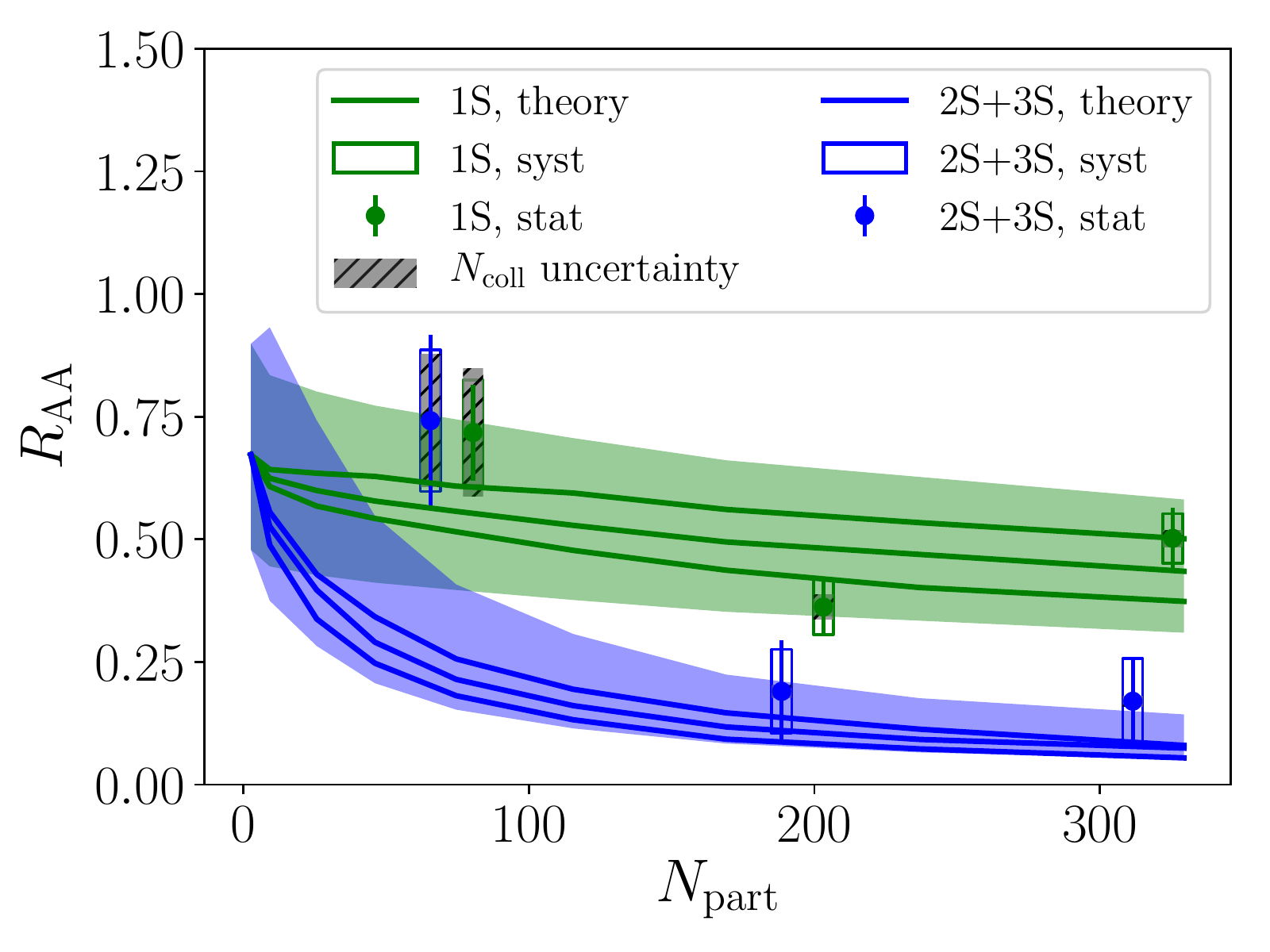}}
\caption{Calculation results of coupled Boltzmann equations~\cite{Yao:2020xzw} on bottomonia $R_\ma{AA}$ as functions of centrality, transverse momentum and momentum rapidity in different collisions. The upper and lower curves correspond to calculations with parameters that differ by $\pm10\%$ respectively from the parameters used in the middle curve. The band is from the uncertainty of the nuclear parton distribution function and is centered at the middle curve. Experimental data are taken from the results of the CMS~\cite{Sirunyan:2018nsz,Khachatryan:2016xxp} and STAR~\cite{Wang:2019vau} collaborations. The plots are taken from Ref.~\cite{Yao:2020xzw}\,.}
\label{fig:raa}
\end{figure}

Phenomenological results of the coupled Boltzmann transport equations~\cite{Yao:2020xzw} on bottomonium nuclear modification factors $R_{\text{AA}}$ are shown in Fig.~\ref{fig:raa}, as functions of centrality (or equivalently, $N_{\rm{part}}$), transverse momentum $p_T$ and momentum rapidity $y$. In the calculations of Ref.~\cite{Yao:2020xzw}\,, the uncertainty caused by the nuclear parton distribution function dominates over the uncertainty of the parameters describing the quarkonium in-medium evolution. The calculations of Ref.~\cite{Yao:2020xzw} showed that it is important to include correlated recombination into the description of quarkonium in-medium transport, which is motivated from the open quantum system studies. Correlated recombination also motivates a new observable that has never been measured before and may dramatically alter our physical understanding of quarkonium transport in the QGP. We will now discuss the concept of correlated recombination, its relation with the open quantum system approach and the new observable.

\subsection{Decoherence and Correlated Recombination}
\begin{figure}[t]
\centering
\subfloat[As a function of centrality.]{\includegraphics[height=1.8in]{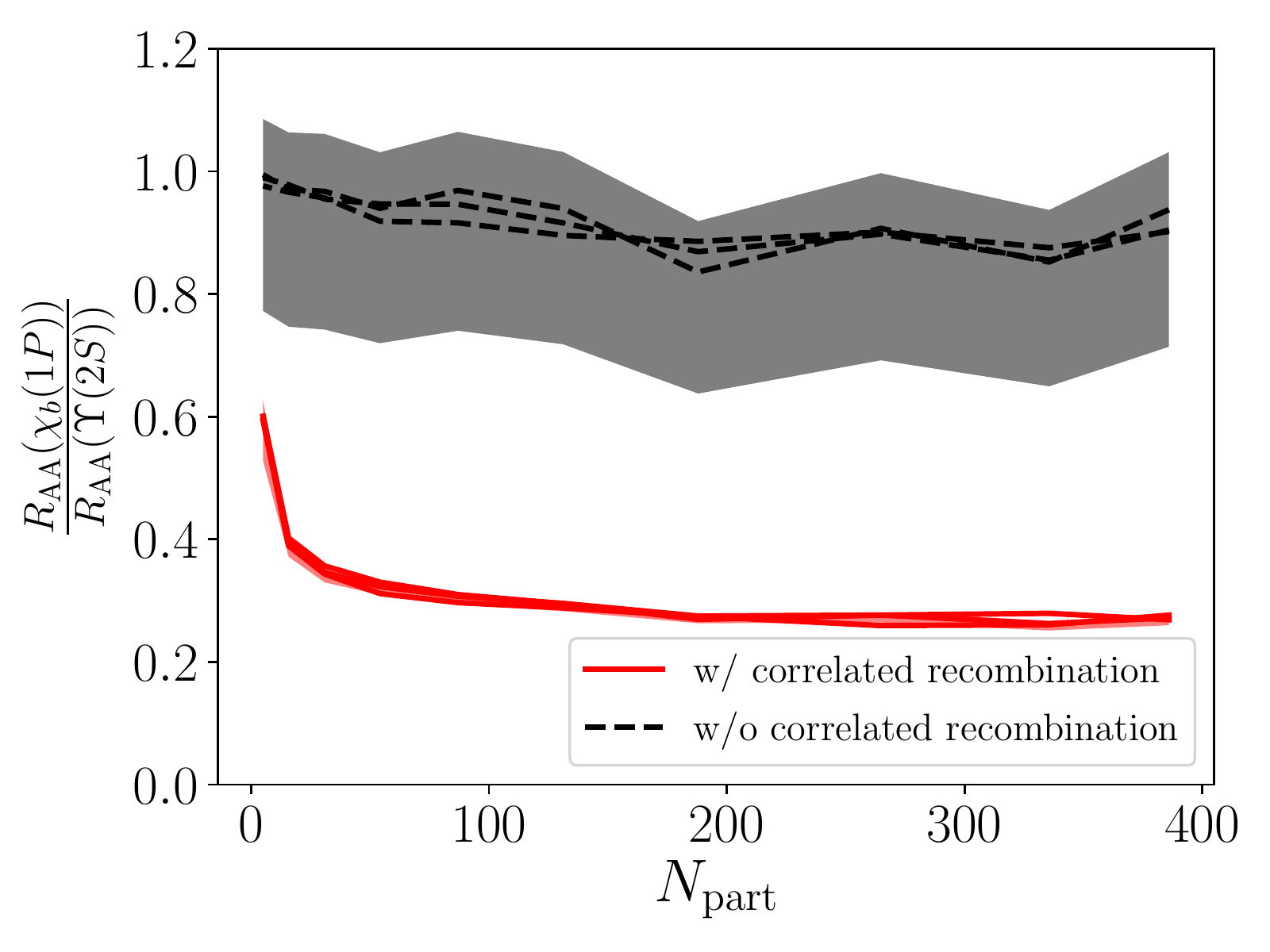}}
\subfloat[As a function of transverse momentum.]{\includegraphics[height=1.8in]{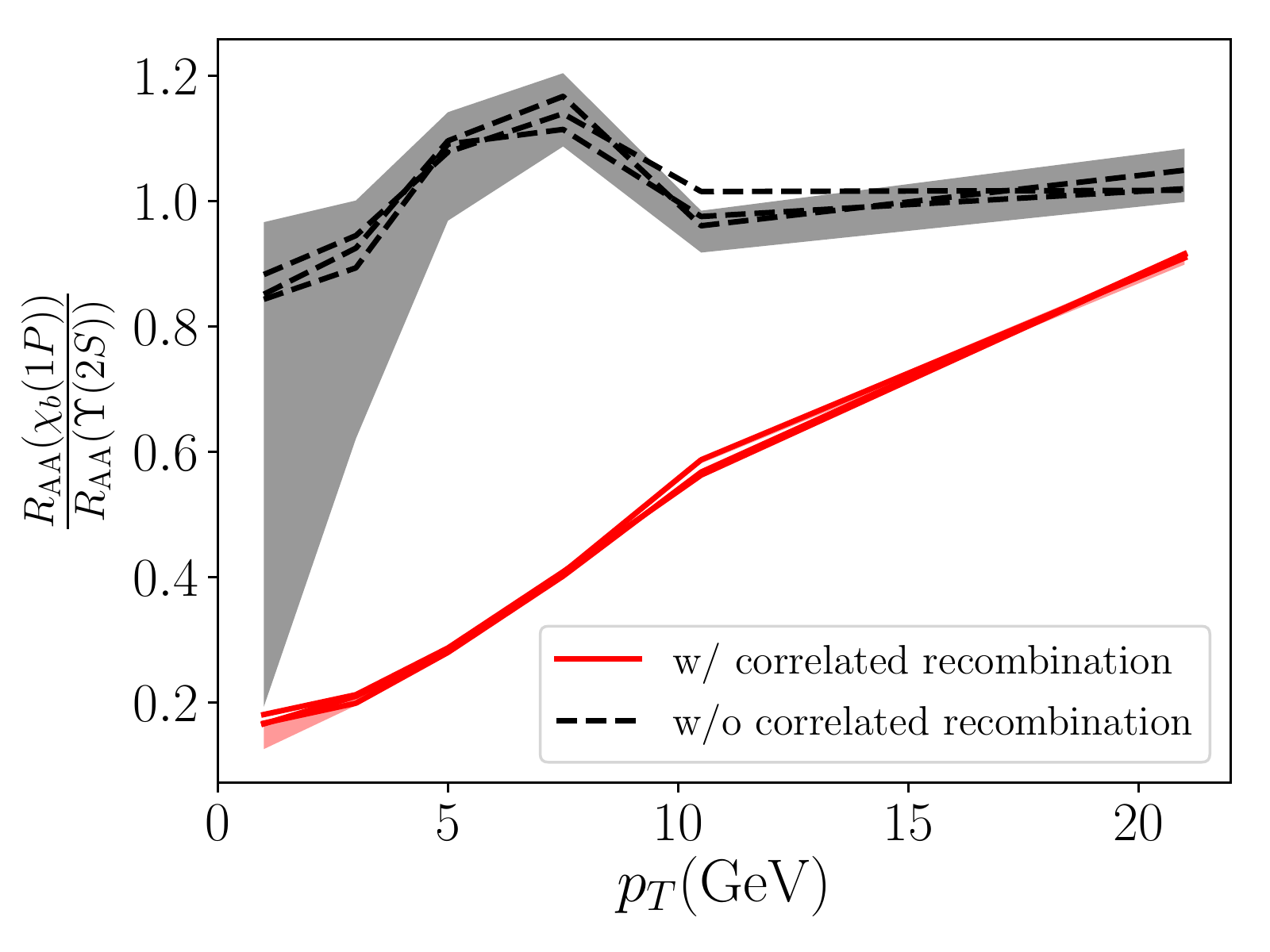}}
\caption{Ratios of $R_{\ma{AA}}(\chi_b(1P))$ and $R_{\ma{AA}}(\Upsilon(2S))$ as functions of centrality and transverse momentum. Different curves correspond to different choices of parameters and the band indicates the uncertainty of the nuclear parton distribution function. The double ratio observable has huge discriminatory power to distinguish calculations with and without correlated recombination. The right plot is taken from Ref.~\cite{Yao:2020xzw}\,.}
\label{fig:chi_b}
\end{figure}

In the quantum optical limit, the dissociation of a quarkonium state appears as a transition from the quarkonium state to an unbound scattering wave. On the other hand, in the quantum Brownian motion limit, the dissociation shows up as a result of the wavefunction decoherence in the $Q\bar{Q}$ relative motion. This can be intuitively explained by using the stochastic Schr\"odinger equation. We suppose the initial wavefunction is $|\psi(t=0)\rangle = |1S\rangle$. Without any stochastic and dissipative terms in the Hamiltonian, the evolution is unitary and the wavefunction stays as the 1S state $|\langle 1S |\psi(t)\rangle|^2 = 1$. However, if stochastic terms are included, the wavefunction will be ``distorted" and become decoherent, which leads to $|\langle 1S |\psi(t)\rangle|^2 < 1$. Furthermore, the wavefunction decoherence can lead to the formation of the 2S state at the same time: $|\langle 2S |\psi(t)\rangle|^2 > 0$, which does not exist initially: $|\langle 2S |\psi(t=0)\rangle|^2 = 0$. This was demonstrated in Refs.~\cite{Akamatsu:2011se,Miura:2019ssi} for some simple models and the probability for a new quarkonium state to form is not small after some time of evolution. This type of quarkonium (re)combination is different from the traditional recombination discussed in the heavy-ion physics community. The traditional recombination involves uncorrelated $Q\bar{Q}$ pairs that are mostly produced from different initial hard collisions. The traditional recombination is enhanced as the collision energy increases since more heavy quarks are produced and the recombination rate depends on the square of the heavy quark density. To distinguish these two types of recombination, the traditional recombination is named uncorrelated recombination while the new type of recombination that is originated from the wavefunction decoherence is named correlated recombination. In the quantum optical and semiclassical limits, correlated recombination can be described by possible recombination of a $Q\bar{Q}$ pair that is from the same dissociated quarkonium state. Ref.~\cite{Yao:2020xzw} shows that correlated recombination will lead to a suppression in the double ratio observable $\frac{R_\text{AA}(\chi_b(1P))}{R_\text{AA}(\Upsilon(2S))}$, which is shown in Fig.~\ref{fig:chi_b} as functions of centrality (or equivalently, $N_{\rm{part}}$) and transverse momentum $p_T$. Since the binding energies of $\Upsilon(2S)$ and $\chi_b(1P)$ are similar, we would expect the ratio to be on the order of unity. However, with correlated recombination, more $\Upsilon(2S)$ states can be regenerated from the dissociated $\chi_b(1P)$ than those $\chi_b(1P)$ states regenerated from the dissociated $\Upsilon(2S)$, since more $\chi_b(1P)$ states are produced initially. As can be seen from Fig.~\ref{fig:chi_b}, the ratio is close to unity without correlated recombination and becomes one third roughly in central collisions after correlated recombination is taken into account. With correlated recombination included, the double ratio increases with the transverse momentum. This prediction of the suppression in the double ratio may be tested in future experiments at RHIC and LHC.

\section{Summary and Outlook}
\label{sect:conclusion}
In this article, I reviewed recent progress of applying the open quantum system framework in the understanding of quarkonium evolution in the QGP. The quantum master equations in both the quantum Brownian motion and the quantum optical limits were explained, together with their semiclassical counterparts. The validity of the time scale hierarchies specifying these two limits was scrutinized by using the separation of energy scales in nonrelativistic effective field theories of QCD. Physical implications for quarkonium transport in the QGP were also discussed. Besides quarkonium transport, the open quantum system framework has also been applied in other areas of high energy physics, such as dark matter formation~\cite{Binder:2020efn}, deeply inelastic reactions~\cite{Braaten:2016sja}, inflation~\cite{Boyanovsky:2015tba,Boyanovsky:2015jen}, jet physics~\cite{Vaidya:2020cyi,Vaidya:2020lih,Vaidya:2021mjl,Neill:2015nya,Balsiger:2018ezi} and small-$x$ physics~\cite{Armesto:2019mna,Li:2020bys}.

Some open questions in the field are worth exploring in the future. The first question is the transition between the quantum Brownian motion limit and the quantum optical limit. The former limit is valid at high temperature while the latter works well at low temperature. The temperature of the QGP created in heavy-ion collisions is changing over time, from high to low. So neither limit is expected to provide a highly accurate description of quarkonium evolution for the whole temperature range and a switch between the two limits is needed. It is not clear how to switch smoothly from one limit to the other limit. However, as mentioned in Section~\ref{sect:open}, these two limits are not always contradictory to each other. In the situation with $\tau_R\gg \tau_S\gg \tau_E$, both limits provide a valid description of quarkonium transport. This situation can be achieved by assuming $Mv \gg T,m_D, Mv^{3/2} \gg Mv^2$. Some qualitative and quantitative studies under this separation of scales will deepen our understanding of the quarkonium dynamics in the transition region between the quantum Brownian motion limit and the quantum optical limit. Quantum transport equations beyond these two limits have not been explored yet for quarkonium traversing the QGP. Quantum computer may provide a way of simulating more general quantum transport equations. Currently only some simple Lindblad equations have been studied on a quantum computer~\cite{hu2020quantum,deJong:2020tvx}\,.

Secondly, the quantum transport equations and their semiclassical counterparts beyond the leading (nontrivial) power in the EFT power counting (of $v$ and $r$) are not known for the moment. Beyond leading power in $v$, the quarkonium wavefunction contains a portion of an octet $Q\bar{Q}$, which has to be taken into account. Also, more terms in the Lagrangian can contribute to the interaction between the $Q\bar{Q}$ pair and the medium, which can lead to more complicated environment correlators. Furthermore, a new type of process, diffusion of quarkonium, which is originated from elastic scattering between quarkonium and the medium, starts to contribute at quadratic power in $r$ (in the amplitude level)~\cite{Yao:2018sgn}. All of these will make the derivation of the Lindblad equations more complicated and may challenge the validity of the Markovian master equations.

Another question is the effect of the QGP flow and anisotropy (viscosity), which should be systematically investigated in the Lindblad equation. The nonrelativistic expansion intensively used by previous studies reviewed here, is only valid in a frame that is close to the rest frame of the $Q\bar{Q}$ pair~\cite{Yao:2018sgn}. However, it is known that the QGP is flowing, and the flow velocity can be as big as $0.7c$. In the rest frame of a $Q\bar{Q}$ pair at low (c.m.) transverse momentum, the medium is moving fast and the effective temperature felt by the $Q\bar{Q}$ pair can be much higher. Then the separation of scales may break down. The EFT developed in Ref.~\cite{Makris:2019ttx} may be useful to derive the Lindblad equation in this situation. Furthermore, if the QGP is anisotropic, the anisotropy has to be included in the calculations of the environment correlators. The c.m. velocity dependence of Debye screening and dissociation has been calculated in Refs.~\cite{Chu:1988wh,Liu:2006nn,Escobedo:2013tca}\,. The anisotropy effects on the screening (both static and dynamical) have also been studied in Refs.~\cite{Dumitru:2007hy,Dumitru:2009ni,Burnier:2009yu,Dumitru:2009fy,Margotta:2011ta,Thakur:2020ifi}\,. It is worth exploring if these studies can be generalized in the open quantum system framework, which unifies Debye screening, dissociation and recombination. 

Finally, we need to carry out nonperturbative calculations of both the momentum independent and momentum dependent chromoelectric structure functions, discussed in Section~\ref{sect:structure_f}. This is important for phenomenological studies since the QGP created in current heavy-ion collisions is strongly-coupled, as mentioned in the Introduction. In the quantum Brownian motion limit, only the zero energy limits of the two momentum independent structure functions (transport coefficients) contribute. So far, only one of them, the heavy quark diffusion coefficient, has been calculated nonperturbatively. (The quoted value of $\gamma$ in Eq.~(\ref{eqn:gamma}) is estimated~\cite{Brambilla:2019tpt} from lattice QCD results of the quarkonium mass shift at finite temperature rather than calculated directly on lattice.) In the quantum optical limit, the finite energy dependence of transport coefficients becomes dominant due to the energy gap between the bound and unbound states. Furthermore, more differential observables of quarkonium suppression will be sensitive to the momentum dependent structure functions. It has been shown in Ref.~\cite{Yao:2020eqy} that the differential reaction rates of quarkonium (dissociation and recombination) depend on the momentum dependent structure function while the inclusive reaction rates depend on the momentum independent structure function. The quantum optical limit may be more important than the quantum Brownian motion limit since the QGP temperature is low in most of its lifetime. (The success of the statistical hadronization model in the description of charmonium production at low transverse momentum in heavy-ion collisions~\cite{BraunMunzinger:2009ih,Andronic:2019wva} also indicates that charmonium production in the late time stage is important, when the temperature is low.) Therefore, precise determination of the momentum independent and momentum dependent structure functions and their finite energy dependence will be useful for phenomenology. On the other hand, these structure functions encode properties of the QGP. If we can extract them from the experimental data of quarkonium suppression, we can deepen our understanding of the QGP. Only in this way, are we really using quarkonium as a probe of the QGP. Precise experimental measurements of high statistics will be greatly helpful for the extraction, which will probably be provided by the sPHENIX program at RHIC and the high-luminosity phase of LHC.

\section*{Acknowledgments}
I would like to thank Yukinao Akamatsu, Nora Brambilla, Miguel Angel Escobedo, Peter Petreczky, Krishna Rajagopal, Michael Strickland and Antonio Vairo for useful comments. This work is supported by the U.S. Department of Energy, Office of Science, Office of Nuclear Physics under grant Contract Number DE-SC0011090.

\appendix

\section{Time Irreversibility, Partial Trace and Relative Entropy}
\label{app:rel_entro}
We provide a short explantion of the time irreversibility of Eq.~(\ref{eqn:reduced}). (This is covered in many textbooks, see e.g. Ref.~\cite{Breuer:2002pc}\,.) To demonstrate this explicitly, we consider the relative entropy between two states of the whole system (subsystem and environment), specified by their density matrices $\rho$ and $\sigma$
\be
S( \rho || \sigma ) \equiv \Tr (\rho \ln\rho) - \Tr( \rho \ln \sigma ) \,.
\ee
The relative entropy is monotonically decreasing under partial trace
\be
S( \rho_S || \sigma_S ) =  S( \Tr_E\rho || \Tr_E\sigma ) \leq S( \rho || \sigma ) \,.
\ee
For simplicity of the discussion, we assume the environment is in thermal equilibrium and the initial total density matrix factorizes $\rho(0) = \rho_S(0)\otimes \rho_E^{\text{eq}}$. We can then define a steady state of the subsystem $\rho_S^{\ma{steady}}$ by
\be
\rho_S^{\ma{steady}} = \Tr_E \big( U(t) ( \rho_S^{\ma{steady}}\otimes \rho_E^\ma{eq}  )  U^\dagger(t)  \big) \,.
\ee
For an arbitrary state of the subsystem $\rho_S(t)$, we find
\be \nn
S( \rho_S(t) || \rho_S^{\ma{steady}} )  &=&  S\big(   \Tr_E( U(t) ( \rho_S(0) \otimes \rho_E^\ma{eq}) U^\dagger(t) ) \big|\big|   \Tr_E( U(t) ( \rho_S^{\ma{steady}} \otimes \rho_E^\ma{eq}) U^\dagger(t) )   \big) \\ \nn
& \leq &  S(  U(t) ( \rho_S(0) \otimes \rho_E^\ma{eq}) U^\dagger(t)  ||    U(t) ( \rho_S^{\ma{steady}} \otimes \rho_E^\ma{eq}) U^\dagger(t)   ) \\ \nn
&=&  S(   \rho_S(0) \otimes \rho_E^\ma{eq}  ||     \rho_S^{\ma{steady}} \otimes \rho_E^\ma{eq}   ) \\ 
&=& S(   \rho_S(0)   ||     \rho_S^{\ma{steady}}   ) \,,
\ee
where in the third line we have used the fact that the relative entropy is invariant under unitary transformations. The inequality implies that the evolution of $\rho_S(t)$ is time-irreversible in general even when the underlying theory respects time reversal symmetry.

\section{Gaussian Smearing of Wigner Transform}
\label{app:smear}
The density matrix is Hermitian and semi-positive definite, so it can be written as
\be
\rho = \sum_n \lambda_n |\psi_n\rangle \langle \psi_n|\,,
\ee
where $\lambda_n \ge 0$. To show the semi-positive definiteness of the Wigner-transformed density matrix, we only need to consider the Wigner transform of a pure state $|\psi\rangle\langle\psi|$. The semi-positive definite Wigner function for a pure state has been constructed by convoluting the Wigner transform with a Gaussian function~\cite{CARTWRIGHT1976210}
\be
&& W_G(x,p) = \frac{\sqrt{ab}}{\pi}\int \frac{\diff x' \diff p'}{2\pi} W(x',p') \exp\pig( -a(x-x')^2 - b(p-p')^2 \pig) \nn\\
&=& \frac{\sqrt{ab}}{2\pi^2} \int \diff x' \diff p' \diff y 
\,\exp\pig( -ip' y -a(x-x')^2 - b(p-p')^2 \pig) \psi\big(x'+\frac{y}{2}\big) \psi^*\big(x'-\frac{y}{2}\big) \nn\\
&=& \frac{\sqrt{\pi a}}{2\pi^2} \int \diff x' \diff y \,\exp\pig( -a(x-x')^2 -ipy - \frac{y^2}{4b}\pig)\psi\big(x'+\frac{y}{2}\big) \psi^*\big(x'-\frac{y}{2}\big) \,,
\ee
where $a>0,b>0$.
Defining $x_1=x'+\frac{y}{2}$, $x_2=x'-\frac{y}{2}$ and
\be
f(x_i) = \psi(x_i)\,\exp\pig( -\frac{a}{4}x_i^2 + axx_i -ipx_i - \frac{1}{4b}x_i^2 \pig)\,,
\ee
we find
\be
W_G(x,p) = \frac{\sqrt{\pi a}}{2\pi^2} e^{-ax^2} \sum_n\frac{1}{n!}\Big( \frac{1}{2b}-\frac{a}{2}\Big)^n
\int \diff x_1 \, x_1^n f(x_1) \int \diff x_2 \, x_2^n f^*(x_2)\,.
\ee
We see that if $ab<1$, $W_G\ge0$. The inequality implies that for semi-positive definiteness, the Gaussian function used in the smearing mush be wider than the minimum phase space volume, which is determined by the uncertainty principle.

\section{Environment Correlators}
\label{app:correlator}
In the limit of quantum Brownian motion, the relevant environment correlators are given by (see Section~\ref{sect:highT})
\be
D^{>ab}(x_1,x_2) = D^{<ba}(x_2,x_1) &=& g^2 \langle A_0^a(t_1,{\bs x}_1) A_0^b(t_2,{\bs x}_2) \rangle_T \\
 \Sigma^{ab}(x_1,x_2) &=& -i g^2 \sign(t_1-t_2) \langle A_0^a(t_1,{\bs x}_1) A_0^b(t_2,{\bs x}_2) \rangle_T \,.\ \ \ 
\ee
If we assume the environment is invariant under spacetime translation, we have $D^{>ab}(x_1,x_2)=D^{>ab}(x_1-x_2,0)= D^{>ab}(x_1-x_2)$ and similarly for $\Sigma^{ab}$. If we Fourier transform only in time, we find
\be
\label{eqn:D>D<}
D^{>ab}(q_0,{\bs x}) = D^{<ab}(-q_0,{\bs x}) \,,
\ee
where we have used $D^{>ab}\propto \delta^{ab}$ and $D^{<ba}\propto \delta^{ab}$.
At thermal equilibrium, the two Wightman functions are related by the Kubo–Martin–Schwinger (KMS) relation:
\be
D^{>ab}(q_0,{\bs x}) = e^{\beta q_0} D^{<ab}(q_0,{\bs x}) \,.
\ee
In the quantum Brownian motion limit, only the zero energy limit contributes, at which the two Wightman functions are the same
\be
\lim_{q_0\to0} D^{>ab}(q_0,{\bs x}) = \lim_{q_0\to0} D^{<ab}(q_0,{\bs x}) \,.
\ee
From the KMS relation, we also find
\be
\frac{\partial D^{>ab}(q_0,{\bs x})}{\partial q_0}\bigg|_{q_0=0} =  \frac{\partial D^{<ab}(q_0,{\bs x})}{\partial q_0}\bigg|_{q_0=0} + \beta D^{<ab}(q_0=0,{\bs x}) \,.
\ee
Combining with Eq.~(\ref{eqn:D>D<}) leads to
\be
\label{eqn:D>final}
\frac{\partial D^{>ab}(q_0,{\bs x})}{\partial q_0}\bigg|_{q_0=0} = \frac{\beta}{2} D^{<ab}(q_0=0,{\bs x})\,.
\ee
Collecting all the formulas, we find the next-leading order (in $H_S$) term of $\widetilde{O}^{(S)}_\alpha$ in Eq.~(\ref{eqn:brown_final}) can be simplified as
\be
\label{eqn:2T}
&& D^{-1}_{\alpha\beta}(\omega=0)\frac{\partial D_{\beta\gamma}(\omega=0) }{\partial \omega}  \big[H_S, O_\gamma^{(S)} \big]
\to \frac{1}{2T} \big[H_S, O_\alpha^{(S)} \big] \,.
\ee

Using the definition of the second correlator, we find $\Sigma^{ab*}(x) = \Sigma^{ba}(-x)$ and $\Sigma^{ab*}(q) = \Sigma^{ba}(q)$. If the environment is spherically symmetric, we have $\Sigma^{ab}(t,{\bs x}) = \Sigma^{ab}(t,-{\bs x}) = \Sigma^{ab}(t,|{\bs x}|)$. So we can write
\be
\Sigma^{ab}(q_0=0,{\bs x}) &=& \int_{-\infty}^{+\infty}\diff t\, \Sigma^{ab}(t,{\bs x}) = \int_{-\infty}^{+\infty}\diff t\, \frac{\Sigma^{ab}(t,{\bs x}) + \Sigma^{ba*}(-t,-{\bs x})}{2} \nn\\
&=& \int_{-\infty}^{+\infty}\diff t\, \frac{\Sigma^{ab}(t,{\bs x}) + \Sigma^{ba*}(t,{\bs x})}{2} \nn\\
&=& g^2 \int_{-\infty}^{+\infty} \diff t\, {\rm Im}\langle \ml{T}A^a_0(x)A^b_0(0) \rangle_T \,,
\ee
where we have flipped the sign of $t$ in $\Sigma^{ba*}$ in the second line and used $\langle A^a_0(x)A^b_0(0) \rangle_T \propto \delta^{ab}$ and the spherical symmetry. Next we evaluate $\frac{\partial}{\partial q_0} \Sigma^{ab}(q_0=0,{\bs x})$. To this end, we write 
\be
\Sigma^{ab}(q_0,{\bs x}) = -ig^2 \int \diff t \,e^{iq_0t}\big( 2\theta(t)-1 \big) \langle A_0^a(t,{\bs x}) A_0^b(0) \rangle_T\,,\ \ \ \ \ 
\ee
where we have used $\sign(t)=2\theta(t)-1$. Using
\be
\theta(t) = \int\frac{\diff k_0}{2\pi} e^{-ik_0t} \frac{i}{k_0+i\epsilon}\,,
\ee
we find $\Sigma^{ab}(q_0,{\bs x})$ can be written as
\be
\Sigma^{ab}(q_0,{\bs x}) &=& 2 \int  \frac{\diff t \diff k_0}{2\pi} e^{iq_0t-ik_0t} \frac{1}{k_0+i\epsilon} D^>(t,{\bs x}) + i \int \diff t\,e^{iq_0t} D^>(t,{\bs x}) \nn\\
&=& 2 \int \frac{\diff k_0}{2\pi} \frac{1}{k_0+i\epsilon} D^>(q_0-k_0,{\bs x}) + i D^>(q_0,{\bs x}) \nn\\
&=& 2\, \mathbb{P} \int\frac{\diff k_0}{2\pi}\frac{1}{k_0} D^>(q_0-k_0,{\bs x}) = 2\, \mathbb{P} \int\frac{\diff k_0}{2\pi}\frac{1}{q_0-k_0} D^>(k_0,{\bs x})\,,\ \ \ \ \ 
\ee
where $\mathbb{P}$ denotes the principal value. The derivative can be written as
\be
\frac{\partial}{\partial q_0} \Sigma^{ab}(q_0=0,{\bs x}) &=& -2\, \mathbb{P} \int\frac{\diff k_0}{2\pi}\frac{1}{k^2_0} D^>(k_0,{\bs x}) \,.
\ee

\section{Projection onto Position Space}
\label{app:projection}
To illustrate the mathematics used when the Lindblad equation in the quantum Brownian motion limit is projected onto the position space, we consider a simple example of a heavy fermion in the U(1) gauge theory. The subsystem and the interaction Hamiltonians are given by
\be
H_S &=& \frac{\hat{\bs p}_f^2}{2M} \\
H_I &=& \int\diff^3x\, \delta^3({\bs x}-{\bs x}_f) \,gA_0({\bs x}) \,.
\ee
The operators $O^{(S)}({\bs x})$, $\widetilde{O}^{(S)}({\bs x})$ and $\widetilde{O}^{(S)\dagger}({\bs x})$ are
\be
O^{(S)}({\bs x}) &=& \delta^3({\bs x}-\hat{\bs x}_f) \\
\widetilde{O}^{(S)}({\bs x}) &=& \delta^3({\bs x}-\hat{\bs x}_f)  -\frac{1}{4T} \big[H_S, \delta^3({\bs x}-\hat{\bs x}_f)\big] \\\nn
&=& \delta^3({\bs x}-\hat{\bs x}_f)  +\frac{1}{4T}\Big( \frac{1}{2M}\pig( \nabla_{{\bs x}_f}^2\delta^3({\bs x}-\hat{\bs x}_f) \pig) + \frac{1}{M} \pig( \nabla_{{\bs x}_f}\delta^3({\bs x}-\hat{\bs x}_f) \pig)  \cdot \nabla_{{\bs x}_f} \Big) \\
\widetilde{O}^{(S)\dagger}({\bs x}) &=& \delta^3({\bs x}-\hat{\bs x}_f)  +\frac{1}{4T} \big[H_S, \delta^3({\bs x}-\hat{\bs x}_f)\big] \\\nn
&=& \delta^3({\bs x}-\hat{\bs x}_f)  -\frac{1}{4T}\Big( \frac{1}{2M}\pig( \nabla_{{\bs x}_f}^2\delta^3({\bs x}-\hat{\bs x}_f) \pig) + \frac{1}{M} \pig( \nabla_{{\bs x}_f}\delta^3({\bs x}-\hat{\bs x}_f) \pig)  \cdot \nabla_{{\bs x}_f} \Big) \,.
\ee
Here by the notation $(\nabla_{{\bs x}_f} \cdots)$, we mean the operator $\nabla_{{\bs x}_f} $ only acts inside the parentheses. If there are no parentheses, $\nabla_{{\bs x}_f}$ acts on everything on its right. The dot product $\cdot$ is originated from the contraction in $(\nabla_i \cdots)\nabla_i$.
We want to show some details in the derivation of the terms that are linear in $H_S$ in
\be
\int\diff^3x \diff^3y \, D({\bs x},{\bs y}) 
\Big( \widetilde{O}^{(S)}({\bs y})\rho_S \widetilde{O}^{(S)\dagger}({\bs x}) 
-\frac{1}{2}\pig\{ \widetilde{O}^{(S)\dagger}({\bs x})\widetilde{O}^{(S)}({\bs y}),\,\rho_S\pig\}
\Big) \,.
\ee
The terms of $\widetilde{O}^{(S)}({\bs y})\rho_S \widetilde{O}^{(S)\dagger}({\bs x}) $ that are linear in $H_S$ can be written as
\be
\label{eqn:oneQ+}
&&\frac{1}{4T}\int \diff^3x \diff^3y \, D({\bs x},{\bs y}) \\
&\times&  \bigg( \Big( \frac{1}{2M}\big( \nabla_{{\bs x}_f}^2\delta^3({\bs y}-\hat{\bs x}_f) \big) + \frac{1}{M} \big( \nabla_{{\bs x}_f}\delta^3({\bs y}-\hat{\bs x}_f) \big) \cdot \nabla_{{\bs x}_f} \Big) 
\rho_S \delta^3({\bs x}-\hat{\bs x}_f) \nn\\
&-& \delta^3({\bs y}-\hat{\bs x}_f) \rho_S
\Big( \frac{1}{2M}\big( \nabla_{{\bs x}_f}^2\delta^3({\bs x}-\hat{\bs x}_f) \big) + \frac{1}{M} \big( \nabla_{{\bs x}_f}\delta^3({\bs x}-\hat{\bs x}_f) \big) \cdot \nabla_{{\bs x}_f} \Big) 
\bigg) \nn\\
&=& \frac{1}{8MT} \int \diff^3x \diff^3y \,
 D({\bs x},{\bs y})
\big(\nabla_{{\bs x}_f}^2 \delta^3({\bs y}-\hat{\bs x}_f) \big) \rho_S
\delta^3({\bs x}-\hat{\bs x}_f) \nn\\
&-& \frac{1}{8MT} \int \diff^3x \diff^3y \,
 D({\bs x},{\bs y})
\delta^3({\bs y}-\hat{\bs x}_f) \rho_S
\big(\nabla_{{\bs x}_f}^2 \delta^3({\bs x}-\hat{\bs x}_f) \big) \nn\\
&+& \frac{1}{4MT}  \int \diff^3x \diff^3y \, D({\bs x},{\bs y})\Big( 
\big( \nabla_{{\bs x}_f}\delta^3({\bs y}-\hat{\bs x}_f) \big)
\rho_S \cdot \big( \nabla_{{\bs x}_f} \delta^3({\bs x}-\hat{\bs x}_f) \big)\nn\\
&+& \big( \nabla_{{\bs x}_f}\delta^3({\bs y}-\hat{\bs x}_f) \big) \cdot \big(\nabla_{{\bs x}_f}\rho_S\big) \delta^3({\bs x}-\hat{\bs x}_f) - \delta^3({\bs y}-\hat{\bs x}_f)\rho_S
\big( \nabla_{{\bs x}_f}\delta^3({\bs x}-\hat{\bs x}_f) \big) \cdot \nabla_{{\bs x}_f}
\Big) \nn
\,.
\ee
For the linear terms in $-\frac{1}{2}\{ \widetilde{O}^{(S)\dagger}({\bs x})\widetilde{O}^{(S)}({\bs y}),\,\rho_S\}$, we have
\be
\label{eqn:oneQ-}
&&-\frac{1}{2}\int \diff^3x \diff^3y \, D({\bs x},{\bs y}) \bigg( \\
&& 
-\frac{1}{4T}\Big( \frac{1}{2M}\pig( \nabla_{{\bs x}_f}^2\delta^3({\bs x}-\hat{\bs x}_f) \pig) + \frac{1}{M} \pig( \nabla_{{\bs x}_f}\delta^3({\bs x}-\hat{\bs x}_f) \pig) \cdot \nabla_{{\bs x}_f} \Big)
\delta^3({\bs y}-\hat{\bs x}_f) \rho_S \nn\\
&&-\frac{1}{4T} \rho_S \Big( \frac{1}{2M}\pig( \nabla_{{\bs x}_f}^2\delta^3({\bs x}-\hat{\bs x}_f) \pig) + \frac{1}{M} \pig( \nabla_{{\bs x}_f}\delta^3({\bs x}-\hat{\bs x}_f) \pig)  \cdot \nabla_{{\bs x}_f} \Big)
\delta^3({\bs y}-\hat{\bs x}_f) \nn\\
&& + \frac{1}{4T} \delta^3({\bs x}-\hat{\bs x}_f)
\Big( \frac{1}{2M}\pig( \nabla_{{\bs x}_f}^2\delta^3({\bs y}-\hat{\bs x}_f) \pig) + \frac{1}{M} \pig( \nabla_{{\bs x}_f}\delta^3({\bs y}-\hat{\bs x}_f) \pig) \cdot \nabla_{{\bs x}_f} \Big) \rho_S \nn\\
&& + \frac{1}{4T} \rho_S \delta^3({\bs x}-\hat{\bs x}_f)
\Big( \frac{1}{2M}\pig( \nabla_{{\bs x}_f}^2\delta^3({\bs y}-\hat{\bs x}_f) \pig) + \frac{1}{M} \pig( \nabla_{{\bs x}_f}\delta^3({\bs y}-\hat{\bs x}_f) \pig) \cdot \nabla_{{\bs x}_f} \Big)
\bigg) \nn\\
&=& \frac{1}{8MT}\int \diff^3x \diff^3y \, D({\bs x},{\bs y}) \bigg( 
 \pig( \nabla_{{\bs x}_f}\delta^3({\bs x}-\hat{\bs x}_f) \pig) \cdot \pig(\nabla_{{\bs x}_f}
\delta^3({\bs y}-\hat{\bs x}_f)\pig) \rho_S  \nn\\
&&+ \rho_S \pig( \nabla_{{\bs x}_f}\delta^3({\bs x}-\hat{\bs x}_f) \pig) \cdot  \pig( \nabla_{{\bs x}_f} 
\delta^3({\bs y}-\hat{\bs x}_f)  \pig)
\bigg) \nn\,,
\ee
where we have used $D({\bs x}, {\bs y}) = D({\bs x} - {\bs y}) = D(|{\bs x} - {\bs y}|) $.

To project onto the position basis, we need the following expressions
\be
&& \langle {\bs r} | \nabla \rho_S |{\bs r}'\rangle = \nabla_{\bs r} \langle {\bs r} |\rho_S |{\bs r}'\rangle \\[4pt]
&& \langle {\bs r} |  \rho_S \nabla |{\bs r}'\rangle = -\big( \langle {\bs r}' | \nabla \rho_S |{\bs r}\rangle \big)^\dagger = - \nabla_{{\bs r}'} \langle {\bs r} |\rho_S |{\bs r}'\rangle \\[4pt]
&& \langle{\bs r} | \big( \nabla_{{\bs x}_f}\delta^3({\bs y}-\hat{\bs x}_f) \big)
\rho_S \cdot \big( \nabla_{{\bs x}_f} \delta^3({\bs x}-\hat{\bs x}_f) \big) | {\bs r}'\rangle \nn\\
&=& \big( \nabla_{{\bs r}}\delta^3({\bs y}-{\bs r} ) \big)
\cdot \big( - \nabla_{{\bs r}'} \delta^3({\bs x}-{\bs r}') \big)
\langle {\bs r}|\rho_S| {\bs r}'\rangle  \nn\\
&=& - \big( \nabla_{{\bs y}}\delta^3({\bs y}-{\bs r} ) \big)
\cdot \big( \nabla_{{\bs x}} \delta^3({\bs x}-{\bs r}') \big)
\langle {\bs r}|\rho_S| {\bs r}'\rangle \\[4pt]
&& \langle{\bs r} | \big( \nabla_{{\bs x}_f}\delta^3({\bs y}-\hat{\bs x}_f) \big) \cdot
\big( \nabla_{{\bs x}_f} \delta^3({\bs x}-\hat{\bs x}_f) \big) \rho_S | {\bs r}'\rangle \nn\\
&=& \big( \nabla_{{\bs r}}\delta^3({\bs y}-{\bs r} ) \big)
\cdot \big( \nabla_{{\bs r}} \delta^3({\bs x}-{\bs r}) \big)
 \langle {\bs r}|\rho_S| {\bs r}'\rangle  \nn\\
&=&  \big( \nabla_{{\bs y}}\delta^3({\bs y}-{\bs r} ) \big)
\cdot \big( \nabla_{{\bs x}} \delta^3({\bs x}-{\bs r}) \big)
\langle {\bs r}|\rho_S| {\bs r}'\rangle  \,.
\ee
Sandwiching Eq.~(\ref{eqn:oneQ+}) between $\langle {\bs r}|$ and $| {\bs r}' \rangle$, we find it becomes
\be
\label{eqn:Dr}
&& \frac{1}{4MT}\Big( - \big(\nabla_{\bs r} \cdot \nabla_{{\bs r}'} D({\bs r}',{\bs r}) \big) \langle {\bs r}| \rho_S | {\bs r}' \rangle
+ \big( \nabla_{{\bs r}} D({\bs r}',{\bs r}) \big) \cdot \nabla_{\bs r}
\langle {\bs r}| \rho_S | {\bs r}' \rangle \\
&+& \big( \nabla_{{\bs r}'} D({\bs r}',{\bs r}) \big) \cdot \nabla_{{\bs r}'} \langle {\bs r}| \rho_S | {\bs r}' \rangle \Big) \nn\\
&=& \frac{1}{4MT} \Big( \big( \nabla_{{\bs r}}^2 D({\bs r}',{\bs r}) \big)  \langle {\bs r}| \rho_S | {\bs r}' \rangle + \big( \nabla_{{\bs r}} D({\bs r}',{\bs r}) \big) 
\cdot \big( \nabla_{{\bs r}} - 
\nabla_{{\bs r}'} \big) \langle {\bs r}| \rho_S | {\bs r}' \rangle
\Big) \nn\,,
\ee
where we have used $D({\bs x},{\bs y}) = D(|{\bs x}-{\bs y}|)$. Similarly we find Eq.~(\ref{eqn:oneQ-}) gives
\be
\label{eqn:D0}
&&\frac{1}{8MT} \Big( \big( \nabla_{\bs x} \nabla_{\bs y} D({\bs x},{\bs y}) \big)\Big|_{{\bs x}={\bs y}={\bs r}} \langle {\bs r}| \rho_S | {\bs r}' \rangle +  \big( \nabla_{\bs x} \nabla_{\bs y} D({\bs x},{\bs y}) \big)\Big|_{{\bs x}={\bs y}={\bs r}'} \langle {\bs r}| \rho_S | {\bs r}' \rangle \Big) \nn\\
&=& - \frac{1}{4MT} \nabla^2 D(0) \langle {\bs r}| \rho_S | {\bs r}' \rangle \,,
\ee
where we have used $D({\bs x},{\bs y}) = D(|{\bs x}-{\bs y}|)$ again and defined 
\be
\nabla^2 D(0) = \nabla^2_{{\bs x}} D({\bs x}) \Big|_{{\bs x}=0} = -\nabla_{{\bs x}} \nabla_{{\bs y}} D({\bs x}-{\bs y}) \Big|_{{\bs x}={\bs y}} \,.
\ee
Eqs.~(\ref{eqn:Dr}, \ref{eqn:D0}) agree with Eq.~(34) in Ref.~\cite{Blaizot:2017ypk} (notice the difference of the sign convention in the definition of environment correlators). The Lindblad equation describing a $Q\bar{Q}$ pair projected onto the position space can be found in Ref.~\cite{Blaizot:2017ypk}\,.

\bibliographystyle{ws-ijmpa}
\bibliography{main.bib}
\end{document}